\RequirePackage{rotating}
\documentclass[a4paper,fleqn,usenatbib]{mnras}

\usepackage[T1]{fontenc}
\usepackage{ae,aecompl}

\usepackage{graphicx}	
\usepackage{amsmath}	
\usepackage{amssymb}	
\usepackage{tikz,graphics,color,float,epsf,caption,subcaption}
\usepackage{rotating}
\usepackage{lscape} 
\usepackage{rotating}
\usepackage{gensymb}
\usepackage{tikz-qtree}
\usepackage{tablefootnote}
\usepackage{soul}

\newcommand{\nsources}{833 }

\newcommand{\nprotocores}{124 }
\newcommand{\nprestellarcores}{604 }
\newcommand{\ntentativecores}{149 }
\newcommand{\nrobustcores}{455 }
\newcommand{\nstarlesscores}{709 }
\newcommand{\nproto}{124 }
\newcommand{\nunboundcores}{105 }

\newcommand{\preOverstarless}{$\sim 85\%$}
\newcommand{\robOverstarless}{$\sim 64\%$}

\newcommand{\get}{{\it getsources} }
\newcommand{\onfil}{$69\%$ }

\newcommand{\onfildistpre}{$81\% $ }
\newcommand{\onfildistun}{$5\%$ }

\newcommand{\onfilall}{$82\%$ }
\newcommand{\onfilpre}{$92\%$ }
\newcommand{\onfilun}{$8\%$ }

\newcommand{\cmfslope}{ $-2.05 \pm 0.34$}
\newcommand{\cmfslopeless}{-1.05 \pm 0.34}
\newcommand{\cmfslopeSM}{- 1.95 \pm 0.58}
\newcommand{\cmfslopelessSM}{-0.95 \pm 0.58}
\newcommand{\cmfslopeAE}{- 1.95 \pm 0.41}
\newcommand{\cmfslopelessAE}{- 0.95 \pm 0.41}

\newcommand{\totalmass}{49400} 

\title[The census of dense cores in the Serpens region from the HGBS]{The census of dense cores in the Serpens region from the \emph{Herschel} Gould Belt Survey}

\author[E. Fiorellino et al.]{E. Fiorellino$^{1,2,3,4}$\thanks{E-mail: eleonora.fiorellino@inaf.it}, D. Elia$^{3}$, Ph. André$^{5}$, A. Men'shchikov$^{5}$, S. Pezzuto$^{3}$, 
\newauthor E. Schisano$^{3}$, V. K\"onyves$^{6}$, D. Arzoumanian$^{7}$, M. Benedettini$^{3}$, D. Ward-
\newauthor Thompson$^{6}$, A. Bracco$^{8}$, J. Di Francesco$^{9,10}$, S. Bontemps$^{11,12}$, J. Kirk$^{6}$,
\newauthor F. Motte$^{13}$, S. Molinari$^{3}$.
\\ \\
$^{1}$Dipartimento di Fisica, Università di Roma `Tor Vergata' Via della Ricerca Scientifica 1, 00133, Roma, Italy\\
$^{2}$INAF - Osservatorio Astronomico di Roma, via di Frascati 33, 00078, Monte Porzio Catone, Italy\\
$^{3}$INAF - Istituto di Astrofisica e Planetologia Spaziali (IAPS), via Fosso del Cavaliere 100, 00133 Roma, Italy\\
$^{4}$ESO/European Southern Observatory, Karl-Schwarzschild-Str. 2, D-85748 Garching bei Munchen, Germany \\
$^{5}$Laboratoire d'Astrophysique (AIM), CEA/DRF, CNRS, Universit\'e Paris-Saclay, Universit\'e Paris Diderot, Sorbonne Paris Cit\'e, \\91191 Gif-sur-Yvette, France\\
$^{6}$Jeremiah Horrocks Institute, University of Central Lancashire, Preston PR12HE, UK \\
$^{7}$Instituto de Astrof\'isica e Ci\^encias do Espa\c{c}o, Universidade do Porto, CAUP, Rua das Estrelas, PT4150-762 Porto, Portugal \\
$^{8}$Rudjer Bošković Institute, Bijenička cesta 54, 10000 Zagreb, Croatia\\
$^{9}$Department of Physics and Astronomy, University of Victoria, PO Box 355, STN CSC, Victoria, BC, V8W 3P6, Canada\\
$^{10}$National Research Council Canada, 5071 West Saanich Road, Victoria, BC, V9E 2E7, Canada\\
$^{11}$Univ. Bordeaux, LAB, UMR5804, 33270 Floirac, France\\
$^{12}$CNRS, LAB, UMR5804, 33270 Floirac, France\\
$^{13}$University Grenoble Alpes, Centre National de la Recherche Scientifique (CNRS), Institut de Plan\'etologie et d'Astrophysique \\
de Grenoble, F-38000 Grenoble, France \\
}
\date{Accepted XXX. Received YYY; in original form ZZZ}

\pubyear{2020}

\begin{document}
\label{firstpage}
\pagerange{\pageref{firstpage}--\pageref{lastpage}}

\maketitle

\begin{abstract}
The \emph{Herschel} Gould Belt survey mapped the nearby ($d < 500$~pc) star-forming regions to understand better how the prestellar phase influences the star formation process.
Here we report a complete census of dense cores in a $\sim 15~{\rm deg}^2$ area of the Serpens star-forming region located between $d \sim 420$~pc and 484~pc.
The PACS and SPIRE cameras imaged this cloud from 70~$\mu$m to 500~$\mu$m. 
With the multi-wavelength source extraction algorithm \get, we extract \nsources sources, of which \nstarlesscores are starless cores and \nproto are candidate proto-stellar cores. 
We obtain temperatures and masses for all the sample, classifying the starless cores in \nprestellarcores prestellar cores and \nunboundcores unbound cores. 
Our census of sources is $80\%$ complete for $M > 0.8$~M$_\odot$ overall. 
We produce the core mass function (CMF) and compare it with the initial mass function (IMF).
The prestellar CMF is consistent with log-normal trend up to $\sim 2$~M$_\odot$, after which it follows a power-law with slope of \cmfslope.
The tail of its CMF is steeper but still compatible with the IMF for the region we studied in this work.
We also extract the filaments network of the Serpens region, finding that 
\onfildistpre of prestellar cores lie on filamentary structures.
The spatial association between cores and filamentary structure supports the paradigm, suggested by other \emph{Herschel} observations, that prestellar cores mostly form on filaments. 
Serpens is confirmed to be a young, low-mass and active star-forming region.
\end{abstract}

\begin{keywords}
 stars: formation - ISM: clouds - ISM: structure - ISM: individual objects (Serpens) - infrared: ISM - submillimetre: ISM
\end{keywords}


\section{INTRODUCTION} \label{intro}

Solar type stars form from the collapse of compact structures called prestellar cores, but how the physical properties of stars are determined during the prestellar core phase is still a matter of debate.
We want to contribute to this debate by analyzing the \emph{Herschel} Gould Belt Survey \citep[HGBS,][]{and10} observations of the Serpens star-forming region, which is deeply described in some detail in Sect.~\ref{serpreg}.

The HGBS is one of the largest projects with the \emph{Herschel} Space Observatory \citep{pil10}, whose main objectives are to take a deep census of prestellar cores in nearby clouds, determine a reliable prestellar CMF, and investigate the link between the CMF and the IMF in detail, based on the observations of the nearest ($d \le 500$~pc) star-forming clouds in the Milky Way. 

\emph{Herschel} observed with its far-infrared and sub-millimeter cameras PACS \citep{pog10} and SPIRE \citep{gri10}, {respectively,} with unprecedented sensitivity and resolution in the range between 70~$\mu$m and 500~$\mu$m. 
In this wavelength range the spectral energy distribution (SED) of cold dust ($T \sim 10$~K) is expected to have its peak ($\lambda \sim 250$~$\mu$m).
The {\it Herschel} sensitivity makes it possible to observe sources never detected before in the Serpens~Main region, allowing a more complete and accurate sampling of the Serpens~Main core population compared to the previous studies based on statistics of few tens of objects. 
Moreover, the \emph{Herschel} multi-wavelength analysis enables to obtain both temperature and mass of each core from its SED, producing more accurate physical values with respect to single-wavelength pre-\emph{Herschel} works that necessarily assumed a common temperature for the entire sample of cores \citep{tes98,eno07}.

The twofold purpose of our work, as a part of the HGBS project, is to characterize the earliest stages of star formation in the Serpens region and to verify whether accompanying results about its prestellar core mass function (CMF) are in agreement or not with ones obtained for other clouds and with previous studies of the Serpens~Main sub-region. 
\\ \\
The paper is organised as follows: in the rest of this section we introduce the Serpens area surveyed by \emph{Herschel} (Sect.~\ref{serpreg}) and the concept of the CMF (Sect.~\ref{sect:introcmf}), respectively.
In Sect.~\ref{observation}, we describe the observations and data reduction. 
In Sect.~\ref{coldenst}, we present temperature and column density maps. 
In Sect.~\ref{source}, we set out the detection of the sources, the production of the catalogue, the fit of a modified black-body to SEDs, and the analysis of the obtained core physical parameters. 
In Sect.~\ref{cmfsection}, we report and discuss the CMF of prestellar cores. 
In Sect.~\ref{comparison}, we focus on the analysis of the Serpens~Main sub-region. 
In Sect.~\ref{filamensSection}, we illustrate the filamentary structure of this region and its relationship with the prestellar core spatial distribution. 
Finally, in Sect.~\ref{conclusions} we draw conclusions from this work.

\subsection{The Serpens Region}
\label{serpreg}
The Serpens region extends over several square degrees around the variable star VV-Ser.
It is part of the Aquila~Rift cloud complex, through the middle of the plane of the Milky Way. 
It was firstly recognized as a star formation site by \citet{str74}.

\begin{figure*}
\includegraphics[width=0.99\textwidth]{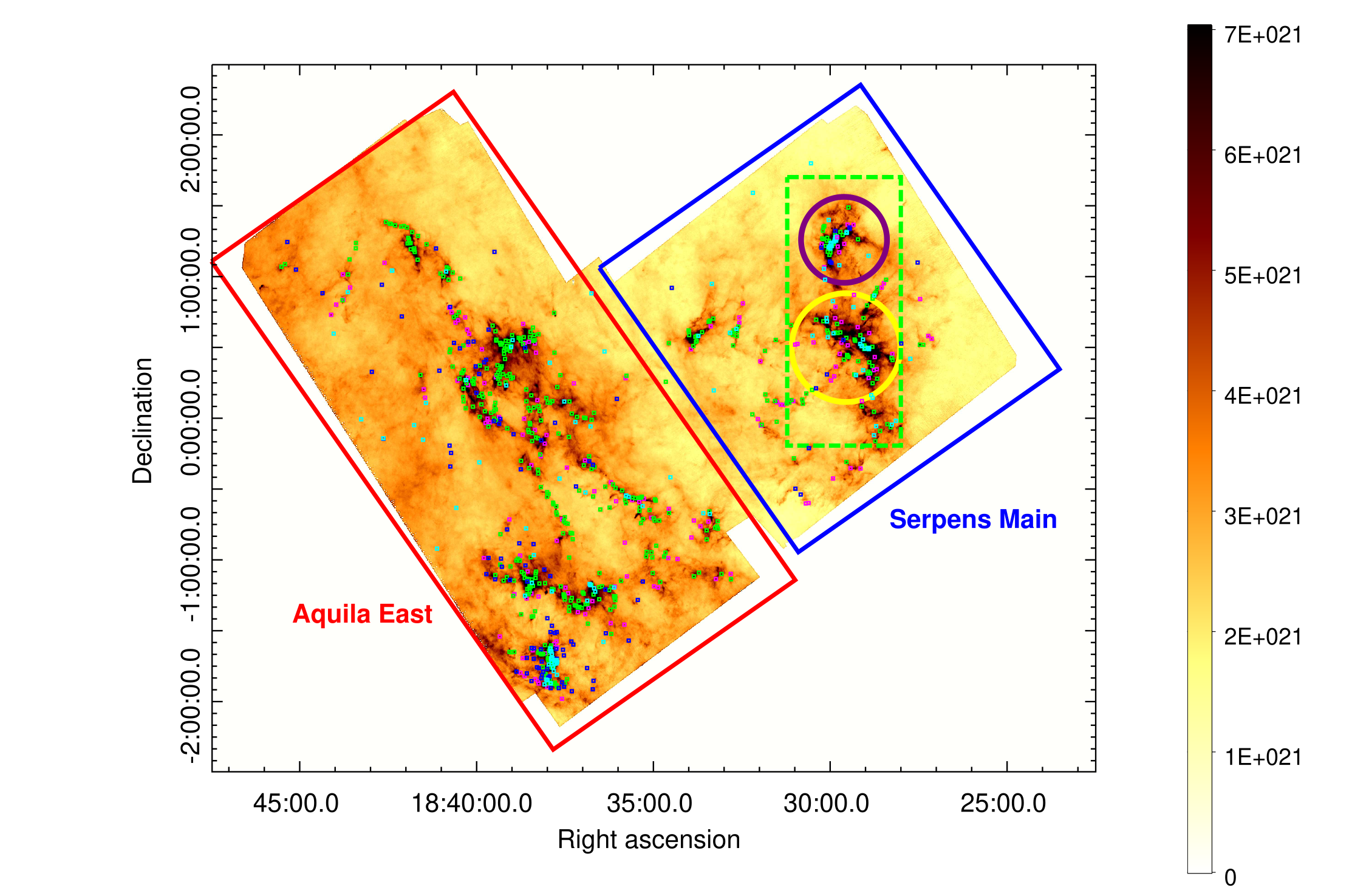}
 \caption{\label{fig:nh2}High-resolution (18.2$\arcsec$) column density map of the Serpens complex observed by \emph{Herschel} (Sect.~\ref{coldenst}), with the sources extracted by \get algorithm (see Sect.~\ref{observation}). 
 Protostellar cores, unbound starless cores, robust and candidate prestellar cores are plotted with cyan, blue, green and magenta symbols, respectively. 
 The color bar reports the column density in cm$^{-2}$. 
 The blue square encompasses the Serpens~Main in which Cluster~A (purple circle) and Cluster~B (yellow circle) are located \citep{dju06,har06,lev13}; the red rectangle contains the Aquila~East. 
 The green dashed box shows the part of the Serpens region we select to compare our results about Serpens~Main with those from previous studies (Sect.~\ref{comparison}).
 }
\end{figure*}

Fig.~\ref{fig:nh2} shows the region mapped by \emph{Herschel}, a wider region (the blue box region) with respect to that ascribed to the Serpens in the literature (i.e., Serpens~Main, the ``Green box" region in Fig.~\ref{fig:nh2}). 
We will refer to this sub-region as Serpens~Main in the following. 
Therefore in this section, when reviewing previous literature about Serpens, we write essentially about Serpens~Main, while the \emph{Herschel} observations presented in this paper also include a large portion the easternmost part of the Aquila~Rift. 
We referred to this last portion of the sky as Aquila~East, which has been poorly studied so far. 
A portion of the Aquila~East region has been analyzed by \citet[][]{her19} in their overview of the structure of Serpens/Aquila~East complex with {\it Gaia} data. This region was also observed as a part of the {\it Spitzer} Gould Belt survey, whose  catalog of young stellar objects (YSOs) was presented by \citet{dun15}. 
Finally, Aquila~East was surveyed in CO by \citet[][]{nak17}.
This work represents an occasion to give a further look to early stages of star formation in this region, thanks to {\it Herschel} far-infrared/sub-mm observations.

Notice that the portion of the Aquila~Rift analyzed in \citet{kon15} is distinct from the field studied in this work. 
In the following we will name this region Aquila~Main to avoid confusion between these two different parts of the area.

The mass of Serpens~Main region has been estimated by several authors, but based on different distance assumptions (see Sect.~\ref{distSerp}), different tracers and corresponding calculation techniques, and, finally, considering different areas. 
The comparison between these mass values and our results is reported in Sect.~\ref{coldenst}.

The Serpens~Main region contains two sub-clumps \citep{tes00,olm02}: the nort-west (NW) one, called Serpens Core or Cluster~A, and the south-east (SE) one, called Serpens G3-66, or Cluster~B \citep{dju06,har06,lev13}. 
Cluster~B has more prominent filaments and more complicated kinematics than Cluster~A. 
Moreover, it has a higher degree of hierarchy \citep{lee14}. 
Between these two clusters, \citet{eno07} found an elongated filament which contains several bright sources.

The Serpens complex is generally classified in the literature as a low-mass star forming region \citep{eno07,eir08}. 
Recently \citet{roc15} suggested that Serpens Main could host high-mass star formation. 
However, \citet{nak17}, considering column density thresholds \citep[][see Sect.~\ref{estimate}]{kru08,kau10}, conclude that none of the clumps presents necessary conditions for the formation of high-mass stars.
According to \citet{cas93} and \citet{kaa04}, star formation in the Serpens~Main region occurred in several steps, because of the presence of evolved stars, young stellar objects, prestellar cores and a relevant amount of dust which in turn traces the presence of molecular gas available for future star formation activity \citep{nak17}. 
Nevertheless, \citet{tes00} described a different scenario in which there is no need for more than one star formation event.

The filamentary structure of Serpens~Main is composed of both gravitationally supercritical and subcritical filaments \citep{lee14}. 
Mapping Serpens~Main in three mm lines, these authors found six filaments in the SE subcluster, whose lengths range from 0.17 to 0.33~pc, and widths from 0.03~pc to 0.05~pc, respectively. 
In addition, several YSOs have formed along two filaments with supercritical masses-per-unit-length, while filaments with nearly critical mass-per-unit-length are not associated with YSOs, suggesting that stars formed preferentially in gravitationally unstable filaments.
\citet{roc15} identified nine filaments in \emph{Herschel} maps, affirming that their simulations could explain the filament morphology but not the kinematics. For example, while the simulated filaments of Serpens~Main are likely turbulence-dominated regions, for which collapse into smaller structures is not expected, the observed filaments are self-gravitating structures that will fragment into cores. 

\subsubsection{The adopted distance} \label{distSerp}
As described in detail by \citet{eir08}, the determination of the distance of the Serpens region is tricky, indeed distances between 200~pc and 700~pc are found in the literature \citep{cha88,zha88,dzi10,dzi11}.

First of all, it makes sense to search for a global distance of the region (with possible secondary differences among sub-regions), as suggested originally by line emission surveys: channel maps presented in \citet{pra08} indeed indicate that CO-emission is kinematically connected throughout the region.

Recently, \citet{ort17, ort18}, based on VLBI astrometry of a number of young stellar objects (YSOs) in the Serpens/Aquila~East complex, concluded that Serpens~Main and Aquila~Rift clouds, and in particular Aquila~East, are physically associated and lie at a common distance of $d = 436.0 \pm 9.2$~pc. 

More recently, \citet{zuc19} reported the distances of several nearby molecular clouds (including the region we study here), by combining line-of-sight extinctions from new near-infrared (NIR) star photometry with \textit{Planck} continuum maps at 353~GHz \citep{abe14} and Gaia parallaxes. 
They conclude that the Serpens/Aquila~Rift clouds form a single coherent complex, at a similar distance, with single regions having possible peculiar distances, resulting in a dispersion along the line of sight of about 50~pc. Having to quote a global distance for such a complex, they determine $d = 484 \pm 28$~pc. 
However, for the Serpens~Main area only, \citet[][]{zuc19} estimate a specific distance of 420~pc (with $5 - 6$\% uncertainty) which is in agreement with \citet[][]{ort17} result.

A quick check we made with the {\it Gaia data release 2} ({\it Gaia-dr2}\footnote{https://gea.esac.esa.int/archive/}) supports this distance estimate for Serpens Main: i.e., the VV-Ser parallax measured is $p=2.38 \pm 0.046$~marcsec, which corresponds to a distance of about $d=1/p \simeq 420.0 \pm 8.1$~pc \citep[in turn consistent, within the errors, also with that of][]{ort18}. 

Summarising, the final decision for this paper consists in studying the whole surveyed area as a unique, coherent, physically connected star-forming complex, composed by two main sub-regions located at two single different distances: 484~pc for Aquila~East, and 420~pc for Serpens~Main, respectively. 
The Serpens~Main sub-region is further discussed separately in Sect.~\ref{comparison}.

\subsection{The Prestellar Core Mass Function} \label{sect:introcmf}
A typical observable useful to investigate the link between the properties of the core population and the characteristics of the star formation in a given region is the statistical mass distribution of cores (core mass function, CMF), which can be compared with the initial mass function (IMF) of stars \citep[e.g.,][]{mot98,tes98,off14}. 

The IMF power-law behaviour was first computed by \citet{sal55}, who found 
\begin{equation}
    \frac{dN}{dM} \propto \left( \frac{M}{\rm M_\odot}\right) ^{-\gamma} 
    \label{eq:imf}
\end{equation}
with $\gamma = 2.35$ for $-0.4 \le \log(\frac{M}{\rm M_\odot}) < 1.0$. 
Decades later, \citet{mot98} and \citet{tes98} found a similarity in shape between the IMF and the CMF in the Ophiucus and Serpens~Main star-forming regions, respectively. 
Indeed, the power-law slopes were consistent, but over a different mass range. 
In particular for Serpens~Main, \citet{tes98} found a slope $\gamma=2.1$, while subsequently \citet{eno07} found $\gamma=2.6$. 
Only the first value is consistent with the observational IMF slope and with that predicted by turbulent fragmentation simulations $\gamma = 2.33$ \citep[e.g.,][]{pad02}.
The similarity between the power-law tail of the IMF and of CMF suggests that a direct link might exist between the star properties and the fragmentation process.
Subsequent works improved the knowledge of the IMF, still observing the relationship with the CMF, not only for the 
power-law tail trend at high masses, but also for the log-normal behaviour of both mass functions at low masses \citep{kro01,cha05,lee14}.
Recent papers based on the HGBS seem to confirm the similarities between CMF and IMF for young star-forming regions \citep{and10,pol13,kon15,mar16}.
One of the purposes of this paper is precisely to verify whether new {\it Herschel} observations are consistent with previous results, both for the specific Serpens region and for general comparison between CMF and IMF (see Sect.~\ref{cmfsection} and \ref{comparison}).

\section{OBSERVATIONS} \label{observation}
\emph{Herschel} observations of the Serpens/Aquila~East complex include two sub-regions: a box of approximately $\sim 2.2 \times 2.5$ square degrees to the east, centered on ${\rm R.A.} = 18{\rm h} \; 29 {\rm m} \; 32 {\rm s}$, ${\rm Dec.} = +0^{\circ}40'42''$ that includes Serpens~Main; the box of approximately $\sim 2.3 \times 4.4$ deg$^2$, centered on ${\rm R.A.} = 18 {\rm h} \;38 {\rm m} \;48 {\rm s}$, ${\rm Dec.} = -0^{\circ} 00'37''$, is the Aquila~East (see Fig.~\ref{fig:nh2}).
Observations were taken on 2010 October 16-18, ObsIDs 1342206676/95 and 1342206694/96 for Serpens~Main and Aquila~East, respectively. 
Data were taken using PACS at 70~$\mu$m and 160~$\mu$m, and SPIRE at 250~$\mu$m, 350~$\mu$m, and 500~$\mu$m. 
Two orthogonal scan maps were performed in parallel mode at 60''/s. Data taken during the turnarounds were included.

The data reduction procedure was different for PACS and SPIRE observations. SPIRE data were reduced entirely with HIPE Version 10.1, provided by \emph{Herschel} Science Center, producing the images using its ``naive'' map-making procedure and destriper module \citep{her11}. Differently, PACS data were reduced with HIPE v10.1 and images were obtained using UNIMAP Version 6.4.1 \citep{pia15} for the map making and destriping phase. 
The pixel size of the maps has been set to 3$\arcsec$ at 70~$\mu$m, 4$\arcsec$ at 160~$\mu$m, 6$\arcsec$ at 250~$\mu$m, 10$\arcsec$ at 350~$\mu$m, and 14$\arcsec$ at 500~$\mu$m, respectively. 
The FWHM beam sizes are 8.4$\arcsec$, 13.5$\arcsec$, 18.2$\arcsec$, 24.9$\arcsec$ and 36.3$\arcsec$ for the maps at 70~$\mu$m, 160~$\mu$m, 250~$\mu$m, 350~$\mu$m, and 500~$\mu$m, respectively\footnote{\url{http://gouldbelt-herschel.cea.fr/archives}. The page at this link contains all the Serpens/Aquila~East products, among other HGBS regions: the FITS maps of the region at individual PACS/SPIRE wavelengths (70$\mu$m - 500$\mu$m), as well as column density and temperature FITS images.}

The \emph{Herschel} maps have an arbitrary zero-level intensity, so the intensity in each pixel is known except for a constant
term. 
\begin{table}
	\centering
		\caption{\label{tab:offset}\emph{\textit{Planck}} and \emph{Herschel} emission medians (for both Serpens~Main and Aquila~East), and final offsets obtained by the difference between them.}
	\label{tab:example_table}
	\begin{tabular}{lccccc} 
		\hline
		\hline
		 Band   & Region & $\left< P \right>$  & $\left< H \right>$  & Offset\\
		 $\mu$m &        & MJy ${\rm sr}^{-1}$ & MJy ${\rm sr}^{-1}$ & MJy ${\rm sr}^{-1}$\\
		\hline
		70 & SM & 25.9 & 14.1 & 11.9\\
		   & AE & 54.0 & 12.6 & 41.4\\
	   160 & SM & 196.0 & 57.0 &139.0\\
		   & AE & 314.0 & 55.2 & 258.8\\
	   250 & SM & 126.4 & 1.1 & 125.3\\
		   & AE & 194.8 & 3.6 & 191.2\\
	   350 & SM & 62.2 & 0.6 & 62.7\\
		   & AE & 95.2 & 1.8 & 93.4\\
	   500 & SM & 26.9 & 0.6 & 26.3\\
	       & AE & 26.9 & 0.6 & 38.9 \\
		\hline
		\hline
	\end{tabular}

\end{table}
This offset was evaluated assuming a dust model exploiting \emph{\textit{Planck}} and IRAS data, using a method similar to the one described by \citet{ber10} to obtain the \emph{\textit{Planck}} median $\left<P\right>$, and subtracting the \emph{Herschel} median $\left<H\right>$. 
The estimated offsets are reported in Table~\ref{tab:offset}.
To analyze the entire region, we composed the calibrated maps of Serpens~Main and Aquila~East into mosaic maps right after the map making by merging the two observations in a single map. 

\begin{figure*}
  \centering
  \includegraphics[width=0.8\textwidth]{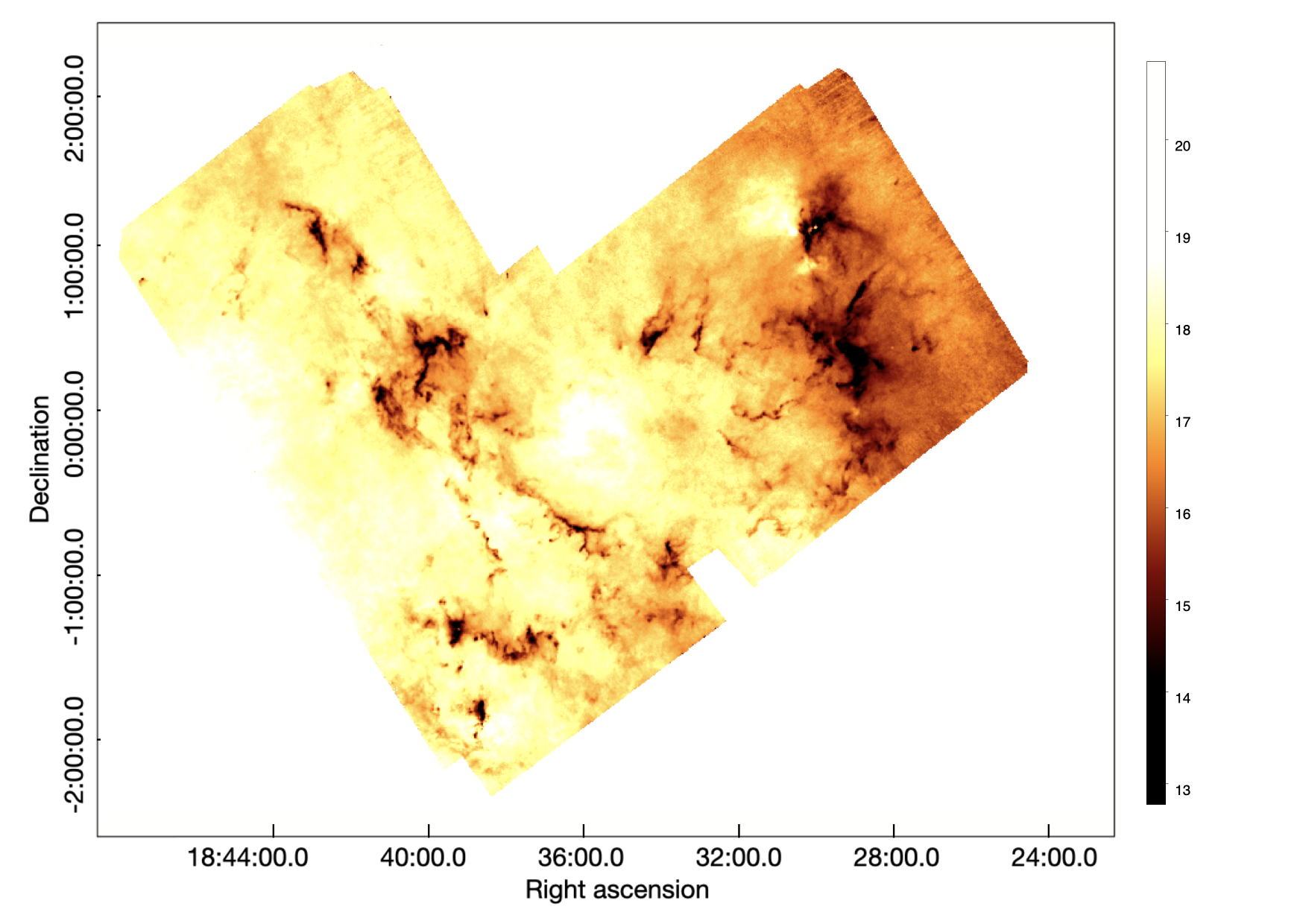}
  \caption{\label{fig:t}Temperature map of the Serpens region. 
  The color bar reports the temperature in K.}
\end{figure*}

\section{PHYSICAL ANALYSIS OF THE SERPENS REGION} \label{coldenst}
\subsection{Column Density \& Temperature Maps} \label{coltemp}
To compute temperature and H$_2$ column density maps, we assumed that the interstellar dust emits like a modified black-body 
in the far-IR:
\begin{equation}
 \label{eq:gb}
 F_\nu = \frac{\kappa_\nu M}{d^2} B_\nu(T)
\end{equation}
\citep[e.g.,][]{eli16}, where $F_\nu$ is the flux emitted by the dust at the frequency $\nu$, $\kappa_\nu$ is the dust opacity, $M$ is the mass of emitting material for a given pixel, $B_\nu(T)$ 
is the \textit{Planck} function at temperature $T$, and $d$ is the source distance.

The dust opacity is usually parametrized as 
\begin{equation}
 \label{eq:op}
 \kappa_\nu = \kappa_0\left( \frac{\nu}{\nu_0} \right)^\beta
\end{equation}
in the far-IR, where $\kappa_0$ is a reference value estimated at $\nu_0$, and $\beta$ is the dust opacity index. 
In this work we use 
$\kappa_0=0.1$~cm$^2$~g$^{-1}$, which already accounts for the gas to dust ratio of 100, at $\lambda_0= c/\nu_0 = 300$~$\mu$m \citep{hil83} and $\beta=2$, for uniformity with other HGBS works \citep{kir13, kon15, ben15}, based on \citet{roy14}, in which it is shown that this assumption is good within better than $\sim 50\%$ in the bulk of molecular clouds.

The choice of $\kappa_0$ and $\beta$ can be somehow critical \citep[see for example,][for a discussion on $\kappa_0$ and $\beta$, respectively]{eli17, alt04} since it has consequences on the column density or masses derived through Eq.~\ref{eq:gb}. 
As $\kappa_0$ increases the mass linearly decreases, while at the large wavelengths, where $(\nu/\nu_0)<1$, if $\beta >2$, then the mass increases with respect the estimate we give in Sect.~\ref{sect:mass_comparison_estimate}.

The temperature and column density maps were obtained pixel-by-pixel by fitting a modified black-body (Eq.~\ref{eq:gb}) to fluxes at 160~$\mu$m, 250~$\mu$m, 350~$\mu$m and 500~$\mu$m, after re-projecting each map over the 500~$\mu$m grid.
The 70~$\mu$m map was not used to produce the column density map because the parametrization in Eq.~\ref{eq:gb} is valid only for $\lambda \gg 100$~$\mu$m, while at shorter wavelengths the dust is not optically thin, and both cold and hot dust components are present.
Note that the same technique is also applied to the SEDs of cores to derive core masses and temperatures (Sect.~\ref{estimate}), which however, differently from this case, are built with background-subtracted and integrated flux densities.
This difference can lead to differences between quantities derived in this way for cores, and those of the pixels corresponding to core locations \citep[see, e.g.,][]{eli13,kir13,kon15}. 
We refer the reader to Sect.~\ref{estimate} for further details.

\begin{figure}
 \includegraphics[width=1.1\columnwidth]{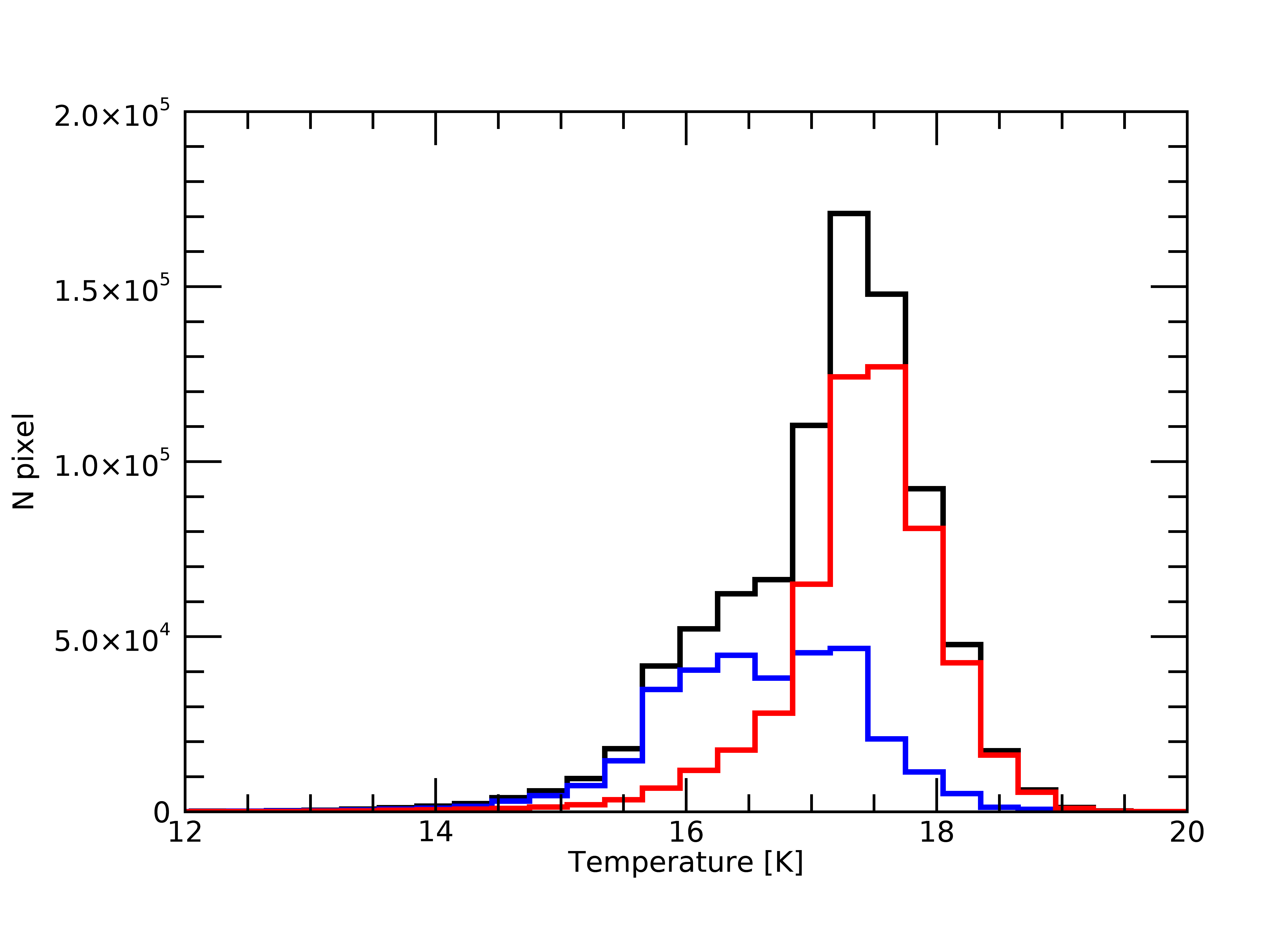}
 \caption{\label{histot}Temperature distributions of the \emph{Herschel} Serpens region (black), see Fig.~\ref{fig:t}.
 The total distribution shows a swelling at about 16-17~K and a peak at about 17-18~K, corresponding to the peaks of the sub-regions: the highest at $\sim 17.5$~K, which is due to the Aquila~East (red) contribution; the lowest at $\sim 16.5$~K, mostly due to the 
 Serpens~Main contribution (blue).}
\end{figure}

The temperature map is shown in Fig.~\ref{fig:t}.
The estimated temperature ranges from 11.5~K and 21.4~K, with remarkable differences between the eastern (Aquila~East region) and western (Serpens~Main region) side of the complex. 
This gradient is probably due to a more efficient radiation shielding by the denser and colder ($T < 16$~K) clusters in Serpens~Main, which is probably a consequence of the fact that the Serpens~Main is farther than the Aquila~East from the Galactic Plane. 
We present the temperature distribution in Fig.~\ref{histot}, splitting it between the two sub-regions. 
The Aquila~Rift is warmer and includes all the regions for which we estimated $T > 19$~K. 
Its average temperature is 17.5~K, dropping to $\sim 16$~K only in the regions at higher density.
We find that, as expected, temperatures are comparable with those derived by \citet{kon15} in the Aquila~Main region, ranging from 14~K to 22~K.

We assumed that the gas composition was primordial and we compute the column density values using
\begin{equation}
 \label{eq:nh2}
N_{\textrm{H}_2} = \frac{M}{\mu_{\rm H_2} \mbox{ }m_\textrm{H}} \frac{1}{A_\textrm{pix}},
\end{equation}
where $\mu_{\rm H_2} = 2.8$ is the mean molecular weight \citep{kau08}, $m_\textrm{H}$ is the hydrogen mass, and $A_\textrm{pix}$ is the pixel area of the $500 \mbox{ }\mu \mbox{m}$ map.

We computed the optical depths in the Serpens region from the column density map whose pixel size was degraded to 5$\arcmin$, namely the resolution of the {\it Planck} map at $\lambda=850$~$\mu$m, using the equation:
\begin{equation}
 \tau_\nu = \kappa_\nu \ \mu \, m_{\rm H} \, N_{{\rm H}_2}
\end{equation}
and compared the results with the \textit{Planck} direct $\tau$ observations. 
We found our values are in agreement within 1$\sigma$, with the mean $\tau_{\rm H}/\tau_{\rm P} = 0.81$, the median $\tau_{\rm H}/\tau_{\rm P}= 0.98$ and the standard deviation $\sigma \left( \tau_{\rm H}/ \tau_{\rm P} \right)= 0.39$, where $\tau_\textrm{H}$ and $\tau_\textrm{P}$ are the optical depths obtained by \emph{Herschel} and {\it Planck} observations, respectively.
This agreement confirms the consistency of the zero-level offsets derived from {\it Planck} data and applied to the {\it Herschel} maps (see Sect.~\ref{observation}).

\begin{figure}
 \centering
  \includegraphics[width=\columnwidth]{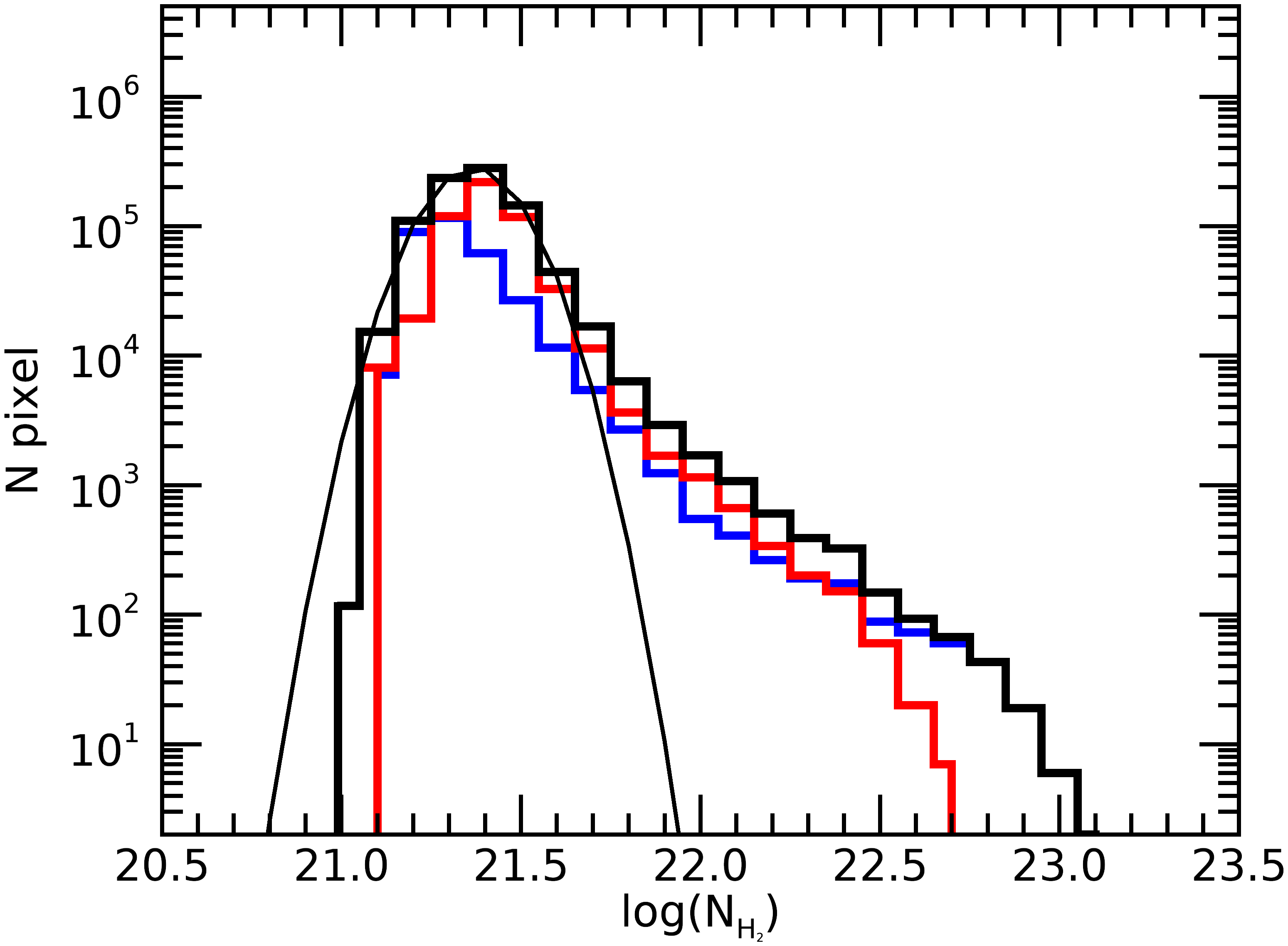}
 \caption{\label{histoCD}Histograms of column density from 18$\arcsec$ column density map for the whole region (black). 
 Note that the main contribution to the peak, which lies at $\sim 2.5 \times 10^{21}$~cm$^{-2}$, is due to the Aquila~East region (blue) while the peak of Serpens~Main region (red) is shifted to lower column densities, $\sim 1.99 \times 10^{21}$cm$^{-2}$.}
\end{figure}

As in previous HGBS works, we also determined a ``high-resolution'' column density map, as shown in Fig.~\ref{fig:nh2}, based on a multi-scale decomposition of the imaging data, as described in Appendix A of \citet{pal13}, and a ``temperature-corrected'' map, based on the color-temperature map of 160~$\mu$m and 250~$\mu$m, built to reduce the effects of strong, anisotropic temperature gradients present in parts of the observed fields \citep{kon15}. 
These two maps are included in the HGBS source extraction process through the \get  algorithm (see Sect.~\ref{source}).
\\ \\
The distribution of pixel column densities for Serpens cloud is displayed in Fig.~\ref{histoCD}. We will refer to this as the probability density function (PDF), even though the pure pixel distribution is not strictly a PDF, since this term is typically used in the context of continuous random variables, while we build it from a discrete set of data. 
The low-density part of the PDF behaves as a log-normal function, which is usually interpreted as the presence of a turbulence-dominated environment \citep{kai11}. 
Accordingly, we fit the low density part with a log-normal, finding that our data are well fitted for $A_\textrm{V} \lesssim 3$. 
In particular, a departure of the PDF from log-normal behaviour is seen around two bins that correspond to $A_\textrm{V} \sim 4$ and  7, respectively. 
The presence of an excess at high column density bins with respect to a log-normal behaviour is generally explained with a gravity-dominated scenario \citep{fed13,sch13},
with presence of overdensities such as filaments and cores. 
This is found also if the two sub-regions are considered separately, with Serpens Main showing a more extended high-density tail (up to $\sim 10^{23}$~cm$^{-2}$)
compared to Aquila~East (up to $\sim 5\times10^{22}$~cm$^{-2}$).

Finally, we also notice that the Serpens PDF is similar to the Aquila~Main PDF \citep{sch13,kon15}, since they insist on similar density ranges, which is not surprising as those regions are part of the same cloud complex \citep{ort17}.


\subsection{Mass and comparison with previous estimates} \label{sect:mass_comparison_estimate}

Integrating the column density map over the entire region, we obtained a total Serpens mass of $\sim \totalmass$~M$_\odot$ (see Tab.~\ref{tab:mass}). 
The two main sources of error on the mass estimate are the background patches which typically affect the HGBS maps \citep[4\%, see e.g.][]{kon15}, and the error on the distance ($\sim 11\%$). Therefore, by propagating these errors we obtain that the uncertainty on the mass estimate is about 12\%. However, the main source of uncertainty is the one in the dust opacity, which is only known to within $\sim 50\%$ in the bulk of the target field (see Sect.~\ref{coltemp}). At very low column densities ($A_v < 1-2$), our adopted $\kappa_0$ value may depart from the “true” $\kappa$ by more than 50\% \citep[see][]{kon20, dif20, pez20}. Therefore, we would quote an uncertainty of ~50\% for the total Serpens mass.
Just $\sim 6\%$ of the mass is contained in regions with $A_V > 7$, roughly considered a transition for the formation of prestellar cores \citep[e.g.,][]{and10,kon15}, if $N_{{\rm H}_2}/A_{\rm V} = 9.4 \times 10^{20}$~cm$^{-2}$ and $R_{\rm V}=3.1$ are assumed \citep{boh78}.
Recently, \citet{nak17} studied the Serpens region in the $^{12}{\rm CO}(J=2-1)$ and $^{13}{\rm CO} (J= 2-1)$. 
If we sum the mass of all the clouds analyzed by \citet{nak17} located in the HGBS Serpens region and rescale to our assumed distance of 420~pc for Serpens~Main and 484~pc for Aquila~East, we obtain 23200 M$_\odot$, $\sim 47\%$ of the mass we obtain.

To compare our estimate to previous works on Serpens~Main, we considered only the Serpens region analyzed in the previous literature (from $\sim 250~\textrm{M}_\odot$ to $\sim 4000$~M$_\odot$), which is indicated in Fig.~\ref{fig:nh2}. 
We derived a total mass of 8844~M$_\odot$ for Serpens~Main. 

As shown in Table~\ref{tab:mass_comp}, mass estimates of Serpens~Main have been computed in the literature by adopting different distances and tracers. The disagreement (10\% - 30\%) between our result and the works of \citet[][]{mcm00,olm02,whi95} might be explained by the different tracers used for estimating the mass. 
Using the C$^{18}\mbox{O}$ transition line, the contribution of the diffuse gas is not detected. 

Thus, we found overall higher mass estimates for the Serpens~Main region with respect to previous works. 
This difference is mainly due to the method we use to compute the sub-region mass. 
Indeed, we estimate the mass from the column density, while other works are based on spectroscopic tracers, which could underestimate the mass value.
However, we can not exclude we are over-estimating the total mass of the cloud. Indeed, the Aquila~East region is very close to the Galactic plane, and the global background emission is large (decreasing away from the Galactic plane), this may end up with large values of the cloud mass, which is affected by the emission uncorrelated with the cloud we are studying. 
On the contrary, when analysing molecular line data, e.g., CO data, since there is an information on the velocity, only the emission corresponding to the cloud, at the velocity of the cloud, is taken into account, thus the masses are lower. 
Moreover, we observed the dust emission which is present even in the diffuse gas in contrast to the CO which is not formed below a few 100~cm$^3$. 
Our estimates are therefore more {\it complete} and it is expected we get larger masses then.

A more direct comparison of the cloud mass estimation is possible with \citet{roc15}, who used the same \emph{Herschel} data we use in this paper, but only for the sub-region we refer to as Serpens~Main. 
They reported a total mass of this sub-region 2.7 times smaller than the estimate we found (see Table~\ref{tab:mass_comp}). 
Comparing the column density histograms, we find a global disagreement of a factor $\sim 4$ \citep[compare our Fig.~\ref{histoCD} and Fig.~4 of][]{roc15}. 
Moreover, \citet{roc15} assumed a different dust opacity law \citep{oss94} with $\beta = 1.9$, implying a dust opacity coefficient about $\sim 2$ times higher than ours, more appropriate for denser protostellar cores than for prestellar cores.
The choices on the dust opacity assumptions should account for most of the difference in the derived masses. 
This difference can be understood, in turn, from the different temperatures found in the sub-region. 
Indeed, systematically higher temperatures typically lead to lower column densities.
\citet{roc15} claimed a minimum temperature of 19~K and an average temperature of about 24~K, while our values are lower, as discussed above. 
Moreover, these authors included 70~$\mu$m emission in their 
modified black-body fits, while, as said above, we do not. 
Note also that \citet{roc15} acknowledged that they found a mass 7 times lower than \citet{whi95} lower limit. 
On the contrary, our analysis satisfies this independent lower limit. 

\begin{table}
	\centering
		\caption{\label{tab:mass}Total mass of the entire Serpens region, Serpens~Main, and East~Aquila region computed from the high-resolution column density map ($18.2\arcsec$). 
	The mass contained in contours of $A_{\rm V} > 4$ and 7, respectively, is also reported.}
	\begin{tabular}{lccc} 
		\hline
		\hline
		M$_{A_V}$ & Serpens & Serp. Main & Aqu. East\\
		& [M$_\odot$] & [M$_\odot$]& [M$_\odot$]\\
		\hline
		M$_{\rm tot}$ & \totalmass & 8800 & 40500 \\
		M$_4$ & 11500 & 2900 & 8500\\
		M$_7$ & 3000 & 1200 & 1800\\
		\hline
		\hline
	\end{tabular}

\end{table}

\begin{table*}
	\centering
	\caption{\label{tab:mass_comp}Mass estimates of the Serpens~Main sub-region available in the previous literature. Masses in column~2 were derived at distances in column~1, and through the tracer(s) in column~3 by authors in column~4. In column~5 these masses are rescaled to the distance of 420~pc adopted in this paper.}
	\begin{tabular}[width=\textwidth]{|lrccr|} 
		\hline
		\hline
		 $d$       & $M_{\rm orig}$ &tracer & reference & $M_{\rm 420}$\\
		
		 pc        &[M$_\odot$]        &       &           & [M$_\odot$]\\
		\hline
		220 & 250      & C$^{18}$O(J=1-0) & \citet{mcm00} & 911\\
		310 & 300      & C$^{18}$O (J=1-0) & \citet{olm02} & 550\\
		311 & $> 1450$ & C$^{18}$O(J=2-1) & \citet{whi95} & $> 2644$\\
		    & $> 1450$ & C$^{17}$O(J=2-1) & \citet{whi95}& $> 2644$\\
		415 & 3213     & N$_{{\rm H}_2}$ & \citet{roc15} & 3290\\
		\hline
		\hline
	\end{tabular}
	
\end{table*}

\section{SOURCE DETECTION AND CLASSIFICATION}  \label{source}
A dense core is a roundish structure that can be considered as the final result of cloud fragmentation, and whose gravitational collapse could give rise to a star or a multiple star system. 
Based on this concept, we generate a catalogue of starless cores for all the star-forming regions of the Gould Belt observed by {\it Herschel}. 
To ensure uniformity with other HGBS works, this catalogue is built following the procedure already described by \citet{kon15} and \citet{mar16} as a two-step process: 
\\ \\
\emph{Step 1: Detection}. Source extraction is carried out by the multi-wavelength, multi-scale software \get \citep{men12} which simultaneously analyzes the \emph{Herschel} images between 70~$\mu$m and 500~$\mu$m, as well as the ``high-resolution column density'' and the ``temperature corrected'' maps (Sect.~\ref{coldenst}), producing a preliminary version of a core catalogue.

A complete summary of the source extraction method can be found in Sect.~4.4 of \citet{kon15}, and full technical details are provided in \citet{men12}. 
After the extraction, properties of detected sources are measured in the original observed images at each wavelength \citep{kon15}.
For producing HGBS first-generation catalogs of starless and protostellar cores, two sets of dedicated \get extractions are performed, optimized for the detection of dense cores and YSOs/protostars, respectively. 
In the former, 
the 160~$\mu$m, 250~$\mu$m, 350~$\mu$m, and 500~$\mu$m maps, and the high-resolution column density image are considered together in a multi-wavelength approach.
In particular, the 160~$\mu$m component of the detection image is “temperature-corrected” to reduce the effects of strong, anisotropic temperature gradients present in parts of the observed fields. 

A second set of \get source extractions is performed to trace the presence of YSOs/protostars. 
In this process, at the detection stage only the PACS 70~$\mu$m data are used. Point-like 70~$\mu$m emission properly traces the internal luminosity, and therefore the presence itself, of a protostar \citep[e.g.][]{dun08}.
We eliminated contamination due to galaxies and other non-YSOs by looking for those sources in Simbad database within 2$\arcsec$ from the centroid coordinates. 
At the end of this process, we count \nprotocores of those objects, named proto-stellar cores. 
Although proto-stellar cores are contained in the catalogue, they are not analyzed in this work, but will be discussed separately in a subsequent paper.
\\ \\
\emph{Step 2: Classification}. 
The raw source list outputted by \get 
in the region studied here, is then filtered to select only dense starless cores, by applying the following criteria:
\begin{enumerate}
 \item column density detection significance higher than 5 in the high-resolution column density map, where significance refers to detection at the 
 relevant single spatial scale, as defined by \citet{men12};
 \item high-resolution column density signal to noise ratio at the peak of the emission, $\left( S/N\right)_{peak}>1$;
 \item global detection significance over all wavelengths \citep{men12} greater than 10;
 \item significance of flux detection greater than 5 for at least two wavelengths between 160~$\mu$m and 500~$\mu$m;
 \item flux measurement with $S/N > 1$ and monochromatic detection significance greater than 5 for at least one band between 160~$\mu$m and 
 500~$\mu$m; 
 \item 70~$\mu$m peak intensity per pixel less than 0, which is typical of noise-dominated areas, or detection significance less than 5 for this band, or source size at this band larger than 1.5 times the band (at this stage, namely the last one based on constraints on detection parameters, the candidate sources remain 1081);
 \item the source is not spatially coincident within 6$\arcsec$ of a known IR source of at least one of the following catalogues: Spitzer Space Telescope (SST, 137 matches found), Widefield Infrared Survey Explorer (WISE, 149 matches found) and Simbad Database (6 matches found) for further infrared observations;
 \item the source is not spatially coincident within 6$\arcsec$ with a known galaxy, i.e., a comparison with NASA Extragalactic Database (NED) found no matches;
 \item the source appears real on visual inspection of the images between 160~$\mu$m and 500~$\mu$m. In this respect, $15\%$ of the dense starless sources were visually rejected.
\end{enumerate}
The resulting \nsources starless cores are contained in the catalog available in the online version of this paper, and from the HGBS website (see the footnote 2 in Sect.~\ref{observation}). 
In this website all the maps discussed in this paper are also available.
The catalog consists of reports for each source, see Table~\ref{tab:es1} and \ref{tab:es2} in App.~\ref{app:cat}.


Figure~\ref{fig:nh2} shows the column density map, in which we notice a large-scale gradient which corresponds to that seen in the temperature map (see Sect.~\ref{coldenst}), although less evident. 
On this column density is plotted the distribution of the \nsources sources of the \get catalogue we obtained after \emph{Classification} step.
We are confident that all \nsources detected cores (starless and proto-stellar) of our census are robust, based on visual inspection checks. 

\subsection{Estimation of Core Masses \& Temperatures} \label{estimate}
Similarly to the approach described in Sect.~\ref{coldenst}, the core masses and the temperatures can be estimated by fitting the 
modified black-body function (Eq.~\ref{eq:gb}) to their SED obtained through the \get software. 
In the case of temperature and column density maps, the SEDs are composed by single and unsubtracted pixel fluxes, while in the case of source fluxes they are background-subtracted and integrated over the entire source area. 
This different treatment of the background, which changes with wavelength, can lead to different estimates of core temperature and mass, whether they are derived from the source SED, or taken from temperature and column density maps \citep[cf., e.g.,][]{kir13}.

Furthermore, we notice that actually the structure of a quiescent core might be more accurately modelled as a series of dust layers, the outermost of which is the warmest, heated by external radiation, and the innermost the coldest \citep{roy14}. 
This is the reason why ``observed'' core masses, obtained from modified black-body fit, can be typically $\sim 50\%$ lower than true core masses, since SED-derived temperatures tend to overestimate the mass-averaged temperatures of starless cores \citep{kon15}. 
Anyway, modified black-body fit remains a reasonable and useful way to study mass and temperature of each single core.

The masses were derived assuming $d=484$~pc for the Aquila~East region, and $d=420$~pc for the Serpens~Main region.
The physical core radius is obtained from the angular FWHM estimated by \get in the high-resolution (18.2$\arcsec$) column density map. 
In our catalog (see Table~\ref{tab:es2}), we provide estimates for both the deconvolved and the observed radius of each core (estimated as the geometrical average between the major and minor FWHM sizes).
For each core, the peak column density, average column density, central-beam volume density, and the average volume density were then derived based on their mass and radius estimates \citep[see Sect.~4.6 in][for details]{kon15}. 
In Table~\ref{tab:es2} all these observables are listed for the whole sample of selected {\it Herschel} cores.
For SEDs composed by only one or two significant flux measurements \citep[see App.~A in][]{pez20}, the fitting procedure was not applied.
We also did not look for the best fit in cases where $F_\nu (350~\mu$m$) < F_\nu(500~\mu$m$)$. 
For all these cases, we adopt a fiducial temperature equal to the median temperature of all the reliable SEDs, excluding the protostars. 
The uncertainty of this median is arbitrarily large (the median $\Delta T/T$ for reliable sources is $\sim 3\%$). 
For these sources, which are the $\sim 20\%$ of the total starless cores, the mass was computed from the measured integrated flux density at the longest significant wavelength, assuming the fiducial temperature. 

We calculated the completeness limit using the same model approach as described in Appendix B.2 of \citet{kon15}. 
In this way, the $\sim 80\%$ completeness level is estimated to be at $\sim 0.8$~M$_\odot$ in \emph{true} prestellar core mass.

\begin{table}
	\caption{\label{tab:cat}Classification of the Serpens dense cores. 
	After the checks described in Sect.~\ref{source} and \ref{estimate}, in the \get catalogue there are \nsources dense cores, \nstarlesscores of them are starless, and \nprotocores of them are protostellar cores. 
	\nunboundcores starless cores are gravitationally unbound cores, the remaining \nprestellarcores gravitationally bound cores are \emph{candidate prestellar} cores. 
	Finally, \nrobustcores prestellar cores are classified as \emph{robust} because their mass fulfils the BE criterion (see text).}
\Tree[.{Detected cores: \nsources} 
[.{Starless: {\nstarlesscores}} {Unbound: {\nunboundcores}} [.{Prestellar: {\nprestellarcores}} {Candidate: \ntentativecores} {Robust: \nrobustcores} ] ] {Protostellar: \nprotocores} ]  
\end{table}
          
To identify dense cores candidates to likely form new stars, one has to verify whether sources are gravitationally bound.
A self gravitating core satisfies the \emph{virial condition}
\begin{equation}
\alpha_{vir} = M_{vir}/M_{obs} \le 2 \; .
\label{eq:vir}
\end{equation}
\citep{kon15}. 
The virial mass ($M_{vir}$) could be computed by measuring the mean velocity dispersion spectroscopically, but such observations are not available for all the sources with a spatial resolution comparable to the \emph{Herschel} one. 
Therefore, we use the Bonnor-Ebert (BE) criterion to select self-gravitating objects. 
According to this criterion, a core whose mass exceeds the BE critical mass will collapse \citep{bon56}.
A BE sphere, whose outer radius is $R_\mathrm{BE}$, has a critical mass given by:
\begin{equation}
\label{becriterion}
 M_\mathrm{BE} = 2.4 \, R_\mathrm{BE} \, a_s^2 / G
\end{equation}
where $G$ is the gravitational constant and $a_s^2=kT/\mu m_H$ is the square of the sound speed, with $k$ being the Boltzmann's constant, $T$ the temperature, and $\mu = 2.33$.
We computed the BE mass starting from the deconvolved core radius $R_\mathrm{deconv}$ measured as the geometric mean of the source FWHM major and minor axes as estimated by \get from the high-resolution column density map, after beam deconvolution, and by using as gas temperature the dust temperature estimated through the SED fit.
In this way, we find that \nrobustcores starless cores are \emph{robust prestellar cores}, i.e., with masses larger than $M_{BE}/2$, where the factor $1/2$ follows the analogy with the criterion of cloud collapse based on the virial condition (Eq.~\ref{eq:vir}). \citet{kon15} show that if they classify as prestellar a core for which $\alpha_\mathrm{BE}=M_\mathrm{BE}/M_{obs} \le 2$, Monte-Carlo simulations performed to assess the mass completeness of the survey suggest that this criterion may be too restrictive. 
Indeed it selects only $\sim 85\%$ of the simulated BE cores detected by \get after the source classification. 
Thus, they proposed a less demanding and size-limiting criterion, based on the results of Monte-Carlo simulations, which produce an envelope containing $>95\%$ of the simulated BE cores after \get extraction:
\begin{equation}
  \alpha_\mathrm{BE} \le 5 \times ({\rm HPBW}_\mathrm{N_{H_2}}/{\rm FWHM}_\mathrm{N_{H_2}})^{0.4}
  \label{hpbw}
\end{equation}
where HPBW$_\mathrm{N_{H_2}}=18.2"$ is the resolution of the high-resolution column density map, and FWHM$_\mathrm{N_{H_2}}$ is the measured FWHM source diameter in the same map. It is important to stress that this is an empirical result found by \get extractions on \emph{Herschel} observations.
Using Eq. \ref{hpbw}, we obtained that \nprestellarcores are \emph{prestellar cores} (i.e. \ntentativecores {\it candidate} cores are added to the {\it robust} ones).
In Table~\ref{tab:cat} the classification scheme is summarized.

For the Serpens star-forming region, the percentage of prestellar cores over the overall starless ones is \preOverstarless, and the percentage of only robust cores is \robOverstarless. 
Note, however, that this fraction can be affected by a distance bias. 
In particular, the mass increases as the square of the distance assumed for the region, while the BE mass increases linearly. 
This implies that the aforementioned fractions generally increase at increasing distance and, as a borderline case, over a certain distance 100\% of sources would have a mass larger than the BE mass \citep[cf.][]{eli13}. 
For this reason, comparisons between regions at distances significantly different based on this observable should be performed with caution.

\begin{figure}
 \includegraphics[width=\columnwidth]{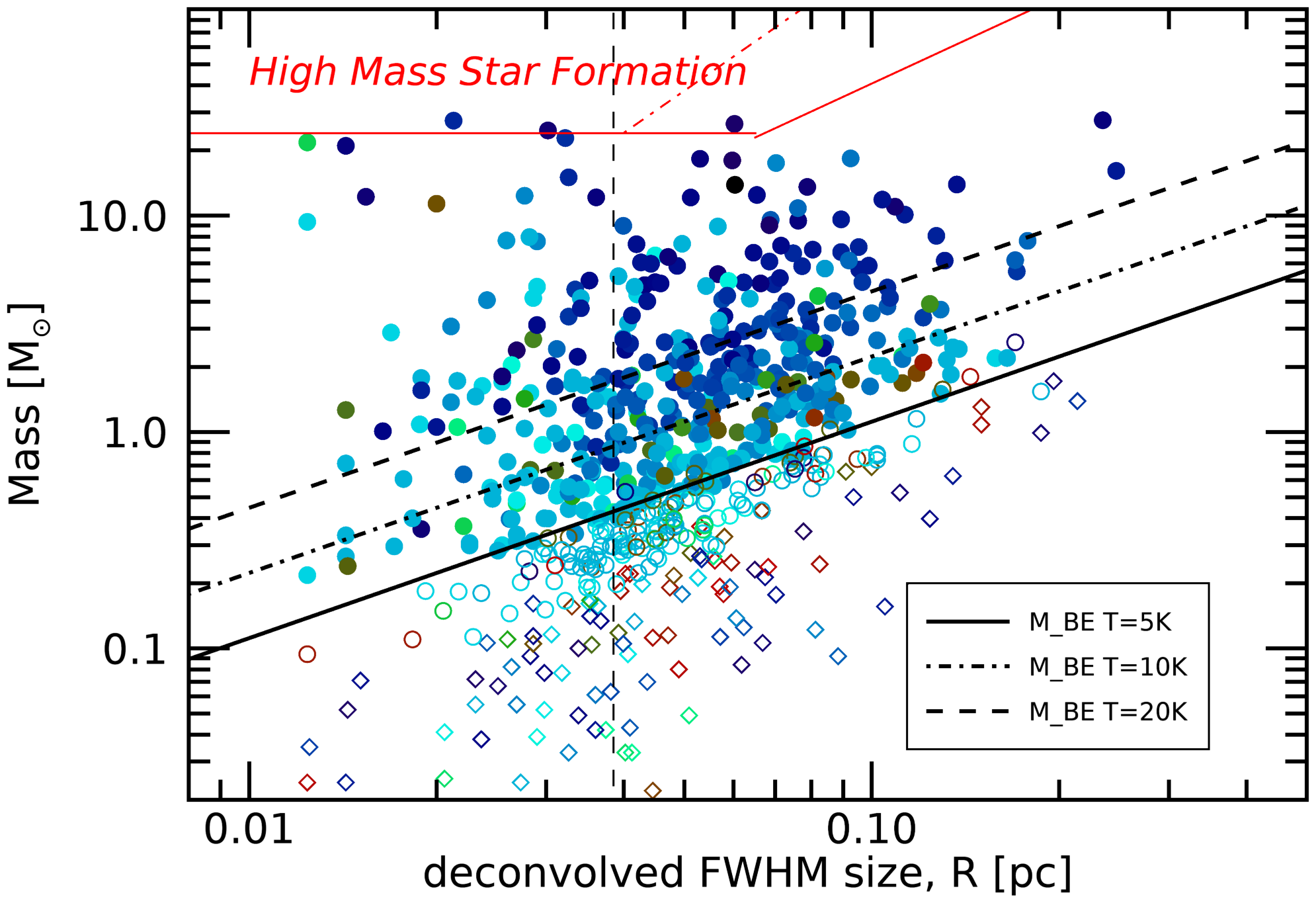}
 \centering
 \includegraphics[height=0.8cm, width=0.7\columnwidth]{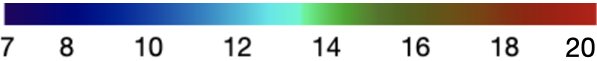}
 \caption{\label{fig:pl6}Mass versus size diagram for the starless cores. 
 The vertical dashed line indicates the corresponding physical resolution at $d=484$~pc. 
 Robust prestellar cores are shown with filled circles, candidate prestellar cores with open circles, and unbound cores with open diamonds, respectively. 
 The color indicates the temperature of the source in K, following the color bar: from blue to red, the temperature linearly increases from 7 to 20~K. 
 The black lines represent the BE mass computed at different temperatures: the solid line at 5~K, the dotted-dashed one at 10~K, and the dashed one at 20~K. 
 The robust cores are all located above the 5~K line, respectively. Red dashed and continuous lines are the \citet{kru08} and \citet{kau10} limits for high-mass star formation, respectively, the former being contained in the latter.}
\end{figure}
\begin{figure}
 \includegraphics[width=\columnwidth]{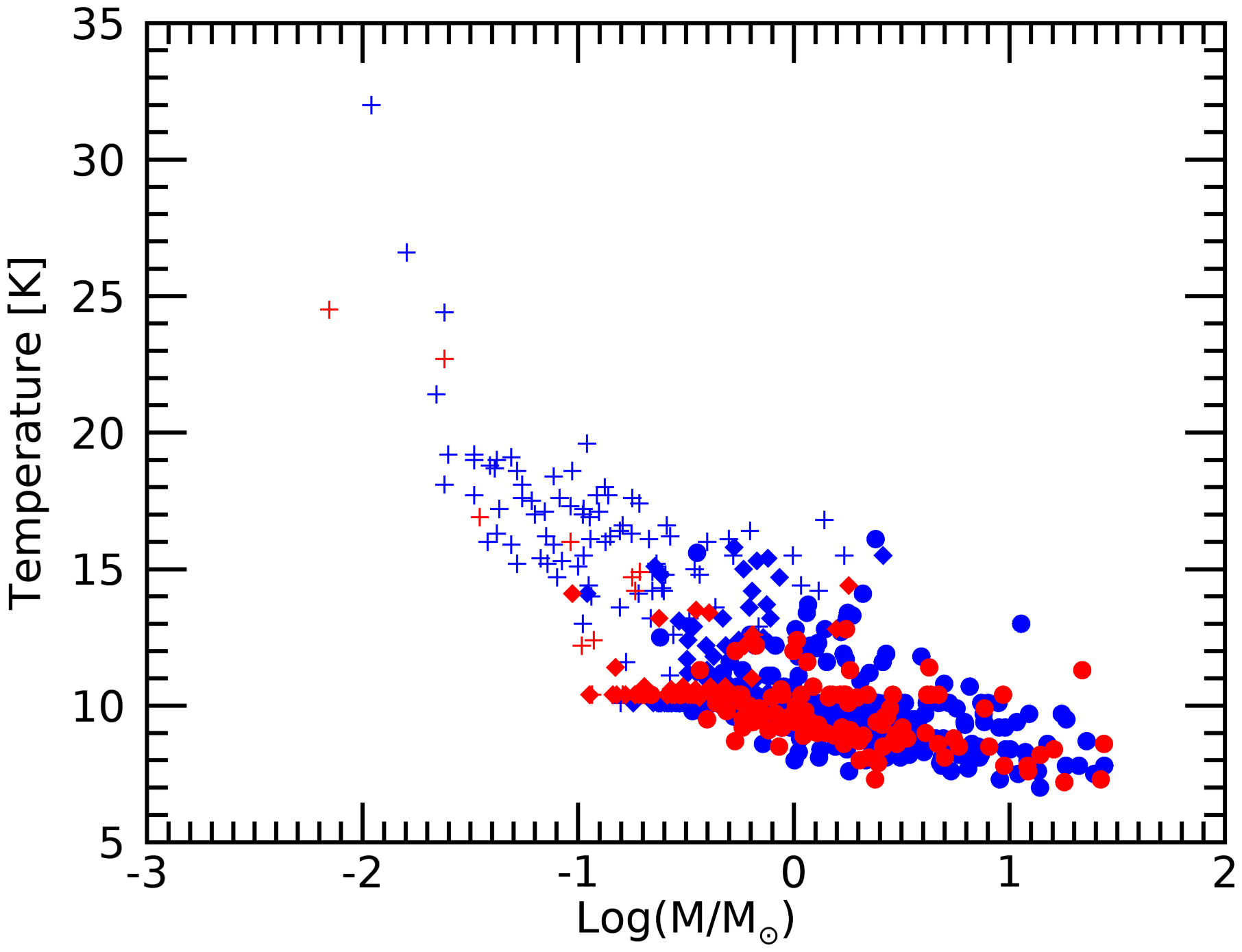}
 \caption{\label{fig:pl_tm}Core temperature versus core mass. 
 Filled circles are robust prestellar cores, filled diamonds are tentative prestellar cores, and crosses are unbound cores. 
 Cores situated in Aquila~East and in Serpens~Main are shown in blue and red.}
\end{figure}

In Fig.~\ref{fig:pl6}, we plot the core mass vs radius relation, which essentially confirms the low-mass nature of the star formation in Serpens/Aquila~East region, in the light of \citet{kru08} and \citet{kau10} requirements for a core to host the formation of massive stars. 
In both these works a power-law threshold for the $M$ vs $r$ relation is set, that we interrupt at the bottom at 24~M$_\odot$ assuming a minimum mass of $M>8$~M$_\odot$ for a high-mass star and a typical core-to-star efficiency of $\varepsilon =1/3$ in the process. 
If we compare our results with the two thresholds, we find 2 and 3 objects, respectively, compatible with the possible formation of a high mass star.
However, in the following we compute a specific core-to-star efficiency for this region, finding $\varepsilon = 0.2$ (see Sect.~\ref{cmfsection}). 
Adopting this value, the lower limit for compatibility with high-mass star formation increases to 40~M$_\odot$, which is not fulfilled by any core in our sample.
This figure also shows that our sample is spread not only in mass, as discussed above, but also in size, denoting a certain relation with the source type: robust bound cores sizes are contained between 0.01~pc and 0.25~pc, candidate bound cores between 0.01~pc and 0.19~pc (therefore, this is the most compact subset), and, finally, unbound core sizes between 0.01~pc and 0.21~pc. 
In Fig.~\ref{fig:pl6} a few sources with size larger than 0.2~pc are seen. 
Taking into account the classification of \citet{ber07, dif07} and \citet{and14}, such sources do not fulfill the definition of core, but rather that of clump, with an underlying but unresolved structure, probably containing multiple cores. 
Although the thresholds on which this classification is based should actually be considered loose, it is undeniable that the distance(s) assumed for the Serpens/Aquila~East region makes it as a slightly particular case with respect to other, generally closer regions surveyed in the HGBS, in which the resolution-distance combination ensures that all detected compact sources are cores.
As we can observe in Fig.~\ref{fig:pl6}, our analysis seems to verify that, as the star formation scenario predicts, the core future collapse depends only on mass and not on the size of the core. 
Indeed there is no correlation between the size range and the cores type. 
The Fig.~\ref{fig:pl_tm} shows the plot of cores temperatures versus mass. 
We find a negative correlation between temperature and mass, as expected according to the scenario in which the dust is heated by the interstellar radiation field \citep{sta07}.
The possibility that this is an artefact due to the analytic relation between these two quantities in the adopted model (modified black-body) is significantly lower with respect to works based on monochromatic observations: here the five-band analysis  in a wavelength range generally containing the peak of the core SED ensures an independent estimate for temperature and mass.
Another source of bias in estimating the mass when using a single-temperature model is the core internal temperature gradient. 
These biases have been estimated to be typically at the 30\% level \citep[see e.g.][]{kon15,mar16}, i.e. smaller compared to the correlation seen in Fig.~\ref{fig:pl_tm}.
We thus conclude that this correlation is real.
Moreover, as expected, unbound cores lie on the upper-left part of the diagram, while prestellar cores are found in the bottom-right part.
We note also that most massive cores are cold, wHereas we can not conclude that least massive cores are typically warm, because our sample is limited by the sensitivity of the instrument. This issue is widely discussed in Appendix A of \citet{mar16}.

\begin{figure}
 \includegraphics[width=\columnwidth]{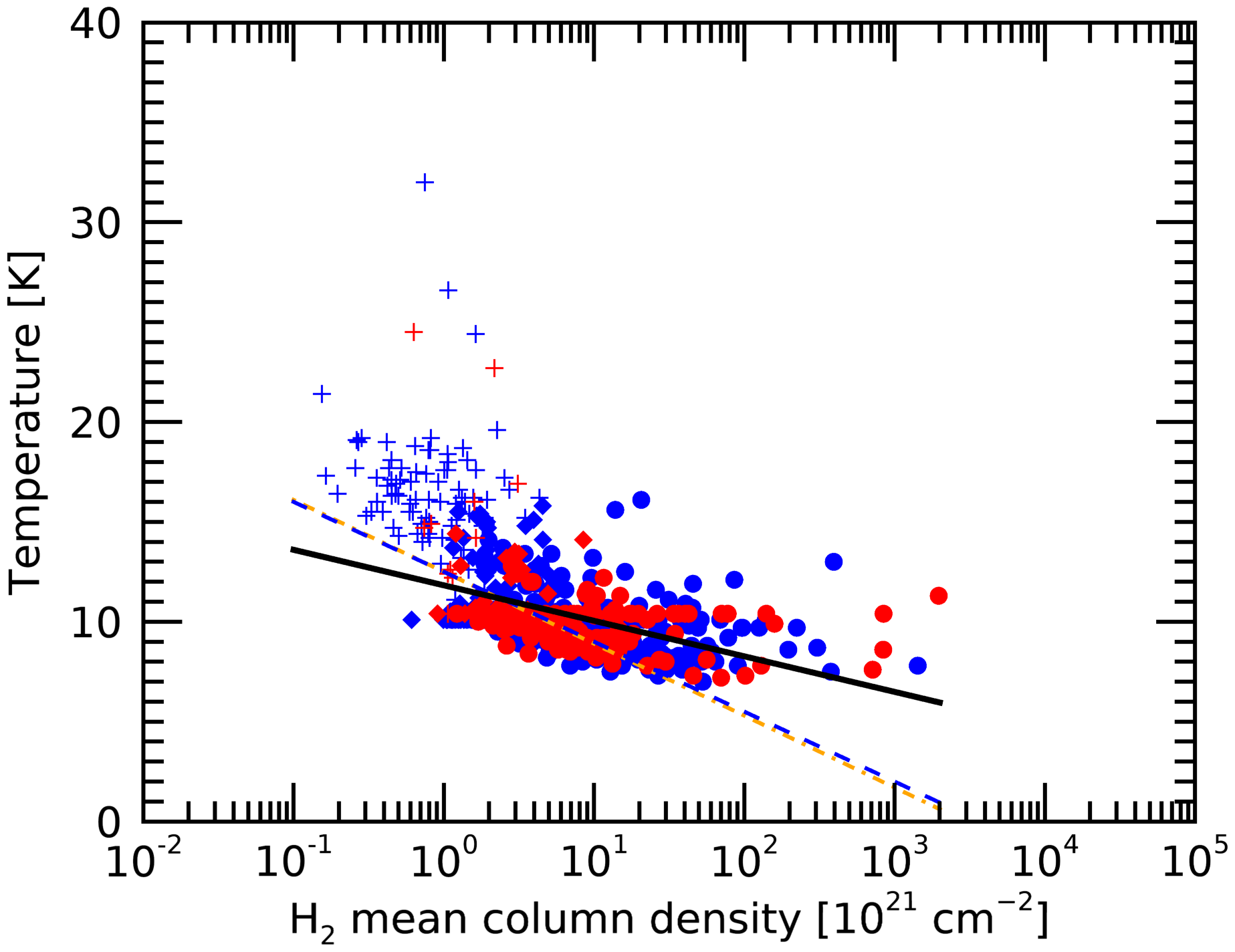}
 \caption{\label{fig:pl_th2}Mean dust temperature of cores plotted as a function of mean column density of cores, $N_{\mathrm{H}_2}$ diagram. 
 Symbols are like in Fig.~\ref{fig:pl_tm}. 
 The blue dashed line is the power-law fit to data of \citet{bre18}, the pink dashed line is the fit to data of \citet{mar16}, and the black solid line is the relation found in this work.}
\end{figure}

To remove any dependence on the distance, still implicit in Fig.~\ref{fig:pl_tm}, we also plotted the dust temperature versus the mean column density for the starless cores (Fig.~\ref{fig:pl_th2}).
To allow a quantitative comparison with \citet{mar16}, we arbitrarily fit a power-law to data 
\begin{equation}
T = a \log_{10}\frac{N_{\mathrm{H}_2}}{10^{21}~\mathrm{cm}^{-2}}+b \;,
\end{equation}
finding $a=-1.8\pm0.1$ and $b=11.8\pm0.1$. 
Similar relations, rescaled to units of $N_{\mathrm{H}_2}/10^{21}~{\rm cm}^{-2}$ to allow comparison, were found for Taurus \citep[$a_{\rm Tau}=-3.6$ and $b_{\rm Tau}=12.5$,][]{mar16} and for Corona Australis \citep[$a_{\rm CorA}=-3.5$ and $b_{\rm CorA}=12.5$,][]{bre18}. 
\citet{mar16} reasonably posited that the coefficients of such fits are not expected to be universal, but depend on the intrinsic conditions of the regions. 
Considering the three regions for which this analysis was carried out so far, Corona Australis and Taurus show quite similar coefficients, while the ones we obtained in Serpens are different. Extending this analysis to further star forming regions it will be possible to depict a more defined scenario on the basis of a larger statistics.


\subsection{Spatial distribution of cores}
\begin{figure}
  \begin{subfigure}{\columnwidth}
	\includegraphics[width=\textwidth]{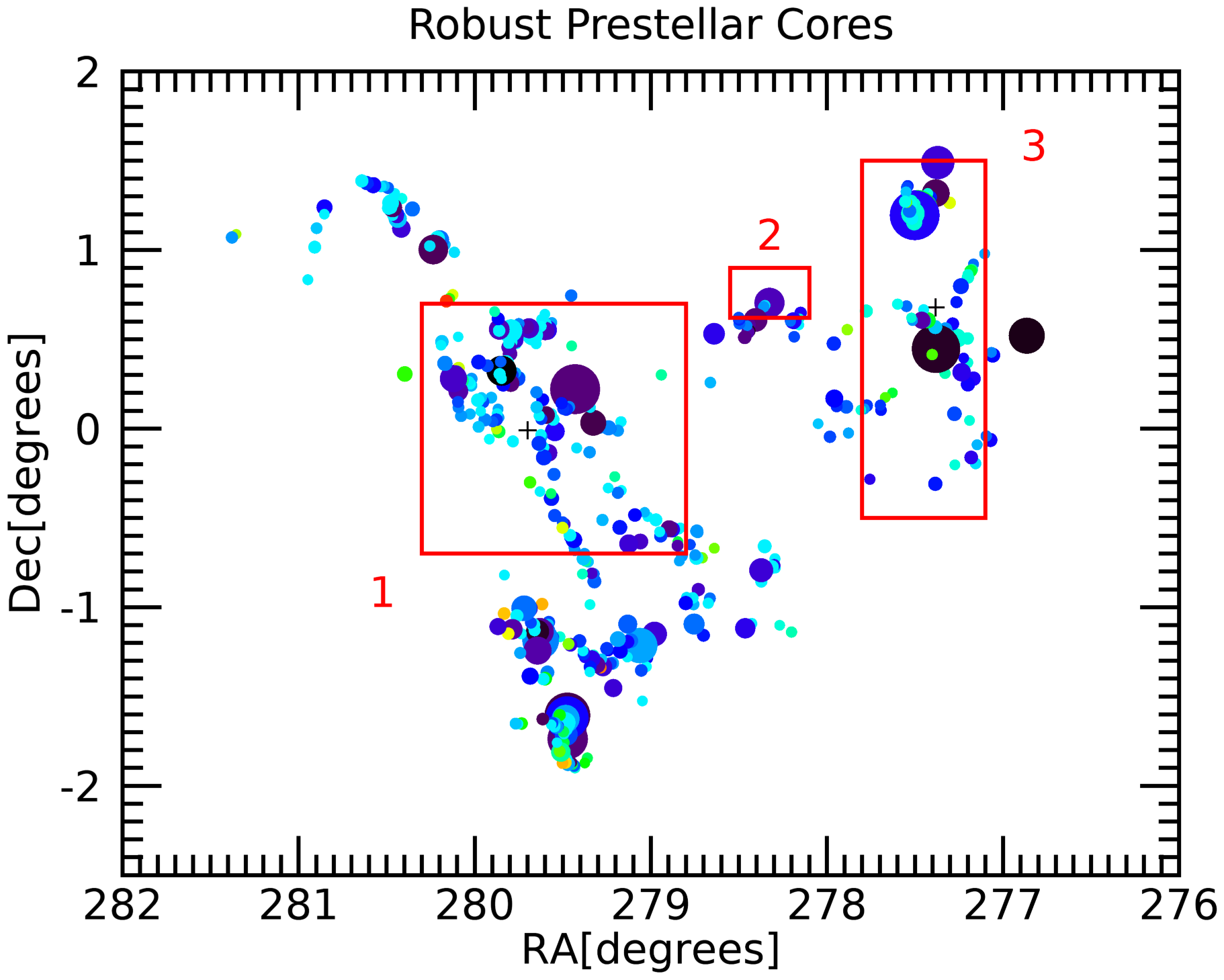}
	\includegraphics[width=\textwidth]{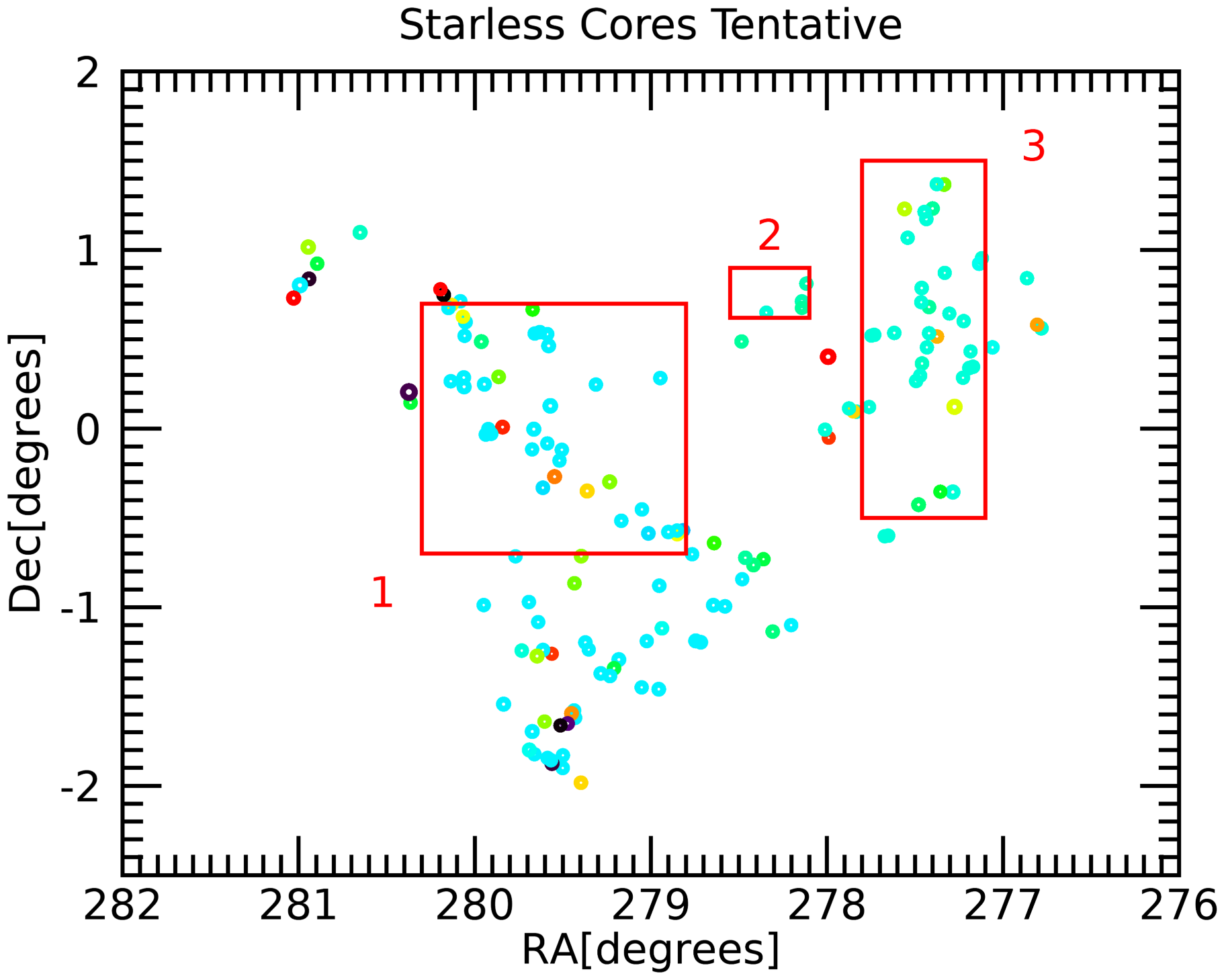}
	
	\includegraphics[width=\textwidth]{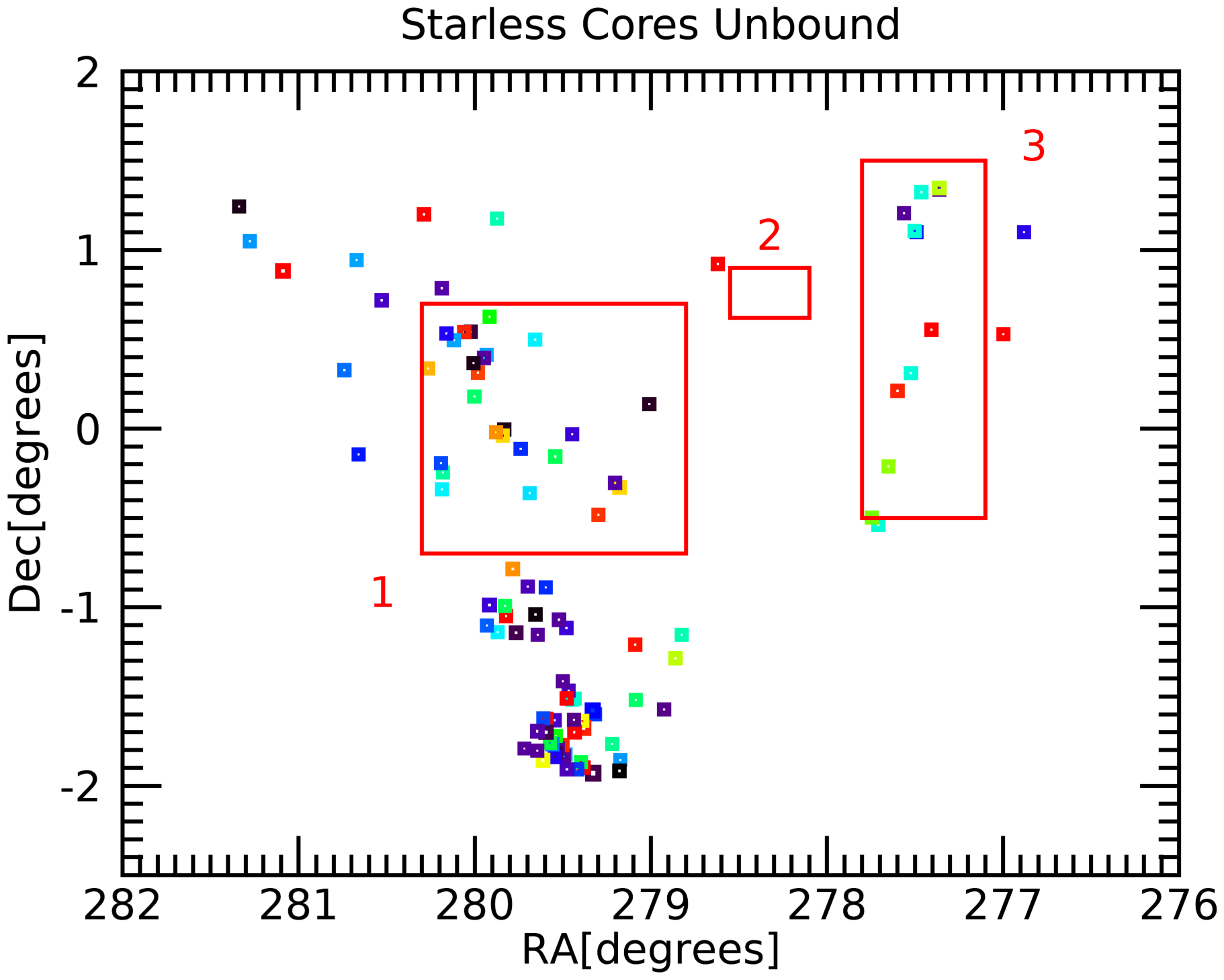}
	\includegraphics[width=1.\textwidth]{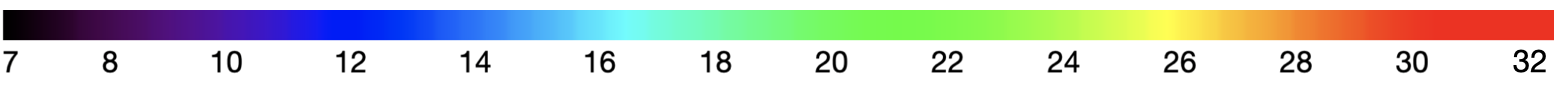}
  \end{subfigure}
    \caption{\label{fig:RaDecMeT} Spatial distribution of robust prestellar (top), candidate bound (middle), and unbound (bottom) cores, respectively. 
    The color, from black to red, represents the temperature from 7~K to 32~K, respectively. 
    The size of the circles and the squares represents the mass, from 0.01 to 27.5~M$_\odot$.} 
\end{figure}

In the following, we analyze the spatial distribution of cores, in the light of the identified sub-classes, with the help of Fig.~\ref{fig:RaDecMeT}, and 
\begin{figure}
\centering
  \begin{subfigure}{\columnwidth}

    \includegraphics[width=1.0\textwidth]{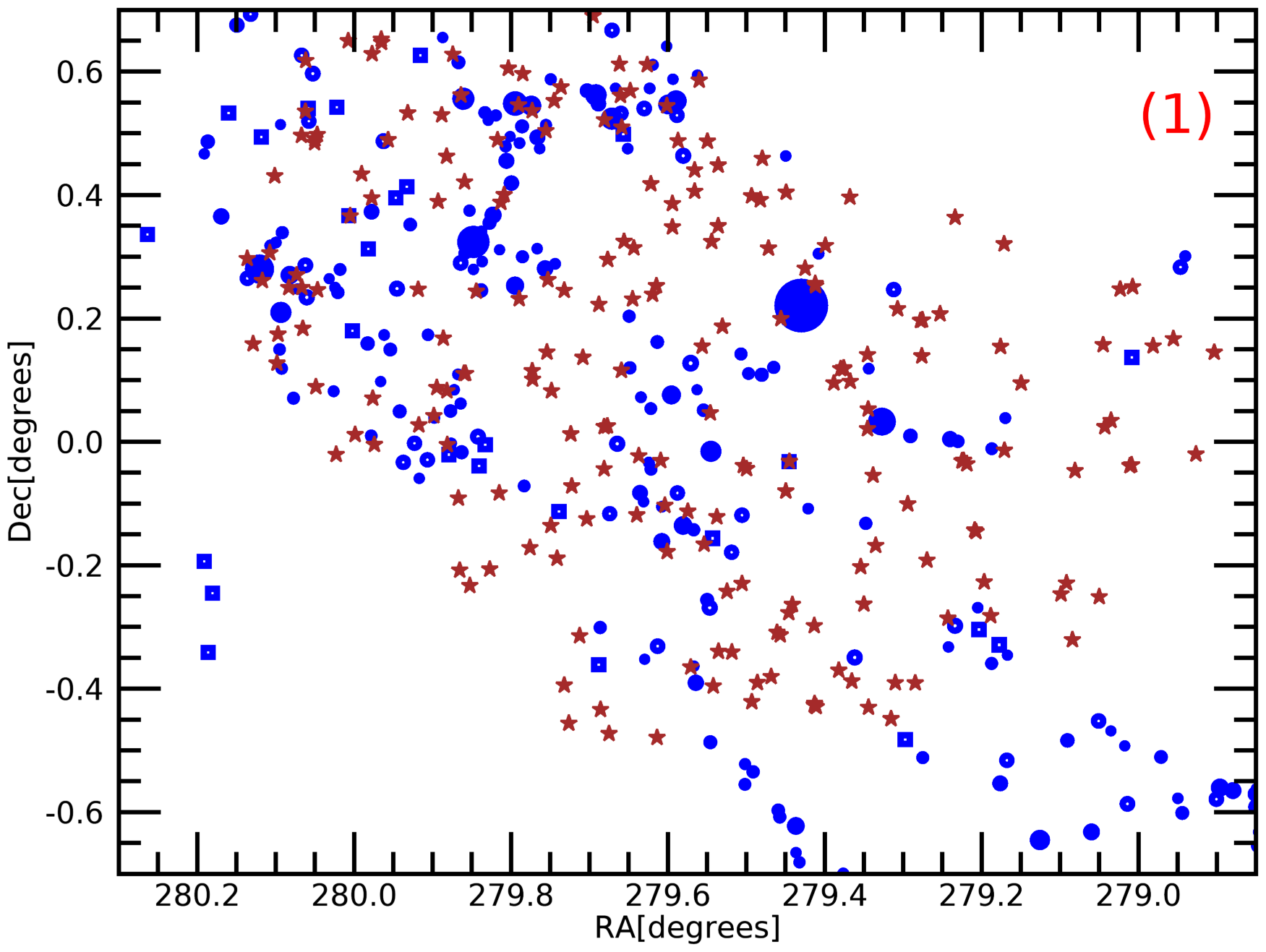}
    \centering
	\includegraphics[width=1.0\textwidth]{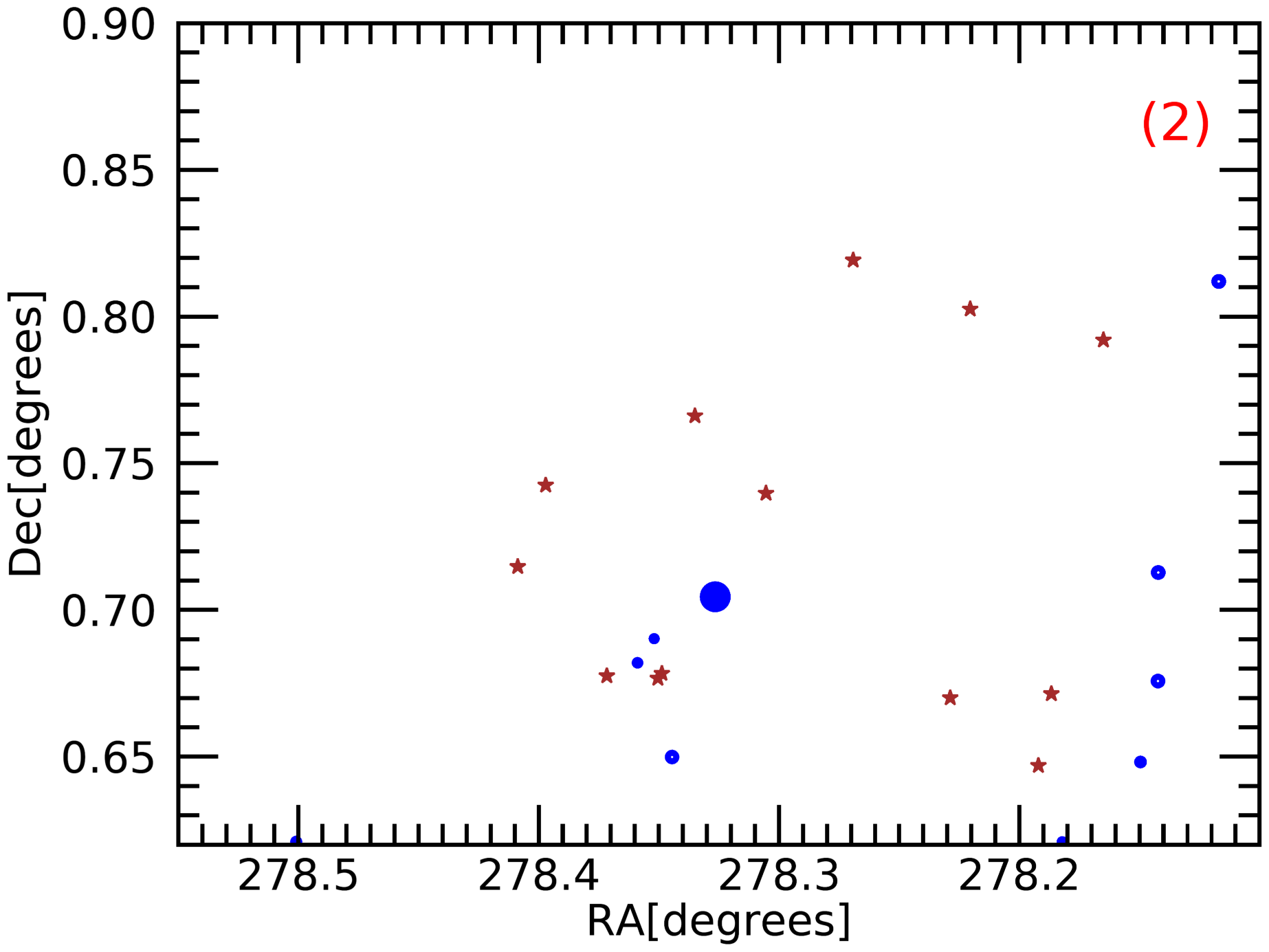}
	
	\includegraphics[width=\textwidth]{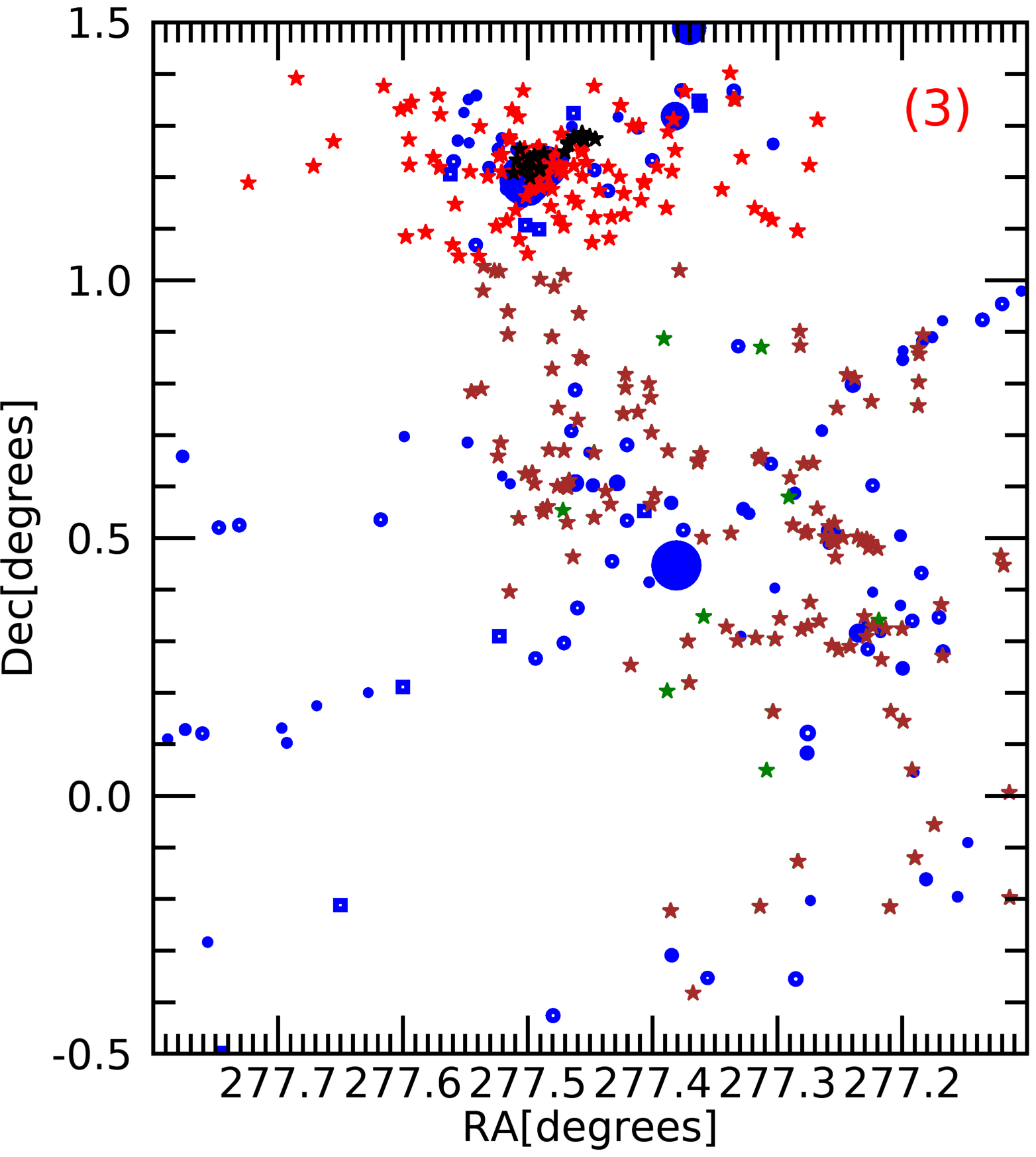}
  \end{subfigure}
    \caption{\label{fig:RaDecMeTzoom} Spatial distribution of cores population (blue circles and squares) and YSOs listed in \citet{win07} (protostars in black and PMS YSOs in red), in \citet{har07} (protostars in grey and PMS YSOs in green), and \citet{dun15} (protostars in brown), respectively. Panels are the red rectangles \#1 (top), \#2 (middle), and \#3 (bottom) shown in Fig.~\ref{fig:RaDecMeT}, corresponding to the area observed by {\it Spitzer}. 
    Notice that axes of each panel have different scales.} 
     
\end{figure}
Table~\ref{tab:RaDecMeT}. 

Mean values of mass and temperature for each sub-sample support the evidence, already found in previous HGBS works and clearly shown also in Fig.~\ref{fig:pl6} and \ref{fig:RaDecMeT}, that more massive cores are generally the coldest, and this is verified for each of the three core classes.
In general, robust bound cores are the most massive and coldest, while, on the other side, unbound cores are the least massive and warmest, with candidate bound prestellar cores lying in the middle between these two populations. 
This behavior is strictly due to the 
modified black-body function (Eq. \ref{eq:gb}) we use to obtain temperature and mass values, but note that the detection sensitivity cutoff does not allow us to know whether there are cold unbound cores, as discussed in \citet{mar16}.

\begin{table}
 \centering
 \caption{\label{tab:RaDecMeT}Average mass and temperature for each sample of cores. 
  Prestellar cores are subdivided here in robust and candidate, cf. Table~\ref{tab:cat}.}
 \begin{tabular}{lcc}
 \hline
 \hline
  Type of Cores & $M$ [M$_\odot$]  & $T$ [K] \\
  \hline
  Robust Prestellar Cores & $2.93 \pm 0.35$ & $9.7 \pm 0.4$\\
  Candidate Prestellar Cores & $0.44 \pm 0.14$ & $11.0 \pm 0.8$ \\
  Unbound Cores &$0.20 \pm 0.02 $ & $15.4 \pm 0.6$ \\
  \hline
  \hline
 \end{tabular}
\end{table}
The spatial distribution of candidate prestellar and unbound cores seems to be more widely distributed than that of robust bound prestellar cores, which follows instead the underlying filamentary structure (as deeply discussed in Sect.~\ref{filamensSection}). 

In the following we discuss the spatial relation between prestellar cores and YSOs (see Fig.~\ref{fig:RaDecMeT} and Fig.~\ref{fig:RaDecMeTzoom}). 
We consider the area observed by Spitzer and analyzed by \citet{win07, har07, dun15}, in which the main clusters of YSOs in Serpens~Main and Aquila~East regions are located, and that is composed by three ``boxes'' (shown in Fig.~\ref{fig:RaDecMeTzoom}).
These areas contain more massive robust bound cores (74 in box~1, 6 in box~2, and 154 in box~3, respectively), than candidate bound (22 in box~1, 4 in box~2, and 39 in box~3, respectively), and unbound cores 11 in box~1, 1 in box~2, and 26 in box~3, respectively).
In other words, about half of the prestellar cores population (43\% of the robust prestellar cores and 52\% of the candidate prestellar cores, 45\% of the prestellar cores in total, respectively), and about one third of the unbound cores population (36\%) is located in these {\it Spitzer} regions.
This indicates the presence of a further mass reservoir in the areas already hosting star formation. 
Moreover, the higher degree of  prestellar core clustering in regions rich of YSOs, suggests that the same event that caused the formation of YSOs in Serpens~Main, also produced the condensation of matter in the cores of this sub-region, which became denser and more massive, and thus ready to collapse in case of a new triggering event, confirming similar considerations by \citet{win07, har07} and \citet{sea12}, based only on analysis of the spatial distribution of YSOs different classes.
According to this scenario, the {\it Spitzer} sources spatially related with cores are the young embedded protostars (Class~0/I), while the sources on the border the more evolved pre-main sequence (PMS, i.e. Class~II/III) YSOs. 

However, looking at shorter scales, there is not strict spatial coincidence, as expected, between massive prestellar cores and YSOs. In Fig.~\ref{fig:RaDecMeTzoom} we can see that in box~1 one of the most massive cores (at the top of the panel) is closely surrounded by protostars and PMS stars in the expected way, while the other massive core (at the center of the panel) appears to have only group of dispersed protostars around it. 
A similar configuration is seen in boxes~2 and 3, where the most massive cores are not related to clustered protostars. 
In fact, a closer spatial correlation with YSOs would be expected for protostellar cores, and this topic will be discussed in a subsequent paper focused on the {\it Herschel} protostellar cores population.




\section{THE PRESTELLAR CORE MASS FUNCTION} \label{cmfsection}
\begin{figure*}
 \includegraphics[width=\textwidth]{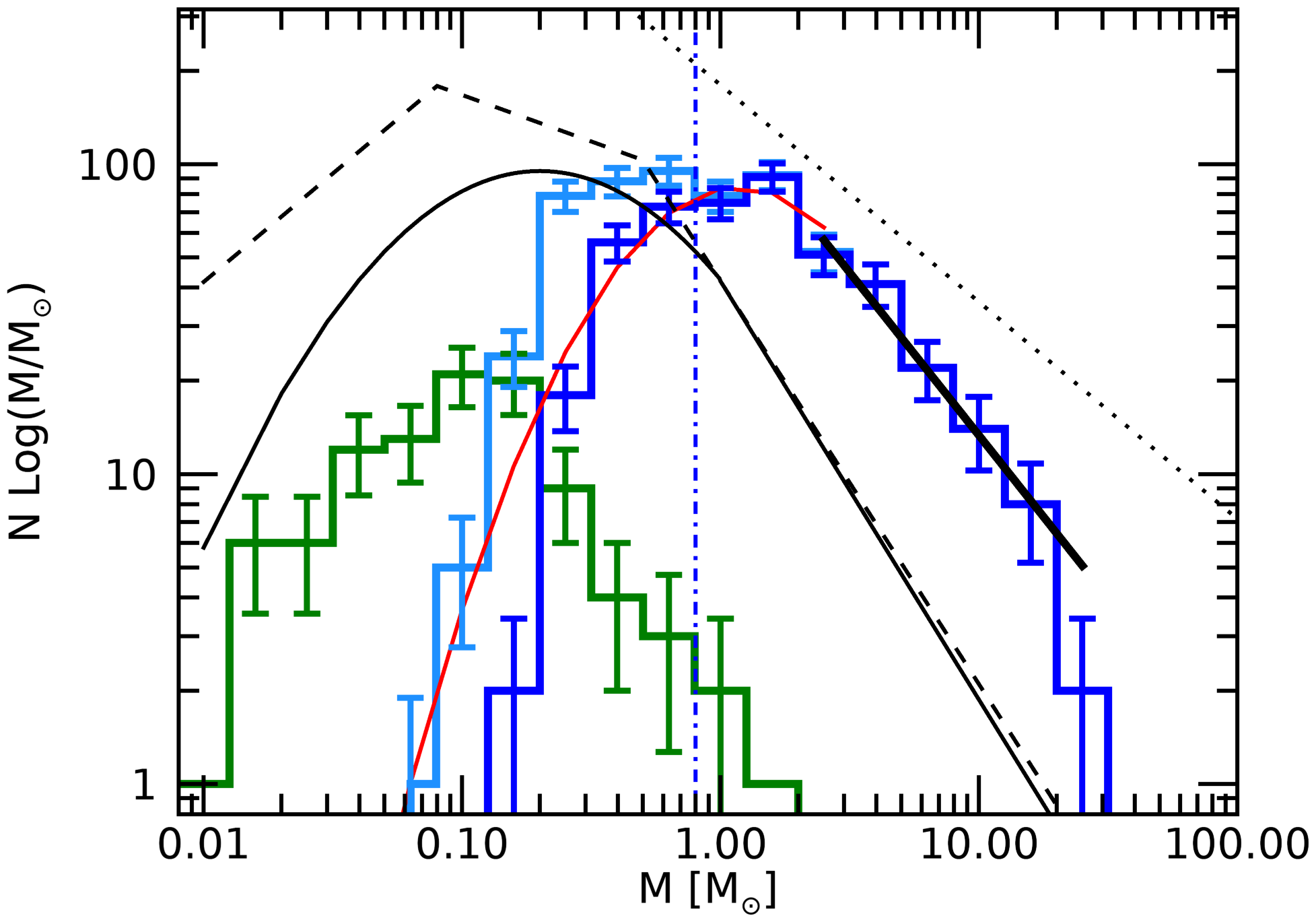}
 \caption{\label{fig:cmf}Core Mass Function (CMF) of Serpens/Aquila~East star-forming region. 
 The CMF of robust prestellar, candidate prestellar, and starless cores are shown as a blue, a light blue and a green solid histogram, respectively, with logarithmic bins.
 The prestellar CMF is well fitted by a log-normal function (red) up to $3-4 {\rm M}_\odot$. 
 At higher masses, the prestellar CMF shape follows a power-law, with slope $\cmfslopeless$ (solid thick black line). 
 For comparison, the CO clump mass function whose slope is $\xi_{\rm CO}=-0.65\pm0.10$ \citep[][]{kra98}, and the \citet{kro01,cha05} stellar IMFs are plotted in dotted, dashed, and solid black, respectively. 
 The \citet{cha05} IMF is normalized to the peak of our CMF. 
 The blue dotted-dashed vertical line represents the mass completeness limit for prestellar sources in Serpens.}
\end{figure*}

We computed the CMF for starless cores (Fig.~\ref{fig:cmf}). 
The Serpens/Aquila~East prestellar CMF is well fitted by a log-normal function at masses lower than $2~{\rm M}_\odot$, and by a power-law at higher masses. 
The completeness limit effectively lies where the mass distributions of candidate vs. prestellar CMFs start to differ, enabling us to locate the peak and discuss the power-law slope, but not the log-normal fit.

In a simple scenario in which each core produces one stellar system, core masses map into stellar masses with a core-to-star efficiency. 
Using as a reference the peak position of the \citet{cha05} IMF for multiple systems, and comparing it with that of the CMF, we estimate an efficiency of $\varepsilon=0.2$ both for the overall region and for Serpens~Main only (see Sect.~\ref{sect:serpmaincmf}), lower than typical estimates for this quantity \citep[$\varepsilon=0.25-0.40$,][]{alv07,eno08,and10}.
However, we want to stress that the core-to-star efficiency computed as we did, strongly depends on the distance. 
In this case, the low efficiency might be seen as a consequence of the larger distance of Serpens/Aquila~East region with respect to other nearby star forming regions. 
For this reason, compact structures detected in the Serpens/Aquila~East region are typically larger than those found in closer regions and, correspondingly, the CMF lies on a range of higher masses. 

About the high-mass tail, we remove from the sample two prestellar cores whose size is larger than 0.2~pc because they probably are not resolved (see Sect.~\ref{estimate}), and find that the best fit power-law slope is $\gamma =$\cmfslope , which is in agreement with both the \citet{sal55} and \citet{cha05} IMFs slopes within the errors, but also with the CO~clump distribution by \citet{kra98}, $\gamma_{\rm CO} = -1.65 \pm 0.10$. 
Since other clouds observed in the HGBS have found similar consistencies, this is a further evidence that statistical mass distribution of main sequence low-mass stars directly depends on the one of the prestellar phase \citep{mot98,tes98,off14, and14}, i.e. it directly results from the fragmentation process of the molecular cloud.
\begin{table*}
 \centering
 \caption{\label{tab:cmf}CMF slope values ($\gamma$), type of starless cores for which the CMF was computed, percentage of prestellar cores which lie on filamentary structure, distance in parsec and percentage of robust on starless cores ($r/s$) for which the $\gamma$ is computed for the HGBS works published until now, respectively. 
The reported $\gamma$ slope of the last six clouds is referred to mostly bound cores, while for the first four it is referred to mostly unbound cores.}

 \begin{tabular}{lcccccl}
  \hline 
  \hline
  Cloud     & cores type & $\gamma$       &on fil [ $\%$]& $d \mbox{ [pc]}$ & $r/s [ \%]$ & Reference \\
  \hline 
  CrA       &mostly unbound     &$-1.59 \pm 0.04$&  88    & 130              & 14         &\citet{bre18}\\
  Ophiuchus &bound + unbound& no power law     & 100    & 139
  & 20      & \citet{lad20} \\
  Taurus    & mostly unbound     & $-1.55\pm0.07$ & 100    & 140              & 10         &\citet{mar16}\\
  Lupus     &bound + unbound &no power law    &  94    & 150-200          & 11         &\citet{ben18}\\
  Aquila Main &bound + unbound &$-2.33 \pm 0.06$&  81    & 260              & 44         &\citet{kon15}\\
  Perseus   &bound + unbound &$-2.30 \pm 0.06$&  70    & 300              & 41         &  \citet{pez20} \\ 
  Cepheus Flare &bound + unbound&$-2.40\pm 0.14$& 80   &358               &
  23         & Di Francesco et al. subm. \\
  Orion-B   &bound + unbound &$-2.27 \pm 0.24$&  90    & 400              &  28       & \citet{kon20}\\
  Orion-A   &bound + unbound &$-2.4\pm0.4$    &  67    & 414              & 77         &\citet{pol13}\\
  Serpens   &bound + unbound &\cmfslope  &    81  & 420-484              & 64         &This work\\
  \hline
  \hline
 \end{tabular}
\end{table*}

From Table~\ref{tab:cmf}, which summarizes results obtained so far for various HGBS regions, we note that the CMF slope strictly can vary from region to region, also depending on the dense core sample used to build the CMF. 
Indeed, Corona Australis \citep{bre18} and Taurus \citep{mar16} CMFs have power-law slopes which are more similar to the CO clump slope \citep{kra98}. 
Noticeably, in these regions the lowest values for the fraction of prestellar cores are found (e.g. in the Taurus region less than 20\% of cores are gravitationaly bound), or even only unbound cores are present (Corona~Australis). 
In addition, the Lupus complex does not show a power-law trend at all \citep{ben18}, because the statistics is quite poor and the uncertainties on the mass estimate are large.
On the contrary, regions in which the CMF is obtained by considering bound starless cores, including Serpens/Aquila~East, have a steeper high-mass power-law slopes $\gamma$ comparable to each other, and are in agreement within the error, with the Salpeter IMF slope. 
However, the core statistics of Ophiucus, Taurus, Corona~Australis and Lupus regions \citep[][, respectively]{lad16, mar16,bre18, ben18} is still not large enough to allow robust statistical conclusions about the power-law tail of the prestellar CMF of these regions, even for the youngest star-forming region. 

\begin{figure}
 \includegraphics[width=\columnwidth]{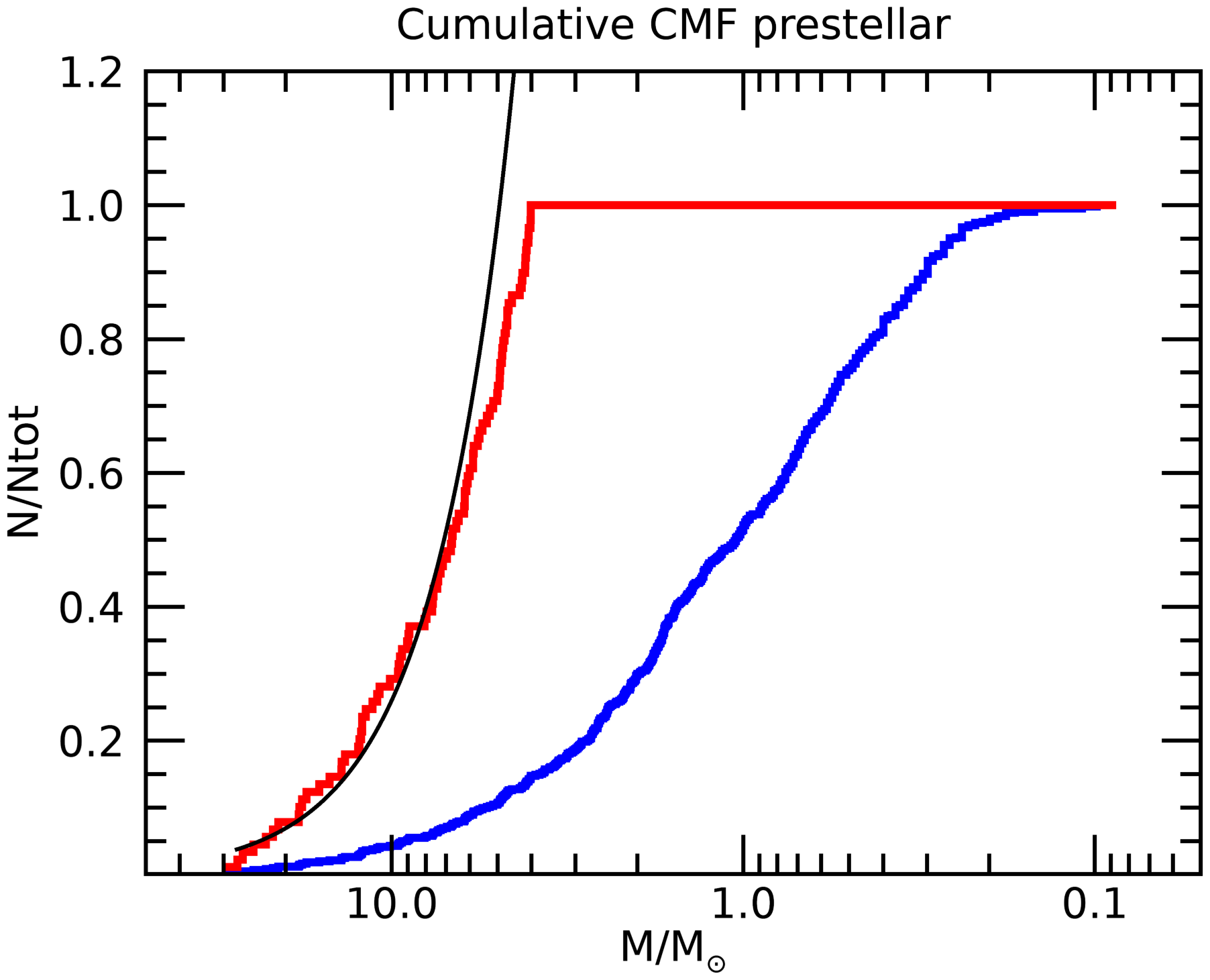}
 \caption{\label{fig:pl7}Inverse cumulative CMF for all prestellar cores (blue) and for those with $M > 2~\mathrm{M}_\odot$ (red), with the cumulative of power-law best fit shown in Fig.~\ref{fig:cmf} (black).}
\end{figure}

The CMF shape estimation can be affected by the adopted binning \citep[e.g.][]{olm13}. 
A way to verify whether or not our choice of the bin size is reasonable is to consider the cumulative function of the CMF, which is not affected by this problem \citep[cf.][]{tes98}. 
Accordingly, we computed it for the prestellar cores population, as shown in Fig.~\ref{fig:pl7}.  
To check the high-mass power-law exponent, note that we computed the cumulative of the sample inversely ordered, i.e. by decreasing mass. 
This is because we want to check the power-law behaviour of the high-mass tail of the distribution.
We compare this curve with the analytic cumulative of the following power-law 
\begin{equation}
\frac{dN}{d\log M} \propto M^{-\xi}
\end{equation}
obtained by fitting the CMF, where $\xi=\gamma - 1$\footnote{Note that in Figures~\ref{fig:cmf}, \ref{fig:cmfserm}, and \ref{fig:cmfaquila} the CMFs are plotted with respect $d \log M$, therefore the slope shown is $\xi$. To be compared with the values presented in literature one must derive $\gamma = \xi +1$}. 
In Fig.~\ref{fig:pl7}, we notice that the CMF shows 
a deficit of sources for $M>20$~M$_\odot$. 
This could be a consequence of the presence of a mass reservoir larger than usually found in a low-mass star forming region \citep{nak17}, suggesting that in Serpens/Aquila~East the mass has accreted onto the most massive sources instead of the least ones. 
This could be explained by the fact that the more massive is the core, the largest (and quicker) is the mass accretion rate onto that source, as we know from high-mass star formation. 

With respect to the previous analysis, HGBS results are more solid due to larger samples and lower limits in mass completeness.
That was achieved thanks to the unprecedented  sensitivity and angular resolution of \textit{Herschel}.
This enables us to probe not only the high-mass part of the CMF but also the presence of a peak CMF mass and, in certain cases \citep{pol13,kon15}, a significant part of the log-normal shape in the low-mass range. 
This is not the case of Serpens/Aquila~East region, being the completeness limit almost coincident with the peak of the prestellar CMF.
Albeit, we confirm the log-normal + power-law shape of the starless and prestellar CMFs. 
Moreover, there are 348 prestellar cores above the completeness limit, and 247 above the peak, which represents a significant improvement of the statistics over previous studies \citep{tes98,eno07}.

\section{FOCUS ON the two subregions} \label{comparison}
In previous sections, we analysed the sample of cores extracted from the whole area of the sky surveyed with \textit{Herschel}, but there are at least two reasons to discuss separately the core populations of Serpens~Main and Aquila~East, as identified in Figure~\ref{fig:nh2}. 
First, the analysis contained in previous sections already highlighted some differences between the two sub-regions (e.g., the presence of larger cores in the Serpens~Main sub-region, together with a higher fraction of bound cores, see Figure~\ref{fig:RaDecMeT}). 
Second, Serpens~Main is by far the more deeply studied of the two sub-regions, thus, considering only the samples of cores from Serpens~Main makes possible more direct comparisons with previous literature.
In this section, we therefore retrace the analyses made in Sections~\ref{coldenst}, \ref{source}, and \ref{cmfsection}, in light of a comparison of the two sub-regions.

\begin{figure}
 \includegraphics[width=\columnwidth]{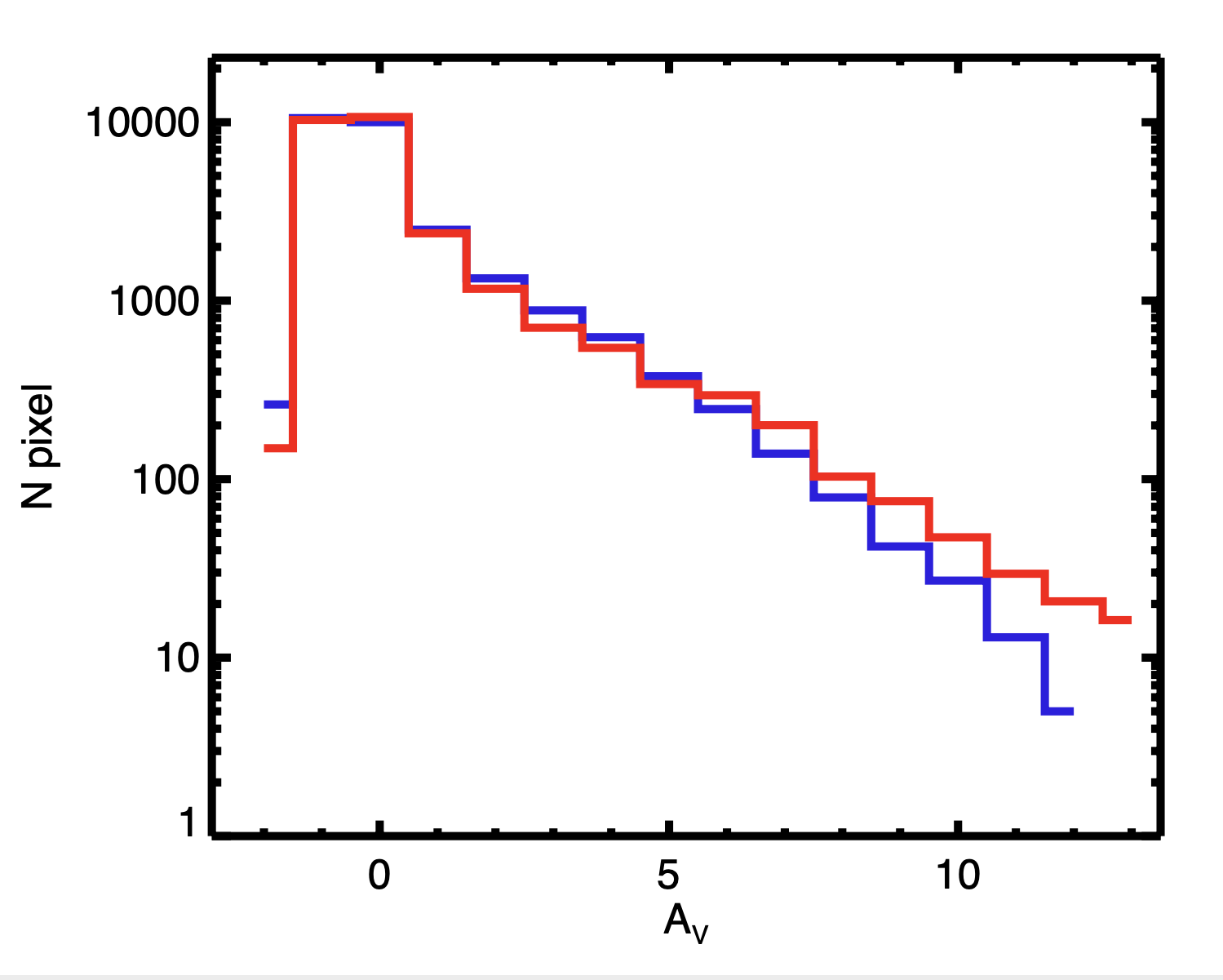}
 \caption{\label{fig:dobashi} Histograms of visual extinctions estimated by \citet{dob11} in the areas corresponding to our definition of Aquila~East (blue) and Serpens~Main (red). 
 The latter histogram is normalized it by the total number of pixels involved in the former. 
 Notice that the \citet{dob11} maps contain negative values, which are responsible for the unrealistic bins before or at $A_V=0$.}
\end{figure}

With Fig.~\ref{histot}, we already highlighted differences between histograms of both temperature and column density for the two sub-regions. 
The former is left-skewed for Aquila~East, and peaks at higher temperatures compared to that of Serpens~Main. 
Correspondingly, larger column densities are found in the Serpens~Main sub-region (Figure~\ref{histoCD}). 
A similar reciprocal behaviour can be recognized in the histograms of visual extinction obtained from the $A_V$ maps of \citet{dob11}, shown in Fig.~\ref{fig:dobashi}.
This behaviour can be easily correlated with the aforementioned presence of the most massive cores in Clusters~A and B of Serpens~Main, and a lower rate of unbound cores, compared with Aquila~East (Fig.~\ref{fig:RaDecMeT}). 
Higher temperatures (on average) in starless cores are generally explained by a more intense local interstellar radiation field \citep{eva01}. 
In the behaviour of dense cores temperature versus mean column density of cores (Fig. \ref{fig:pl_th2}) and mass (Fig.~\ref{fig:pl_tm}), it is possible to note further differences between the two regions. 
First, the spread in mean column density and mass is smaller for the Serpens~Main region than for the Aquila~East one. 
This difference is due not only to the larger number of cores in the Aquila~East, but also to the different composition of the two sub-populations. 
Indeed, Serpens~Main starless cores are 176 in total, 15 (8\%) of which are unbound cores, and 161 (92\%) are prestellar cores (111 of which are in turn robust prestellar cores). 
In comparison Aquila~East starless cores are 533 in total, 90 (17\%) of which are unbound cores and 443 (83\%) are prestellar cores, (344 of which are in turn robust prestellar cores).
Thus, the Serpens~Main sub-region has higher percentage of prestellar cores than Aquila~East, further indicating the former as a more active star forming region than the latter.
Besides, the distribution of the Serpens/Aquila~East population in Fig.~\ref{fig:pl_tm} is flattened at the instrument sensitivity limit, suggesting that a significant percentage of Serpens/Aquila~East cores might be less massive and colder but not detected, thus our results on mass and temperature could be overestimated. 
However, we cannot exclude that this difference in the amount of irradiation between Serpens~Main and Aquila~East might affect the core detection and extraction, which has been performed with a single and overall setup for \get. 

\begin{figure}
 \includegraphics[width=\columnwidth]{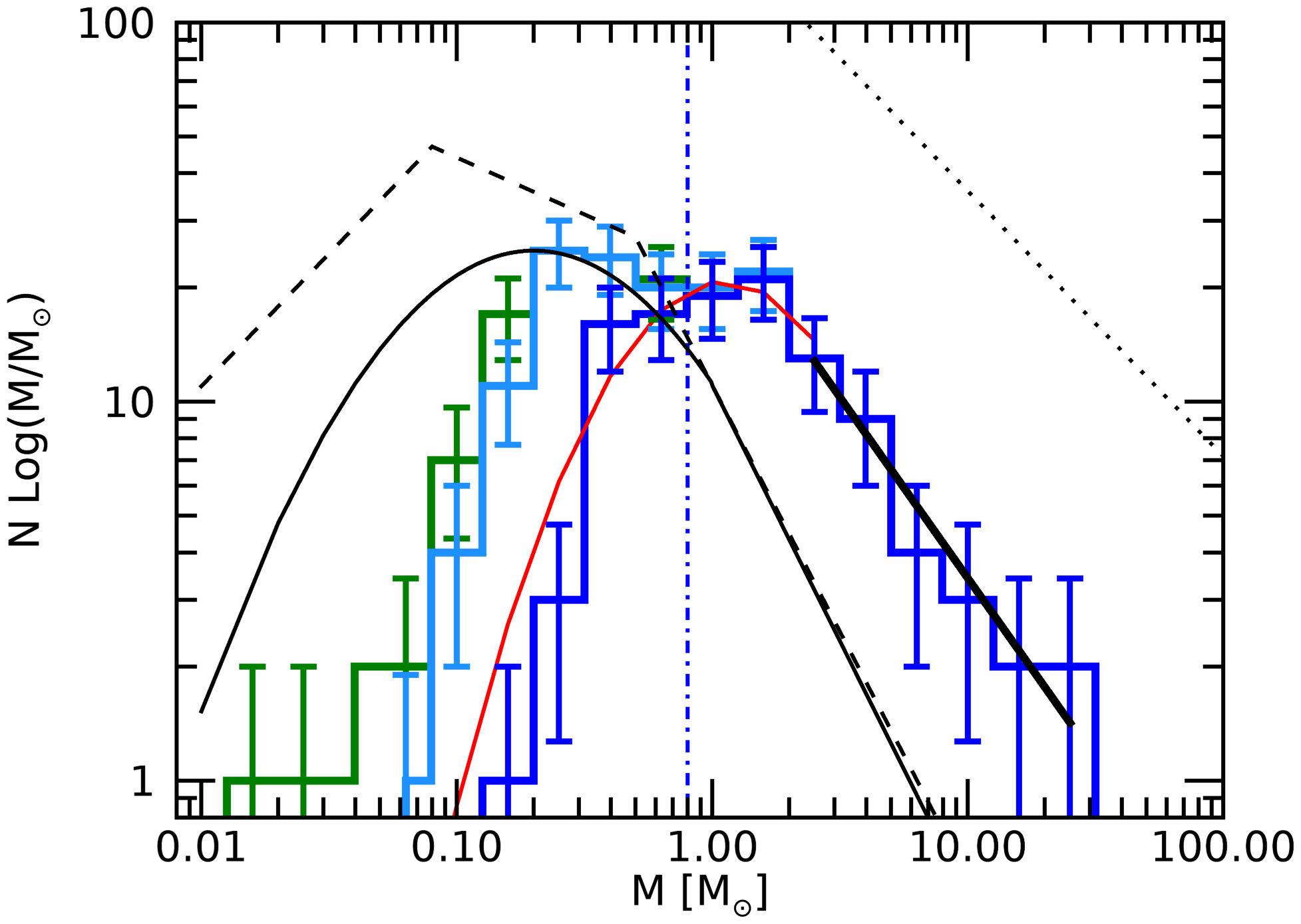}
 \caption{\label{fig:cmfserm}
 The same as Fig.~\ref{fig:cmf}, but for the Serpens~Main region only.
 The  high-mass tail of the prestellar CMF shape follows a power-law, with slope $\xi_{SM} =$$\cmfslopelessSM$ (solid thick black line). 
 One source with size larger than 0.2~pc is ruled out from the sample.}
\end{figure}

\subsection{Serpens Main CMF}\label{sect:serpmaincmf}

\begin{table}
 \centering
 \caption{\label{tab:cmfserp} Statistics of cores detected in Serpens~Main in the previous literature (TS98 = \citet{tes98}; E07=\citet{eno07}),  compared to this work. 
The number of cores effectively used to compute the CMF is highlighted in boldface.}

 \begin{tabular}{lccc}
  \hline 
  \hline
&TS98 & E07 & This work \\
  \hline 
  Protostellar & 26 & 14 &  58\\
  Starless & 6 & 21& 176 \\
  (Unbound) &&& (15) \\
  (Prestellar) &&& ({\bf 161}) \\
  Total &{\bf 32} & {\bf 35} & 234 \\
 
  \hline
  \hline
 \end{tabular}
\end{table}
It is interesting to analyze the CMF obtained only for the Serpens~Main sub-region, since, as said before, it is directly comparable with previous works, although based on a much larger statistics. 

The Serpens~Main CMF (Fig.~\ref{fig:cmfserm}) shows a nearly log-normal behaviour up to $\sim 1 - 2$~M$_\odot$, and a power-law trend at higher masses.
The estimated slope of the power-law fit is $\gamma_{SM} = \cmfslopeSM$, which is in agreement with that found for the entire region ($\gamma =$\cmfslope, Sect.~\ref{cmfsection}), although associated to a large error bar. 
In particular, the power-law fit fails in reproducing the rightmost bin. 

To compare the CMF calculated for Serpens~Main here and by \citet{tes98} and \citet{eno07}, we note that the way we built our CMF differs in some aspects from those of the earlier studies. 
First, a larger statistics is involved in our case, due to better sensitivity of \textit{Herschel} observations.
Second, here the CMF is coherently built only with prestellar cores, while 14 out of 35 cores observed by \citet{eno07} and 26 out of the 32 observed by \citet{tes98} are proto-stellar cores, therefore not directly comparable with our result.

However, due to the large error, the slope of the power-law fit for the Serpens~Main region is intermediate between the IMF and the CO clump mass function slopes.
Probably, this might be interpreted as a statistical effect: the Serpens~Main corresponds to a particular sub-region in which a concentration of relatively high-mass cores is present, but to recover a good fit of the typical power-law CMF it is necessary to complete the statistical basis with the masses from the remaining portion of the region. 

\begin{figure}
 \includegraphics[width=\columnwidth]{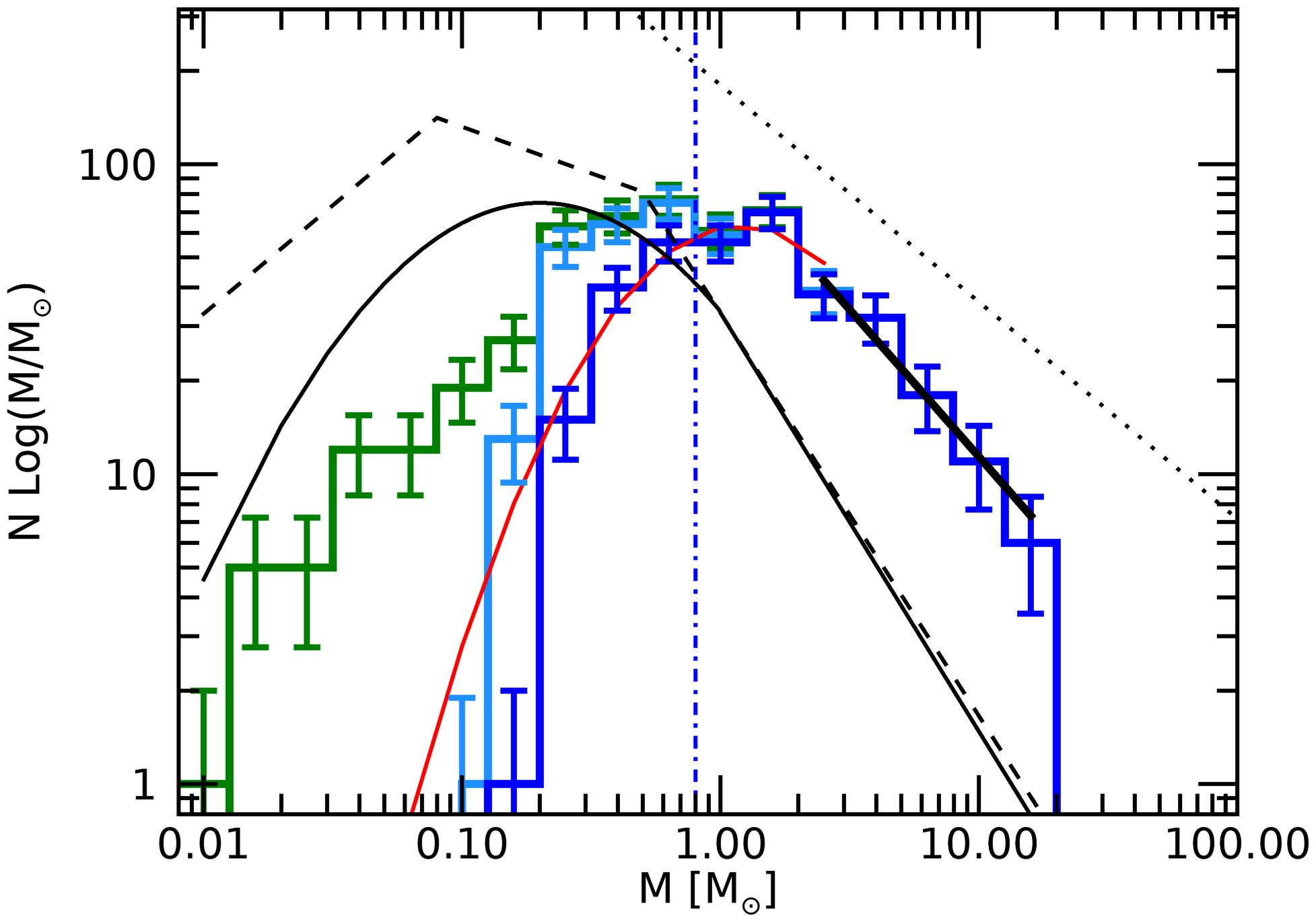}
 \caption{\label{fig:cmfaquila}
 The same as Fig.~\ref{fig:cmf}, but for the Aquila~East region only.
 The high-mass tail of the prestellar CMF shape follows a power-law, with slope $\xi_{AE}=\cmfslopelessAE$ (solid thick black line). 
 One source with size larger than 0.2~pc is ruled out from the sample.}
\end{figure}

\subsection{Aquila~East CMF}
The CMF of the Aquila~East sub-region (Fig.~\ref{fig:cmfaquila}) is obtained here for the first time.
As the CMF of the Serpens~Main sub-region, it shows a nearly log-normal behaviour up to $\sim 1 - 2$~M$_\odot$, and a power-law trend at higher masses.
The estimated slope of the power-law fit is $\gamma_{AE} = \cmfslopeAE$, which is the same as the Serpens~Main one, but with a smaller error bar (see Sect.~\ref{cmfsection}), and is also in agreement with that found for the entire region ($\gamma =$\cmfslope, Sect.~\ref{cmfsection}). 
Therefore, the comparison with other Gould Belt regions in Sect. 5 also applies to Aquila East.


The right tail of this CMF is closer to a power law with respect to that of Serpens Main, as indicated by the smaller uncertainty on the slope, which is probably related to the larger statistics: 443 prestellar cores are present in the Aquila~East region, compared to 161 in the Serpens~Main region. 

Finally we notice that also in the case of the Aquila~East CMF, the slope of the power-law fit is intermediate between the IMF and the CO clump mass function slopes.

\section{RELATIONSHIP BETWEEN CORES AND FILAMENTARY STRUCTURE IN SERPENS} \label{filamensSection}

 \begin{figure*}
  \centering

 \includegraphics[width=\textwidth]{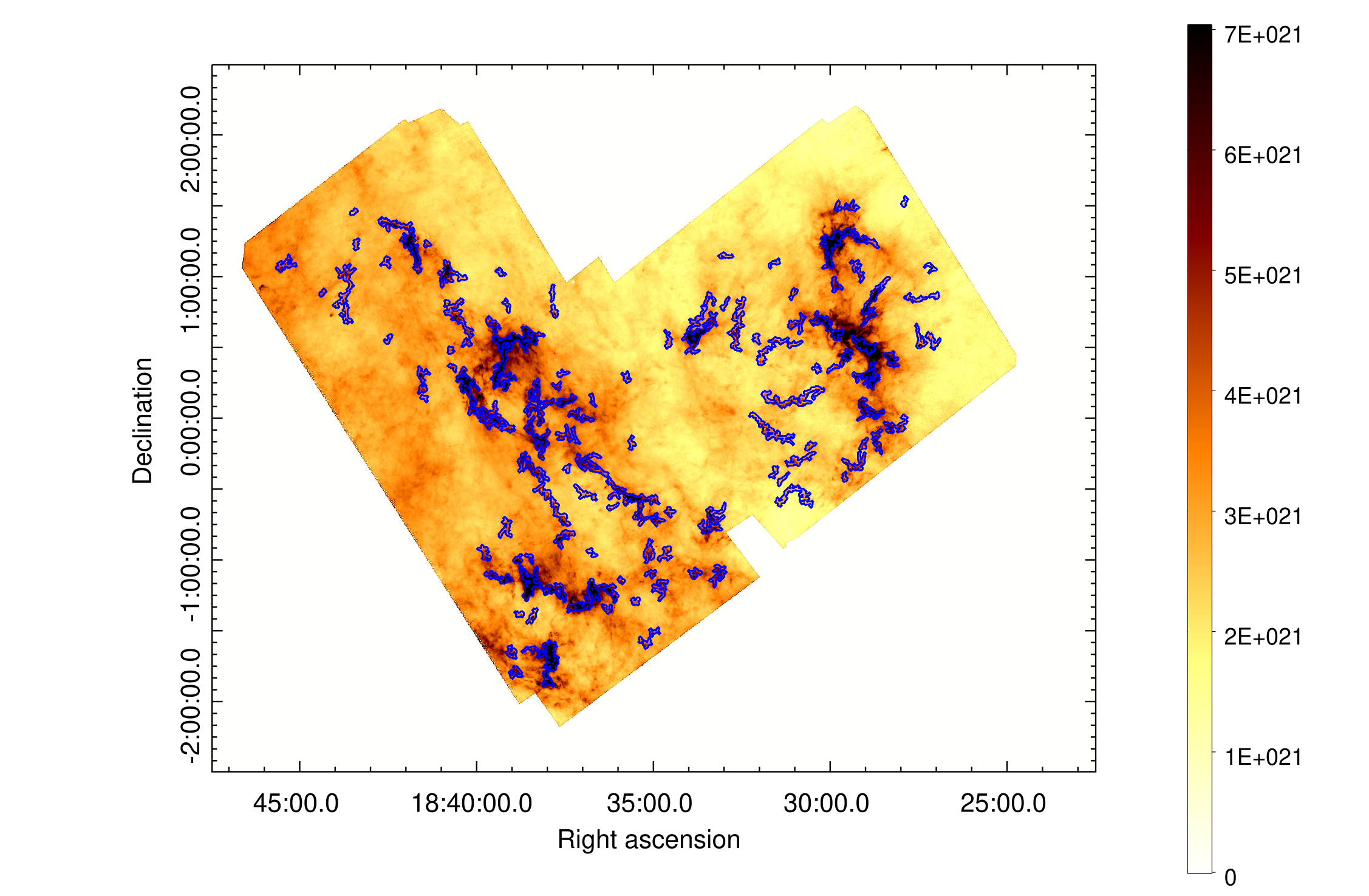}
 \caption{Boundaries of filamentary structures extracted as described in the text, overplotted on the column density map. The error bar displays values in cm$^{-2}$.\label{mapfil}}
 \end{figure*}
 \begin{figure}
  \centering
 \includegraphics[width=\columnwidth]{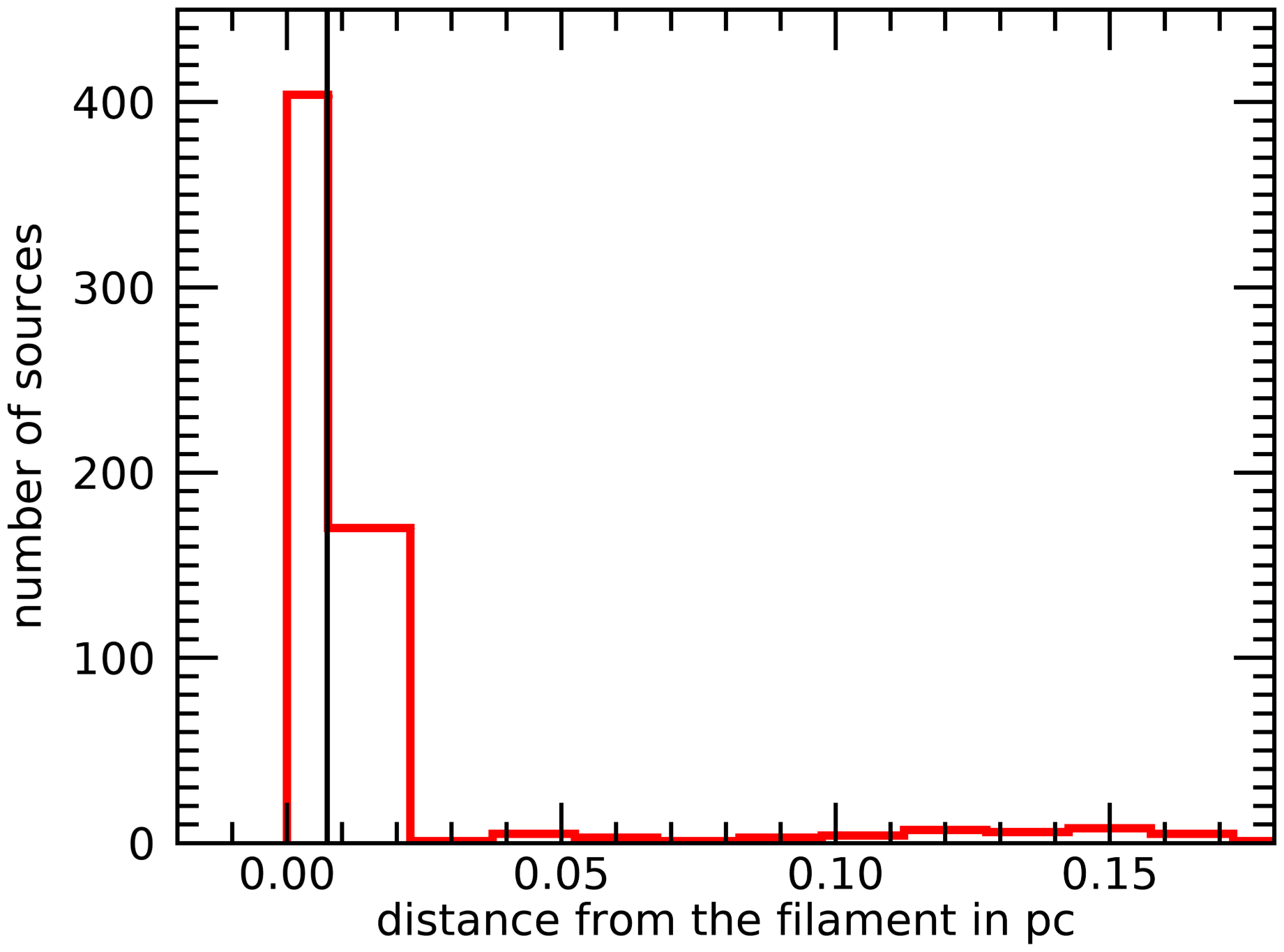}
 \caption{\label{histofil}Number of sources at a given distance from the nearest filament. The majority of the on-filament sources appear to be quite close to the spine of the filament. The vertical line corresponds to the resolution element of the column density map.}
 \end{figure}

 \citet{and10,and14} and \citet{kon15} described the relationship observed in Aquila~Main between supercritical filaments and prestellar cores. 
The fact that most prestellar cores were seen to be along supercritical filaments and, on the contrary, unbound cores were found to be situated along subcritical filaments, suggests that star formation predominantly occurs in filaments, making them a key structure in the global star formation scenario. 
Although we do not include a specific filament analysis here, we want to address this topic by briefly discussing the spatial association between cores and filaments in the overall Serpens/Aquila~East region.

To detect filaments in the Serpens/Aquila~East data, we applied to the column density map the algorithm presented in \citet{sch14} and \citet{sch20}, to which we refer the reader for a complete description of the algorithm. 
Here we just summarise basic details. 
Among several definitions of {\it filaments}, given in the recent literature , the one adopted by \citet{sch20} is ``any extended two-dimensional, cylindrical-like feature that is elongated and shows a higher brightness contrast with respect to its surroundings". 
Starting from this definition, such structures are identified by using the Hessian Matrix $H(x,y)$ of an image, its eigenvalues $\lambda_a, \lambda_b$, and their linear combination. 
The algorithm is applied to the column density map $N_{{\rm H}_2}(x,y)$. 
During the filament extraction phase, $H(x,y)$ is diagonalized and its eigenvalues computed. 
The diagonalization of the Hessian Matrix corresponds to the rotation of the axes towards the direction in which $N_{{\rm H}_2}(x,y)$ has the maximum and minimum variation. 
These variations are measured by $\lambda_a$ and $\lambda_b$, respectively. 
In this way it is possible to properly identify a cylindrical shape.
 
This algorithm requests two input parameters: a threshold value $Tr$ and a dilation parameter, that permits to extend the initial mask (essentially based on the filament spine) in order to include the entire filament area with its borders. 
In Appendix~B of \citet{sch20} the range of variability for these parameters is widely investigated, and the choice of default values for them is justified. 
Here we use the default value to set the dilation parameter, and $Tr=3.5 \times \sigma$ as threshold value, different from the default value $Tr=3 \times \sigma$, because on the one hand it corresponds to a better definition and separation of filaments in the Serpens~Main region (especially in Clusters~A and B), and, on the other hand, it makes no appreciable changes on filaments extracted in the Aquila~East region.
 
This algorithm was already used in HGBS papers \citep{ben15,ben18}. 
They determine that a source is spatially associated with a filament if its central position falls within the boundary of the filament, as it is evaluated by the algorithm. 
Following this method, we find that \onfilall of the entire catalogue sources are on filaments. 
Of these, \onfilpre are prestellar cores and \onfilun are unbound cores. 
 
This analysis of the core-filament relationship can be extended by adopting a further definition of spatial association, inspired by the approach of \citet{sew19} for associating YSOs to filaments, i.e. computing the minimum distance from each source to the nearest filament. 
From the histogram of these distances (Fig.~\ref{histofil}), we note there is a remarkable amount of cores concentrated within 0.03~pc from the spines of filaments, which is the physical distance corresponding to 1 pixel. 
Therefore, we decide, as an alternative, to associate a core to a filament if it is located at  most 0.03~pc from a filament spine. %

With the second operational definition, we found that \onfil of all detected cores are on-filament within 0.03~pc. 
 \begin{figure}
  \centering
 \includegraphics[width=\columnwidth]{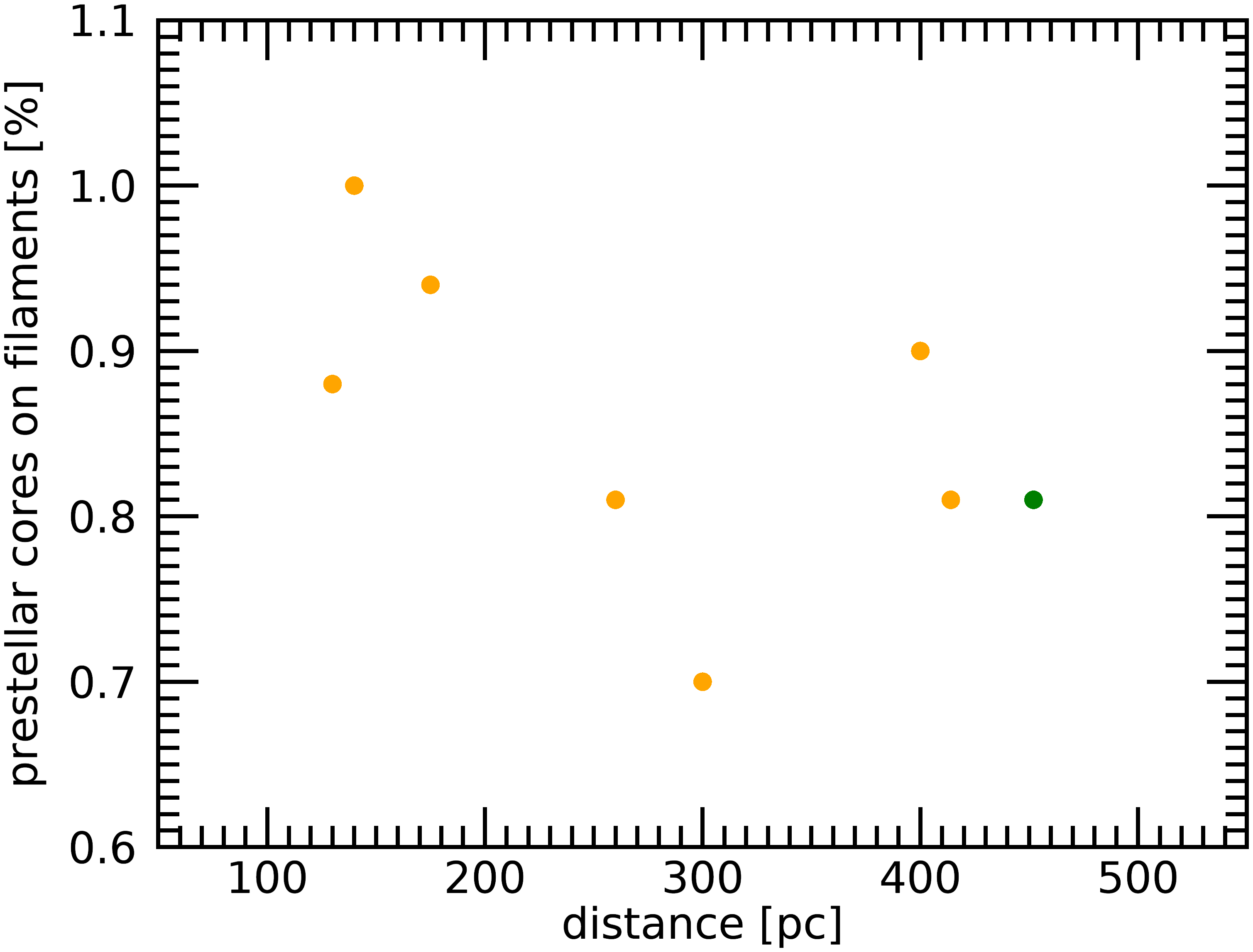}
 \caption{\label{corfil}Percentage of on-filament prestellar cores for the HGBS star-forming regions reported in Table~\ref{tab:cmf}. 
 We use the average value of 175~pc for Lupus cloud and 452~pc for Serpens. 
 The value corresponding to Serpens is in green.}
 \end{figure}
Note that the number of on-filament sources estimated through the first method (\onfilall) is larger than that from the second (\onfil).
In the first method, we also labelled as on-filament those sources that are close to the border of the filament, which are ruled out if their distances from the filament spine are considered.
We consider the results of the second method more robust, and use them in the following discussion.

The majority of the cores which lie on to filamentary structures are prestellar cores (467 out of 574, 81\%), about 14\% (77) are protostellar, and only 5\% (30) on-filament sources are unbound. This confirm the scenario in which most cores found to be associated with filaments are the ones that will collapse to form stars; instead, cores located outside the filamentary web are more likely to be gravitationally unbound. 

To make a comparison between our result and the HGBS previous ones, in Table~\ref{tab:cmf} the percentage of on-filament prestellar cores is reported for each region catalogued so far. 
For all the analyzed regions, this quantity is larger or equal than $67 \%$, being in agreement with the scenario in which stars preferentially form in the filamentary structure. 
Anyway, there is a spread in this percentage up to $100\%$, depending on the region and on the different techniques to detect filaments and associate cores to them.
This spread, instead, does not depend on the distance, since no correlation is seen in Fig.~\ref{corfil}, nor on the type of cores. 
Therefore, the percentage of on-filament sources is likely to be a combination of a peculiar feature of each specific region and of the filament detection technique adopted, which is different in different works. 
For this reason, such analysis should be considered only from the qualitative point of view.

In this respect, the Serpens/Aquila~East region is configured as one with more on-filament prestellar cores. 
Fig.~\ref{corfil} also shows that a relevant fraction of the prestellar population is found to lie on filaments. 
This supports the idea that filamentary structures are preferential places for star formation.

\section{Conclusions} \label{conclusions}

In this article, one of the HGBS ``first generation papers'', we presented the \textit{Herschel} photometric survey (at 70~$\mu$m, 160~$\mu$m, 250~$\mu$m, 350~$\mu$m and 500~$\mu$m) of the Serpens/Aquila~East star-forming region, in which two sub-regions can be easily identified: Serpens~Main, already studied in the past in different spectral ranges, and Aquila~East, so far poorly studied and characterized.

Column density and temperature maps were produced, and their analysis revealed that:
\begin{itemize}
    \item The temperature map contains values between $\sim 14$~K and 19~K. 
    The temperature distributions of the two sub-regions appear quite different, however, with Aquila~East being on average warmer than the Serpens~Main, which is probably more shielded from the interstellar radiation field coming from the Galactic plane. 
    \item The total mass of the region has been computed from the column density map, amounting to about \totalmass~M$_\odot$, larger with respect to estimates in previous works. 
\end{itemize}
 
From the \textit{Herschel} maps, we also extracted the first complete catalog of prestellar sources. Over the observed field, \nsources dense cores are detected, of which \nstarlesscores are starless and \nproto are proto-stellar. 
We analyzed the physical properties of this sub-sample (mass, temperature, and size) and their position with respect the filamentary structure of the cloud. 
The main results of this analysis are: 
\begin{itemize}
    \item From the gravitational stability analysis, \nunboundcores cores are found to be gravitationally unbound, and \nprestellarcores are bound (prestellar).
    \item The $M$ versus $r$ diagram confirms that Serpens/Aquila~East is a low-mass star-forming region. 
    Only 3 prestellar cores might be characterized as compatible with high-mass star formation, but this classification strongly depends on the adopted core-to-star efficiency.
    \item We find that 45\% of the prestellar cores lie in regions populated by {\it Spitzer} YSOs, indicating the presence of a mass reservoir for further star formation. 
    \item A negative correlation is found between cores temperature and mean column density, but the coefficients are not consistent with previous estimates for the other HGBS regions.
    \item The prestellar CMF of the Serpens overall region is well fitted by a log-normal function at masses lower than $\sim 2$~M$_\odot$, and, at higher masses, by a power-law whose slope $\gamma =$~\cmfslope.
    \item Both the CMFs restricted to the Serpens~Main and Aquila~East sub-regions are well fitted in turn by a log-normal function at low masses, and by a power-law at high-masses, respectively, with a slope similar to that of the overall region: $\cmfslopeSM$, and $\cmfslopeAE$, respectively. 
    \item We performed the first comparison between CMFs of the HGBS star-forming regions. 
    We notice that the CMF shape strictly depends on the region and the type of cores contained in it. 
    The CMF of unbound cores (Corona~Australis and Taurus) shows a shape similar to that of CO clumps. 
    On the contrary, all the regions with a reasonable number of prestellar cores (Aquila~Main, Perseus, Orion A and B, Serpens, Cepheus) show a log-normal form in the low-mass regime and the power-law slope at high masses in agreement with the IMF one. 
    This compatibility between prestellar CMF and IMF slopes suggests that the latter is determined by the former, which in turn depends on the cloud fragmentation process.
    \item We extract the filamentary structure with the tool described in \citet{sch20}, finding that \onfil of the Serpens cores lie on-filaments. 
    More specifically, \onfildistpre of them are prestellar cores, 13\% are protostellar cores, and \onfildistun are unbound cores. 
    This strong correspondence between prestellar core locations and filamentary structures supports the idea that filaments are the structures of the molecular clouds in which stars preferentially form.
\end{itemize}

\section*{Acknowledgements}
SPIRE has been developed by a consortium of institutes led by Cardiff Univ. (UK) and including: Univ. Lethbridge (Canada); NAOC (China); CEA, LAM (France); IFSI, Univ. Padua (Italy); IAC (Spain); Stockholm Observatory (Sweden); Imperial College London, RAL, UCL-MSSL, UKATC, Univ. Sussex (UK); and Caltech, JPL, NHSC, Univ. Colorado (USA). This development has been supported by national funding agencies: CSA (Canada); NAOC (China); CEA, CNES, CNRS (France); ASI (Italy); MCINN (Spain); SNSB (Sweden); STFC, UKSA (UK); and NASA (USA). PACS has been developed by a consortium of institutes led by MPE (Germany) and includ- ing UVIE (Austria); KUL, CSL, IMEC (Belgium); CEA, OAMP (France); MPIA (Germany); IFSI, OAP/AOT, OAA/CAISMI, LENS, SISSA (Italy); IAC (Spain). This development has been supported by the funding agencies BMVIT (Austria), ESA-PRODEX (Belgium), CEA/CNES (France), DLR (Germany), ASI (Italy), and CICT/MCT (Spain). \newline
D.A. acknowledges support by FCT/MCTES through national funds (PIDDAC) by the grants UID/FIS/04434/2019 \& UIDB/04434/2020. \newline
The computational needs of this project were supported by the GENESIS cluster, which was funded by the project PRIN-INAF 2016 The Cradle of Life - GENESIS-SKA (General Conditions in Early Planetary Systems for the rise of life with SKA). \newline
This work has received support from the European Research Council under the European Union's Seventh Framework Program (ERC Advanced Grant Agreement no. 291294 - ``ORISTARS") and from the French national programs of CNRS/INSU on stellar and ISM physics (PNPS and PCMI).
\newline
A.B. acknowledges the support of the European Union’s Horizon 2020 research and innovation program under the Marie Skłodowska-Curie Grant agreement No. 843008 (MUSICA).

\section*{Data availability}
All data are incorporated into the article and its online supplementary material.



\bibliographystyle{mnras}
\bibliography{biblio} 

\begin{thebibliography}{}
\makeatletter
\relax
\def\mn@urlcharsother{\let\do\@makeother \do\$\do\&\do\#\do\^\do\_\do\%\do\~}
\def\mn@doi{\begingroup\mn@urlcharsother \@ifnextchar [ {\mn@doi@}
  {\mn@doi@[]}}
\def\mn@doi@[#1]#2{\def\@tempa{#1}\ifx\@tempa\@empty \href
  {http://dx.doi.org/#2} {doi:#2}\else \href {http://dx.doi.org/#2} {#1}\fi
  \endgroup}
\def\mn@eprint#1#2{\mn@eprint@#1:#2::\@nil}
\def\mn@eprint@arXiv#1{\href {http://arxiv.org/abs/#1} {{\tt arXiv:#1}}}
\def\mn@eprint@dblp#1{\href {http://dblp.uni-trier.de/rec/bibtex/#1.xml}
  {dblp:#1}}
\def\mn@eprint@#1:#2:#3:#4\@nil{\def\@tempa {#1}\def\@tempb {#2}\def\@tempc
  {#3}\ifx \@tempc \@empty \let \@tempc \@tempb \let \@tempb \@tempa \fi \ifx
  \@tempb \@empty \def\@tempb {arXiv}\fi \@ifundefined
  {mn@eprint@\@tempb}{\@tempb:\@tempc}{\expandafter \expandafter \csname
  mn@eprint@\@tempb\endcsname \expandafter{\@tempc}}}

\bibitem[\protect\citeauthoryear{{Alton}, {Xilouris}, {Misiriotis}, {Dasyra}
  \& {Dumke}}{{Alton} et~al.}{2004}]{alt04}
{Alton} P.~B.,  {Xilouris} E.~M.,  {Misiriotis} A.,  {Dasyra} K.~M.,   {Dumke}
  M.,  2004, \mn@doi [\aap] {10.1051/0004-6361:20040438}, \href
  {https://ui.adsabs.harvard.edu/abs/2004A&A...425..109A} {425, 109}

\bibitem[\protect\citeauthoryear{{Alves}, {Lombardi}  \& {Lada}}{{Alves}
  et~al.}{2007}]{alv07}
{Alves} J.,  {Lombardi} M.,   {Lada} C.~J.,  2007, \mn@doi [\aap]
  {10.1051/0004-6361:20066389}, \href
  {http://adsabs.harvard.edu/abs/2007A%26A...462L..17A} {462, L17}

\bibitem[\protect\citeauthoryear{{Andr{\'e}} et~al.,}{{Andr{\'e}}
  et~al.}{2010}]{and10}
{Andr{\'e}} P.,  et~al., 2010, \mn@doi [\aap] {10.1051/0004-6361/201014666},
  \href {http://adsabs.harvard.edu/abs/2010A%26A...518L.102A} {518, L102}

\bibitem[\protect\citeauthoryear{{Andr{\'e}}, {Di Francesco}, {Ward-Thompson},
  {Inutsuka}, {Pudritz}  \& {Pineda}}{{Andr{\'e}} et~al.}{2014}]{and14}
{Andr{\'e}} P.,  {Di Francesco} J.,  {Ward-Thompson} D.,  {Inutsuka} S.~I.,
  {Pudritz} R.~E.,   {Pineda} J.~E.,  2014, in {Beuther} H.,  {Klessen} R.~S.,
  {Dullemond} C.~P.,   {Henning} T.,  eds, Protostars and Planets VI. p.~27
  (\mn@eprint {arXiv} {1312.6232}),
  \mn@doi{10.2458/azu_uapress_9780816531240-ch002}

\bibitem[\protect\citeauthoryear{{Benedettini}, {Schisano}, {Pezzuto}  \& {et
  al.}}{{Benedettini} et~al.}{2015}]{ben15}
{Benedettini} M.,  {Schisano} E.,  {Pezzuto} S.,   {et al.} 2015, \mn@doi
  [\mnras] {10.1093/mnras/stv1750}, \href
  {http://adsabs.harvard.edu/abs/2015MNRAS.453.2036B} {453, 2036}

\bibitem[\protect\citeauthoryear{{Benedettini}, {Pezzuto}, {Schisano}  \& {et
  al.}}{{Benedettini} et~al.}{2018}]{ben18}
{Benedettini} M.,  {Pezzuto} S.,  {Schisano} E.,   {et al.} 2018, \mn@doi
  [\aap] {10.1051/0004-6361/201833364}, \href
  {https://ui.adsabs.harvard.edu/abs/2018A&A...619A..52B} {619, A52}

\bibitem[\protect\citeauthoryear{{Bergin} \& {Tafalla}}{{Bergin} \&
  {Tafalla}}{2007}]{ber07}
{Bergin} E.~A.,  {Tafalla} M.,  2007, \mn@doi [\araa]
  {10.1146/annurev.astro.45.071206.100404}, \href
  {https://ui.adsabs.harvard.edu/abs/2007ARA&A..45..339B} {45, 339}

\bibitem[\protect\citeauthoryear{{Bernard} et~al.,}{{Bernard}
  et~al.}{2010}]{ber10}
{Bernard} J.-P.,  et~al., 2010, \mn@doi [\aap] {10.1051/0004-6361/201014540},
  \href {http://adsabs.harvard.edu/abs/2010A\%\ 26A...518L..88B} {518, L88}

\bibitem[\protect\citeauthoryear{{Bohlin}, {Savage}  \& {Drake}}{{Bohlin}
  et~al.}{1978}]{boh78}
{Bohlin} R.~C.,  {Savage} B.~D.,   {Drake} J.~F.,  1978, \mn@doi [\apj]
  {10.1086/156357}, \href {http://adsabs.harvard.edu/abs/1978ApJ...224..132B}
  {224, 132}

\bibitem[\protect\citeauthoryear{{Bonnor}}{{Bonnor}}{1956}]{bon56}
{Bonnor} W.~B.,  1956, \mn@doi [\mnras] {10.1093/mnras/116.3.351}, \href
  {http://adsabs.harvard.edu/abs/1956MNRAS.116..351B} {116, 351}

\bibitem[\protect\citeauthoryear{{Bresnahan}, {Ward-Thompson}, {Kirk}  \& {et
  al.}}{{Bresnahan} et~al.}{2018}]{bre18}
{Bresnahan} D.,  {Ward-Thompson} D.,  {Kirk} J.~M.,   {et al.} 2018, \mn@doi
  [\aap] {10.1051/0004-6361/201730515}, \href
  {http://adsabs.harvard.edu/abs/2018A%26A...615A.125B} {615, A125}

\bibitem[\protect\citeauthoryear{{Casali}, {Eiroa}  \& {Duncan}}{{Casali}
  et~al.}{1993}]{cas93}
{Casali} M.~M.,  {Eiroa} C.,   {Duncan} W.~D.,  1993, \aap, \href
  {http://adsabs.harvard.edu/abs/1993A%26A...275..195C} {275, 195}

\bibitem[\protect\citeauthoryear{{Chabrier}}{{Chabrier}}{2005}]{cha05}
{Chabrier} G.,  2005, in {Corbelli} E.,  {Palla} F.,   {Zinnecker} H.,  eds,
  Astrophysics and Space Science Library Vol. 327, The Initial Mass Function 50
  Years Later. p.~41 (\mn@eprint {} {astro-ph/0409465}),
  \mn@doi{10.1007/978-1-4020-3407-7_5}

\bibitem[\protect\citeauthoryear{{Chavarria-K.}, {de Lara}, {Finkenzeller},
  {Mendoza}  \& {Ocegueda}}{{Chavarria-K.} et~al.}{1988}]{cha88}
{Chavarria-K.} C.,  {de Lara} E.,  {Finkenzeller} U.,  {Mendoza} E.~E.,
  {Ocegueda} J.,  1988, \aap, \href
  {http://adsabs.harvard.edu/abs/1988A%26A...197..151C} {197, 151}

\bibitem[\protect\citeauthoryear{{Di Francesco}, {Evans}, {Caselli}, {Myers},
  {Shirley}, {Aikawa}  \& {Tafalla}}{{Di Francesco} et~al.}{2007}]{dif07}
{Di Francesco} J.,  {Evans} N.~J. I.,  {Caselli} P.,  {Myers} P.~C.,  {Shirley}
  Y.,  {Aikawa} Y.,   {Tafalla} M.,  2007, in {Reipurth} B.,  {Jewitt} D.,
  {Keil} K.,  eds, Protostars and Planets V. p.~17 (\mn@eprint {arXiv}
  {astro-ph/0602379})

\bibitem[\protect\citeauthoryear{{Di Francesco}, {Keown}, {Fallscheer}  \& {et
  al.}}{{Di Francesco} et~al.}{2020}]{dif20}
{Di Francesco} J.,  {Keown} J.,  {Fallscheer} C.,   {et al.} 2020, \apj

\bibitem[\protect\citeauthoryear{{Djupvik}, {Andr{\'e}}, {Bontemps}, {Motte},
  {Olofsson}, {G{\aa}lfalk}  \& {Flor{\'e}n}}{{Djupvik} et~al.}{2006}]{dju06}
{Djupvik} A.~A.,  {Andr{\'e}} P.,  {Bontemps} S.,  {Motte} F.,  {Olofsson} G.,
  {G{\aa}lfalk} M.,   {Flor{\'e}n} H.-G.,  2006, \mn@doi [\aap]
  {10.1051/0004-6361:20065533}, \href
  {http://adsabs.harvard.edu/abs/2006A%26A...458..789D} {458, 789}

\bibitem[\protect\citeauthoryear{{Dobashi}}{{Dobashi}}{2011}]{dob11}
{Dobashi} K.,  2011, \mn@doi [\pasj] {10.1093/pasj/63.sp1.S1}, \href
  {https://ui.adsabs.harvard.edu/abs/2011PASJ...63S...1D} {63, S1}

\bibitem[\protect\citeauthoryear{{Dunham}, {Crapsi}, {Evans}, {Bourke},
  {Huard}, {Myers}  \& {Kauffmann}}{{Dunham} et~al.}{2008}]{dun08}
{Dunham} M.~M.,  {Crapsi} A.,  {Evans} Neal~J. I.,  {Bourke} T.~L.,  {Huard}
  T.~L.,  {Myers} P.~C.,   {Kauffmann} J.,  2008, \mn@doi [\apjs]
  {10.1086/591085}, \href
  {https://ui.adsabs.harvard.edu/abs/2008ApJS..179..249D} {179, 249}

\bibitem[\protect\citeauthoryear{{Dunham} et~al.,}{{Dunham}
  et~al.}{2015}]{dun15}
{Dunham} M.~M.,  et~al., 2015, \mn@doi [\apjs] {10.1088/0067-0049/220/1/11},
  \href {https://ui.adsabs.harvard.edu/abs/2015ApJS..220...11D} {220, 11}

\bibitem[\protect\citeauthoryear{{Dzib}, {Loinard}, {Mioduszewski}, {Boden},
  {Rodr{\'{\i}}guez}  \& {Torres}}{{Dzib} et~al.}{2010}]{dzi10}
{Dzib} S.,  {Loinard} L.,  {Mioduszewski} A.~J.,  {Boden} A.~F.,
  {Rodr{\'{\i}}guez} L.~F.,   {Torres} R.~M.,  2010, \mn@doi [\apj]
  {10.1088/0004-637X/718/2/610}, \href
  {http://adsabs.harvard.edu/abs/2010ApJ...718..610D} {718, 610}

\bibitem[\protect\citeauthoryear{{Dzib}, {Loinard}, {Mioduszewski}, {Boden},
  {Rodr{\'{\i}}guez}  \& {Torres}}{{Dzib} et~al.}{2011}]{dzi11}
{Dzib} S.,  {Loinard} L.,  {Mioduszewski} A.~J.,  {Boden} A.~F.,
  {Rodr{\'{\i}}guez} L.~F.,   {Torres} R.~M.,  2011, in Revista Mexicana de
  Astronomia y Astrofisica Conference Series. pp 231--232

\bibitem[\protect\citeauthoryear{{Eiroa}, {Djupvik}  \& {Casali}}{{Eiroa}
  et~al.}{2008}]{eir08}
{Eiroa} C.,  {Djupvik} A.~A.,   {Casali} M.~M.,  2008, {The Serpens Molecular
  Cloud}.
{Reipurth}, B., p.~693

\bibitem[\protect\citeauthoryear{{Elia} \& {Pezzuto}}{{Elia} \&
  {Pezzuto}}{2016}]{eli16}
{Elia} D.,  {Pezzuto} S.,  2016, \mn@doi [\mnras] {10.1093/mnras/stw1399},
  \href {http://adsabs.harvard.edu/abs/2016MNRAS.461.1328E} {461, 1328}

\bibitem[\protect\citeauthoryear{{Elia} et~al.,}{{Elia} et~al.}{2013}]{eli13}
{Elia} D.,  et~al., 2013, \mn@doi [\apj] {10.1088/0004-637X/772/1/45}, \href
  {http://adsabs.harvard.edu/abs/2013ApJ...772...45E} {772, 45}

\bibitem[\protect\citeauthoryear{{Elia} et~al.,}{{Elia} et~al.}{2017}]{eli17}
{Elia} D.,  et~al., 2017, \mn@doi [\mnras] {10.1093/mnras/stx1357}, \href
  {https://ui.adsabs.harvard.edu/abs/2017MNRAS.471..100E} {471, 100}

\bibitem[\protect\citeauthoryear{{Enoch}, {Glenn}, {Evans}, {Sargent}, {Young}
  \& {Huard}}{{Enoch} et~al.}{2007}]{eno07}
{Enoch} M.~L.,  {Glenn} J.,  {Evans} II N.~J.,  {Sargent} A.~I.,  {Young}
  K.~E.,   {Huard} T.~L.,  2007, \mn@doi [\apj] {10.1086/520321}, \href
  {http://adsabs.harvard.edu/abs/2007ApJ...666..982E} {666, 982}

\bibitem[\protect\citeauthoryear{{Enoch}, {Evans}, {Sargent}, {Glenn},
  {Rosolowsky}  \& {Myers}}{{Enoch} et~al.}{2008}]{eno08}
{Enoch} M.~L.,  {Evans} II N.~J.,  {Sargent} A.~I.,  {Glenn} J.,  {Rosolowsky}
  E.,   {Myers} P.,  2008, \mn@doi [\apj] {10.1086/589963}, \href
  {http://adsabs.harvard.edu/abs/2008ApJ...684.1240E} {684, 1240}

\bibitem[\protect\citeauthoryear{{Evans}, {Rawlings}, {Shirley}  \&
  {Mundy}}{{Evans} et~al.}{2001}]{eva01}
{Evans} Neal~J. I.,  {Rawlings} J. M.~C.,  {Shirley} Y.~L.,   {Mundy} L.~G.,
  2001, \mn@doi [\apj] {10.1086/321639}, \href
  {https://ui.adsabs.harvard.edu/abs/2001ApJ...557..193E} {557, 193}

\bibitem[\protect\citeauthoryear{{Federrath}}{{Federrath}}{2013}]{fed13}
{Federrath} C.,  2013, \mn@doi [\mnras] {10.1093/mnras/stt1644}, \href
  {http://adsabs.harvard.edu/abs/2013MNRAS.436.1245F} {436, 1245}

\bibitem[\protect\citeauthoryear{{Griffin} et~al.,}{{Griffin}
  et~al.}{2010}]{gri10}
{Griffin} M.~J.,  et~al., 2010, \mn@doi [\aap] {10.1051/0004-6361/201014519},
  \href {http://adsabs.harvard.edu/abs/2010A%26A...518L...3G} {518, L3}

\bibitem[\protect\citeauthoryear{{Harvey} et~al.,}{{Harvey}
  et~al.}{2006}]{har06}
{Harvey} P.~M.,  et~al., 2006, \mn@doi [\apj] {10.1086/503520}, \href
  {http://adsabs.harvard.edu/abs/2006ApJ...644..307H} {644, 307}

\bibitem[\protect\citeauthoryear{{Harvey}, {Mer{\'\i}n}, {Huard}, {Rebull},
  {Chapman}, {Evans}  \& {Myers}}{{Harvey} et~al.}{2007}]{har07}
{Harvey} P.,  {Mer{\'\i}n} B.,  {Huard} T.~L.,  {Rebull} L.~M.,  {Chapman} N.,
  {Evans} Neal~J. I.,   {Myers} P.~C.,  2007, \mn@doi [\apj] {10.1086/518646},
  \href {https://ui.adsabs.harvard.edu/abs/2007ApJ...663.1149H} {663, 1149}

\bibitem[\protect\citeauthoryear{{Herczeg} et~al.,}{{Herczeg}
  et~al.}{2019}]{her19}
{Herczeg} G.~J.,  et~al., 2019, \mn@doi [\apj] {10.3847/1538-4357/ab1d67},
  \href {https://ui.adsabs.harvard.edu/abs/2019ApJ...878..111H} {878, 111}

\bibitem[\protect\citeauthoryear{{Herschel Science Ground Segment
  Consortium}}{{Herschel Science Ground Segment Consortium}}{2011}]{her11}
{Herschel Science Ground Segment Consortium} 2011, {HIPE: Herschel Interactive
  Processing Environment}, Astrophysics Source Code Library (\mn@eprint {ascl}
  {1111.001})

\bibitem[\protect\citeauthoryear{{Hildebrand}}{{Hildebrand}}{1983}]{hil83}
{Hildebrand} R.~H.,  1983, \qjras, \href
  {http://adsabs.harvard.edu/abs/1983QJRAS..24..267H} {24, 267}

\bibitem[\protect\citeauthoryear{{Kaas} et~al.,}{{Kaas} et~al.}{2004}]{kaa04}
{Kaas} A.~A.,  et~al., 2004, \mn@doi [\aap] {10.1051/0004-6361:20035775}, \href
  {http://adsabs.harvard.edu/abs/2004A%26A...421..623K} {421, 623}

\bibitem[\protect\citeauthoryear{{Kainulainen}, {Beuther}, {Banerjee},
  {Federrath}  \& {Henning}}{{Kainulainen} et~al.}{2011}]{kai11}
{Kainulainen} J.,  {Beuther} H.,  {Banerjee} R.,  {Federrath} C.,   {Henning}
  T.,  2011, \mn@doi [\aap] {10.1051/0004-6361/201016383}, \href
  {http://adsabs.harvard.edu/abs/2011A%26A...530A..64K} {530, A64}

\bibitem[\protect\citeauthoryear{{Kauffmann} \& {Pillai}}{{Kauffmann} \&
  {Pillai}}{2010}]{kau10}
{Kauffmann} J.,  {Pillai} T.,  2010, \mn@doi [\apjl]
  {10.1088/2041-8205/723/1/L7}, \href
  {https://ui.adsabs.harvard.edu/abs/2010ApJ...723L...7K} {723, L7}

\bibitem[\protect\citeauthoryear{{Kauffmann}, {Bertoldi}, {Bourke}, {Evans}  \&
  {Lee}}{{Kauffmann} et~al.}{2008}]{kau08}
{Kauffmann} J.,  {Bertoldi} F.,  {Bourke} T.~L.,  {Evans} II N.~J.,   {Lee}
  C.~W.,  2008, \mn@doi [\aap] {10.1051/0004-6361:200809481}, \href
  {http://adsabs.harvard.edu/abs/2008A%26A...487..993K} {487, 993}

\bibitem[\protect\citeauthoryear{{Kirk}, {Ward-Thompson}, {Palmeirim}  \& {et
  al.}}{{Kirk} et~al.}{2013}]{kir13}
{Kirk} J.~M.,  {Ward-Thompson} D.,  {Palmeirim} P.,   {et al.} 2013, \mn@doi
  [\mnras] {10.1093/mnras/stt561}, \href
  {https://ui.adsabs.harvard.edu/abs/2013MNRAS.432.1424K} {432, 1424}

\bibitem[\protect\citeauthoryear{{K{\"o}nyves}, {Andr{\'e}}, {Men'shchikov}  \&
  {et al.}}{{K{\"o}nyves} et~al.}{2015}]{kon15}
{K{\"o}nyves} V.,  {Andr{\'e}} P.,  {Men'shchikov} A.,   {et al.} 2015, \mn@doi
  [\aap] {10.1051/0004-6361/201525861}, \href
  {http://adsabs.harvard.edu/abs/2015A%26A...584A..91K} {584, A91}

\bibitem[\protect\citeauthoryear{{K{\"o}nyves}, {Andr{\'e}}, {Arzoumanian}  \&
  {et al.}}{{K{\"o}nyves} et~al.}{2020}]{kon20}
{K{\"o}nyves} V.,  {Andr{\'e}} P.,  {Arzoumanian} D.,   {et al.} 2020, \mn@doi
  [\aap] {10.1051/0004-6361/201834753}, \href
  {https://ui.adsabs.harvard.edu/abs/2020A&A...635A..34K} {635, A34}

\bibitem[\protect\citeauthoryear{{Kramer}, {Stutzki}, {Rohrig}  \&
  {Corneliussen}}{{Kramer} et~al.}{1998}]{kra98}
{Kramer} C.,  {Stutzki} J.,  {Rohrig} R.,   {Corneliussen} U.,  1998, \aap,
  \href {http://adsabs.harvard.edu/abs/1998A%26A...329..249K} {329, 249}

\bibitem[\protect\citeauthoryear{{Kroupa}}{{Kroupa}}{2001}]{kro01}
{Kroupa} P.,  2001, \mn@doi [\mnras] {10.1046/j.1365-8711.2001.04022.x}, \href
  {http://adsabs.harvard.edu/abs/2001MNRAS.322..231K} {322, 231}

\bibitem[\protect\citeauthoryear{{Krumholz} \& {McKee}}{{Krumholz} \&
  {McKee}}{2008}]{kru08}
{Krumholz} M.~R.,  {McKee} C.~F.,  2008, \mn@doi [\nat] {10.1038/nature06620},
  \href {https://ui.adsabs.harvard.edu/abs/2008Natur.451.1082K} {451, 1082}

\bibitem[\protect\citeauthoryear{{Ladjelate}, {Andr{\'e}}, {K{\"o}nyves}  \&
  {et al.}}{{Ladjelate} et~al.}{2016}]{lad16}
{Ladjelate} B.,  {Andr{\'e}} P.,  {K{\"o}nyves} V.,   {et al.} 2016, in
  {Jablonka} P.,  {Andr{\'e}} P.,   {van der Tak} F.,  eds,  IAU Symposium Vol.
  315, From Interstellar Clouds to Star-Forming Galaxies: Universal Processes?.
  p.~E46, \mn@doi{10.1017/S1743921316008073}

\bibitem[\protect\citeauthoryear{{Ladjelate}, {Andr{\'e}}, {K{\"o}nyves}  \&
  {et al.}}{{Ladjelate} et~al.}{2020}]{lad20}
{Ladjelate} B.,  {Andr{\'e}} P.,  {K{\"o}nyves} V.,   {et al.} 2020, \mn@doi
  [\aap] {10.1051/0004-6361/201936442}, \href
  {https://ui.adsabs.harvard.edu/abs/2020A&A...638A..74L} {638, A74}

\bibitem[\protect\citeauthoryear{{Lee} et~al.,}{{Lee} et~al.}{2014}]{lee14}
{Lee} K.~I.,  et~al., 2014, \mn@doi [\apj] {10.1088/0004-637X/797/2/76}, \href
  {http://adsabs.harvard.edu/abs/2014ApJ...797...76L} {797, 76}

\bibitem[\protect\citeauthoryear{{Levshakov}, {Henkel}, {Reimers}, {Wang},
  {Mao}, {Wang}  \& {Xu}}{{Levshakov} et~al.}{2013}]{lev13}
{Levshakov} S.~A.,  {Henkel} C.,  {Reimers} D.,  {Wang} M.,  {Mao} R.,  {Wang}
  H.,   {Xu} Y.,  2013, \mn@doi [\aap] {10.1051/0004-6361/201220354}, \href
  {http://adsabs.harvard.edu/abs/2013A%26A...553A..58L} {553, A58}

\bibitem[\protect\citeauthoryear{{Marsh}, {Kirk}, {Andr{\'e}}  \& {et
  al.}}{{Marsh} et~al.}{2016}]{mar16}
{Marsh} K.~A.,  {Kirk} J.~M.,  {Andr{\'e}} P.,   {et al.} 2016, \mn@doi
  [\mnras] {10.1093/mnras/stw301}, \href
  {http://adsabs.harvard.edu/abs/2016MNRAS.459..342M} {459, 342}

\bibitem[\protect\citeauthoryear{{McMullin}, {Mundy}, {Blake}, {Wilking},
  {Mangum}  \& {Latter}}{{McMullin} et~al.}{2000}]{mcm00}
{McMullin} J.~P.,  {Mundy} L.~G.,  {Blake} G.~A.,  {Wilking} B.~A.,  {Mangum}
  J.~G.,   {Latter} W.~B.,  2000, \mn@doi [\apj] {10.1086/308952}, \href
  {http://adsabs.harvard.edu/abs/2000ApJ...536..845M} {536, 845}

\bibitem[\protect\citeauthoryear{{Men'shchikov}, {Andr{\'e}}, {Didelon},
  {Motte}, {Hennemann}  \& {Schneider}}{{Men'shchikov} et~al.}{2012}]{men12}
{Men'shchikov} A.,  {Andr{\'e}} P.,  {Didelon} P.,  {Motte} F.,  {Hennemann}
  M.,   {Schneider} N.,  2012, \mn@doi [\aap] {10.1051/0004-6361/201218797},
  \href {http://adsabs.harvard.edu/abs/2012A%26A...542A..81M} {542, A81}

\bibitem[\protect\citeauthoryear{{Motte}, {Andre}  \& {Neri}}{{Motte}
  et~al.}{1998}]{mot98}
{Motte} F.,  {Andre} P.,   {Neri} R.,  1998, \aap, \href
  {http://adsabs.harvard.edu/abs/1998A%26A...336..150M} {336, 150}

\bibitem[\protect\citeauthoryear{{Nakamura}, {Dobashi}, {Shimoikura}, {Tanaka}
  \& {Onishi}}{{Nakamura} et~al.}{2017}]{nak17}
{Nakamura} F.,  {Dobashi} K.,  {Shimoikura} T.,  {Tanaka} T.,   {Onishi} T.,
  2017, \mn@doi [\apj] {10.3847/1538-4357/aa5ea6}, \href
  {http://adsabs.harvard.edu/abs/2017ApJ...837..154N} {837, 154}

\bibitem[\protect\citeauthoryear{{Offner}, {Clark}, {Hennebelle}, {Bastian},
  {Bate}, {Hopkins}, {Moraux}  \& {Whitworth}}{{Offner} et~al.}{2014}]{off14}
{Offner} S.~S.~R.,  {Clark} P.~C.,  {Hennebelle} P.,  {Bastian} N.,  {Bate}
  M.~R.,  {Hopkins} P.~F.,  {Moraux} E.,   {Whitworth} A.~P.,  2014, \mn@doi
  [Protostars and Planets VI] {10.2458/azu_uapress_9780816531240-ch003}, \href
  {http://adsabs.harvard.edu/abs/2014prpl.conf...53O} {pp 53--75}

\bibitem[\protect\citeauthoryear{{Olmi} \& {Testi}}{{Olmi} \&
  {Testi}}{2002}]{olm02}
{Olmi} L.,  {Testi} L.,  2002, \mn@doi [\aap] {10.1051/0004-6361:20020959},
  \href {http://adsabs.harvard.edu/abs/2002A%26A...392.1053O} {392, 1053}

\bibitem[\protect\citeauthoryear{{Olmi} et~al.,}{{Olmi} et~al.}{2013}]{olm13}
{Olmi} L.,  et~al., 2013, \mn@doi [\aap] {10.1051/0004-6361/201220409}, \href
  {http://adsabs.harvard.edu/abs/2013A%26A...551A.111O} {551, A111}

\bibitem[\protect\citeauthoryear{{Ortiz-Le{\'o}n} et~al.,}{{Ortiz-Le{\'o}n}
  et~al.}{2017}]{ort17}
{Ortiz-Le{\'o}n} G.~N.,  et~al., 2017, \mn@doi [\apj]
  {10.3847/1538-4357/834/2/143}, \href
  {http://adsabs.harvard.edu/abs/2017ApJ...834..143O} {834, 143}

\bibitem[\protect\citeauthoryear{{Ortiz-Le{\'o}n} et~al.,}{{Ortiz-Le{\'o}n}
  et~al.}{2018}]{ort18}
{Ortiz-Le{\'o}n} G.~N.,  et~al., 2018, \mn@doi [\apjl]
  {10.3847/2041-8213/aaf6ad}, \href
  {http://adsabs.harvard.edu/abs/2018ApJ...869L..33O} {869, L33}

\bibitem[\protect\citeauthoryear{{Ossenkopf} \& {Henning}}{{Ossenkopf} \&
  {Henning}}{1994}]{oss94}
{Ossenkopf} V.,  {Henning} T.,  1994, \aap, \href
  {https://ui.adsabs.harvard.edu/abs/1994A&A...291..943O} {291, 943}

\bibitem[\protect\citeauthoryear{{Padoan} \& {Nordlund}}{{Padoan} \&
  {Nordlund}}{2002}]{pad02}
{Padoan} P.,  {Nordlund} {\AA}.,  2002, \mn@doi [\apj] {10.1086/341790}, \href
  {http://adsabs.harvard.edu/abs/2002ApJ...576..870P} {576, 870}

\bibitem[\protect\citeauthoryear{{Palmeirim} et~al.,}{{Palmeirim}
  et~al.}{2013}]{pal13}
{Palmeirim} P.,  et~al., 2013, \mn@doi [\aap] {10.1051/0004-6361/201220500},
  \href {http://adsabs.harvard.edu/abs/2013A%26A...550A..38P} {550, A38}

\bibitem[\protect\citeauthoryear{{Pezzuto}, {Benedettini}, {Di Francesco}  \&
  {et al.}}{{Pezzuto} et~al.}{2020}]{pez20}
{Pezzuto} S.,  {Benedettini} M.,  {Di Francesco} J.,   {et al.} 2020, arXiv
  e-prints, \href {https://ui.adsabs.harvard.edu/abs/2020arXiv201000006P} {p.
  arXiv:2010.00006}

\bibitem[\protect\citeauthoryear{{Piazzo}, {Calzoletti}, {Faustini},
  {Pestalozzi}, {Pezzuto}, {Elia}, {di Giorgio}  \& {Molinari}}{{Piazzo}
  et~al.}{2015}]{pia15}
{Piazzo} L.,  {Calzoletti} L.,  {Faustini} F.,  {Pestalozzi} M.,  {Pezzuto} S.,
   {Elia} D.,  {di Giorgio} A.,   {Molinari} S.,  2015, \mn@doi [\mnras]
  {10.1093/mnras/stu2453}, \href
  {http://adsabs.harvard.edu/abs/2015MNRAS.447.1471P} {447, 1471}

\bibitem[\protect\citeauthoryear{{Pilbratt} et~al.,}{{Pilbratt}
  et~al.}{2010}]{pil10}
{Pilbratt} G.~L.,  et~al., 2010, \mn@doi [\aap] {10.1051/0004-6361/201014759},
  \href {http://adsabs.harvard.edu/abs/2010A%26A...518L...1P} {518, L1}

\bibitem[\protect\citeauthoryear{{Planck Collaboration} et~al.,}{{Planck
  Collaboration} et~al.}{2014}]{abe14}
{Planck Collaboration} et~al., 2014, \mn@doi [\aap]
  {10.1051/0004-6361/201323195}, \href
  {https://ui.adsabs.harvard.edu/abs/2014A&A...571A..11P} {571, A11}

\bibitem[\protect\citeauthoryear{{Poglitsch} et~al.,}{{Poglitsch}
  et~al.}{2010}]{pog10}
{Poglitsch} A.,  et~al., 2010, \mn@doi [\aap] {10.1051/0004-6361/201014535},
  \href {http://adsabs.harvard.edu/abs/2010A%26A...518L...2P} {518, L2}

\bibitem[\protect\citeauthoryear{{Polychroni}, {Schisano}, {Elia}  \& {et
  al.}}{{Polychroni} et~al.}{2013}]{pol13}
{Polychroni} D.,  {Schisano} E.,  {Elia} D.,   {et al.} 2013, \mn@doi [\apjl]
  {10.1088/2041-8205/777/2/L33}, \href
  {http://adsabs.harvard.edu/abs/2013ApJ...777L..33P} {777, L33}

\bibitem[\protect\citeauthoryear{{Prato}, {Rice}  \& {Dame}}{{Prato}
  et~al.}{2008}]{pra08}
{Prato} L.,  {Rice} E.~L.,   {Dame} T.~M.,  2008, {Where are all the Young
  Stars in Aquila?}.
p.~18

\bibitem[\protect\citeauthoryear{{Roccatagliata} et~al.,}{{Roccatagliata}
  et~al.}{2015}]{roc15}
{Roccatagliata} V.,  et~al., 2015, \mn@doi [\aap]
  {10.1051/0004-6361/201425253}, \href
  {http://adsabs.harvard.edu/abs/2015A%26A...584A.119R} {584, A119}

\bibitem[\protect\citeauthoryear{{Roy} et~al.,}{{Roy} et~al.}{2014}]{roy14}
{Roy} A.,  et~al., 2014, \mn@doi [\aap] {10.1051/0004-6361/201322236}, \href
  {https://ui.adsabs.harvard.edu/abs/2014A&A...562A.138R} {562, A138}

\bibitem[\protect\citeauthoryear{{Salpeter}}{{Salpeter}}{1955}]{sal55}
{Salpeter} E.~E.,  1955, \mn@doi [\apj] {10.1086/145971}, \href
  {http://adsabs.harvard.edu/abs/1955ApJ...121..161S} {121, 161}

\bibitem[\protect\citeauthoryear{{Schisano} et~al.,}{{Schisano}
  et~al.}{2014}]{sch14}
{Schisano} E.,  et~al., 2014, \mn@doi [\apj] {10.1088/0004-637X/791/1/27},
  \href {http://adsabs.harvard.edu/abs/2014ApJ...791...27S} {791, 27}

\bibitem[\protect\citeauthoryear{{Schisano} et~al.,}{{Schisano}
  et~al.}{2020}]{sch20}
{Schisano} E.,  et~al., 2020, \mn@doi [\mnras] {10.1093/mnras/stz3466}, \href
  {https://ui.adsabs.harvard.edu/abs/2020MNRAS.492.5420S} {492, 5420}

\bibitem[\protect\citeauthoryear{{Schneider} et~al.,}{{Schneider}
  et~al.}{2013}]{sch13}
{Schneider} N.,  et~al., 2013, \mn@doi [\apjl] {10.1088/2041-8205/766/2/L17},
  \href {http://adsabs.harvard.edu/abs/2013ApJ...766L..17S} {766, L17}

\bibitem[\protect\citeauthoryear{{Seale}, {Looney}, {Wong}, {Ott}, {Klein}  \&
  {Pineda}}{{Seale} et~al.}{2012}]{sea12}
{Seale} J.~P.,  {Looney} L.~W.,  {Wong} T.,  {Ott} J.,  {Klein} U.,   {Pineda}
  J.~L.,  2012, \mn@doi [\apj] {10.1088/0004-637X/751/1/42}, \href
  {https://ui.adsabs.harvard.edu/abs/2012ApJ...751...42S} {751, 42}

\bibitem[\protect\citeauthoryear{{Sewi{\l}o} et~al.,}{{Sewi{\l}o}
  et~al.}{2019}]{sew19}
{Sewi{\l}o} M.,  et~al., 2019, \mn@doi [\apjs] {10.3847/1538-4365/aaf86f},
  \href {https://ui.adsabs.harvard.edu/abs/2019ApJS..240...26S} {240, 26}

\bibitem[\protect\citeauthoryear{{Stamatellos}, {Whitworth}  \&
  {Ward-Thompson}}{{Stamatellos} et~al.}{2007}]{sta07}
{Stamatellos} D.,  {Whitworth} A.~P.,   {Ward-Thompson} D.,  2007, \mn@doi
  [\mnras] {10.1111/j.1365-2966.2007.11999.x}, \href
  {https://ui.adsabs.harvard.edu/abs/2007MNRAS.379.1390S} {379, 1390}

\bibitem[\protect\citeauthoryear{{Strom}, {Grasdalen}  \& {Strom}}{{Strom}
  et~al.}{1974}]{str74}
{Strom} S.~E.,  {Grasdalen} G.~L.,   {Strom} K.~M.,  1974, \mn@doi [\apj]
  {10.1086/152948}, \href {http://adsabs.harvard.edu/abs/1974ApJ...191..111S}
  {191, 111}

\bibitem[\protect\citeauthoryear{{Testi} \& {Sargent}}{{Testi} \&
  {Sargent}}{1998}]{tes98}
{Testi} L.,  {Sargent} A.~I.,  1998, \mn@doi [\apjl] {10.1086/311724}, \href
  {http://adsabs.harvard.edu/abs/1998ApJ...508L..91T} {508, L91}

\bibitem[\protect\citeauthoryear{{Testi}, {Sargent}, {Olmi}  \&
  {Onello}}{{Testi} et~al.}{2000}]{tes00}
{Testi} L.,  {Sargent} A.~I.,  {Olmi} L.,   {Onello} J.~S.,  2000, \mn@doi
  [\apjl] {10.1086/312858}, \href
  {http://adsabs.harvard.edu/abs/2000ApJ...540L..53T} {540, L53}

\bibitem[\protect\citeauthoryear{{White}, {Casali}  \& {Eiroa}}{{White}
  et~al.}{1995}]{whi95}
{White} G.~J.,  {Casali} M.~M.,   {Eiroa} C.,  1995, \aap, \href
  {http://adsabs.harvard.edu/abs/1995A%26A...298..594W} {298, 594}

\bibitem[\protect\citeauthoryear{{Winston} et~al.,}{{Winston}
  et~al.}{2007}]{win07}
{Winston} E.,  et~al., 2007, \mn@doi [\apj] {10.1086/521384}, \href
  {https://ui.adsabs.harvard.edu/abs/2007ApJ...669..493W} {669, 493}

\bibitem[\protect\citeauthoryear{{Zhang}, {Laureijs}, {Clark}  \&
  {Wesselius}}{{Zhang} et~al.}{1988}]{zha88}
{Zhang} C.~Y.,  {Laureijs} R.~J.,  {Clark} F.~O.,   {Wesselius} P.~R.,  1988,
  \aap, \href {http://adsabs.harvard.edu/abs/1988A%26A...199..170Z} {199, 170}

\bibitem[\protect\citeauthoryear{{Zucker}, {Speagle}, {Schlafly}, {Green},
  {Finkbeiner}, {Goodman}  \& {Alves}}{{Zucker} et~al.}{2019}]{zuc19}
{Zucker} C.,  {Speagle} J.~S.,  {Schlafly} E.~F.,  {Green} G.~M.,  {Finkbeiner}
  D.~P.,  {Goodman} A.~A.,   {Alves} J.,  2019, \mn@doi [\apj]
  {10.3847/1538-4357/ab2388}, \href
  {https://ui.adsabs.harvard.edu/abs/2019ApJ...879..125Z} {879, 125}

\makeatother
\end{thebibliography}




\appendix
\section{A catalogue of dense cores identified with \emph{Herschel} in the Serpens region} \label{app:cat}
Based on our \emph{Herschel} SPIRE/PACS parallel-mode imaging survey of the Serpens~Main and Aquila~East star-forming region, we identified a total of \nsources dense cores, see Table~\ref{tab:cat}.
The catalogue of observed properties of all of these \emph{Herschel} cores is available in online Table~\ref{tab:es1}. 
A template of this online catalog is provided below to illustrate its form and content.
The derived properties (physical radius, mass, SED dust temperature, peak column density at the resolution of the 500~$\mu$m data, average column density, peak volume density, and average density) are given in online Table~\ref{tab:es2} for each core. 
A portion of this online table is also provided below. 
The derived properties of the \emph{Herschel}-detected protostars and YSOs will be published in a forthcoming paper.

\begin{sidewaystable*}[htb]\tiny\setlength{\tabcolsep}{2.5pt}
{\renewcommand{\arraystretch}{0.5}
\begin{tabular}{|r|c|c c c c c c c c c c} 
\hline
\hline
 rNO      & Core name         &  RA$_{\rm 2000}$ &  Dec$_{\rm 2000}$        & Sig$_{\rm 070}$ &  $S^{\rm peak}_{\rm 070}$ &  $S^{\rm peak}_{\rm 070}$/$S_{\rm bg}$ &  $S^{\rm conv,500}_{\rm 070}$ &  $S^{\rm tot}_{\rm 070}$ &  FWHM$^{\rm a}_{\rm 070}$ &  FWHM$^{\rm b}_{\rm 070}$ &  PA$_{\rm 070}$  \\ 
          & HGBS\_J*          &  (h m s)         &  (\degr~\arcmin~\arcsec) &                 & (Jy/beam)                 &                                        & (Jy/beam$_{\rm 500}$)         &  (Jy)                    &  (\arcsec)                &  (\arcsec)                &  (\degr)         \\         
 (1)      & (2)               &  (3)             &  (4)                     &  (5)            &      (6) ~ $\pm$ ~ (7)    &  (8)                                   & (9)                           &    (10) ~ $\pm$ ~ (11)   &  (12)                     &  (13)                     &  (14)            \\         
\hline
$\cdots$  &                   &                  &                          &                 &                           &                                        &                               &                          &                           &                           &                  \\
    2	 & 182713.5+003453 & 18:27:13.55 &+00:34:52.9             &    0.0          &   3.01e-02 ~ 1.7e-02     &     0.90                   &  4.22e-01          & 4.86e-01 ~ 2.8e-01       &  57                   & 49                 & 43 \\      
    $\cdots$  &                   &                  &                          &                 &                           &                                        &                               &                          &                           &                           &                  \\

    6	 & 182731.6+010601 & 18:27:31.61 &+01:06:01.3             &  15.0           &  3.17e-01 ~ 2.5e-02      &     4.47                   &  4.40e-01            & 4.04e-01 ~ 3.1e-02      &  11                   &  8                & $-46$ \\                     
$\cdots$  &                   &                  &                          &                 &                           &                                        &                               &                          &                           &                           &                  \\
     
    9	  & 182808.5-000107&  18:28:08.50 &-00:01:07.0             & 16.3           &  4.04e-01 ~ 2.3e-02      &  18.49                     &  5.45e-01           & 4.48e-01 ~ 2.5e-02      &  11                    &  8                & 36\\      

\hline
\end{tabular}
}
\scalebox{1.2}{$\sim$}
\vspace{0.2cm}

\scalebox{1.2}{$\sim$}
{\renewcommand{\arraystretch}{0.5}
\begin{tabular}{c c c c c c c c c  c c c c c c c} 
\hline
\hline
 Sig$_{\rm 160}$ &  $S^{\rm peak}_{\rm 160}$    &  $S^{\rm peak}_{\rm 160}$/$S_{\rm bg}$ &  $S^{\rm conv,500}_{\rm 160}$ &  $S^{\rm tot}_{\rm 160}$ &  FWHM$^{\rm a}_{\rm 160}$ &  FWHM$^{\rm b}_{\rm 160}$ &  PA$_{\rm 160}$  &  Sig$_{\rm 250}$ &  $S^{\rm peak}_{\rm 250}$ &  $S^{\rm peak}_{\rm 250}$/$S_{\rm bg}$ &  $S^{\rm conv,500}_{\rm 250}$ &  $S^{\rm tot}_{\rm 250}$ &  FWHM$^{\rm a}_{\rm 250}$ &  FWHM$^{\rm b}_{\rm 250}$ &  PA$_{\rm 250}$  \\ 
                 &  (Jy/beam)                   &                                        & (Jy/beam$_{\rm 500}$)         &  (Jy)		    &  (\arcsec)		&  (\arcsec)		    &  (\degr)         &		  & (Jy/beam)		      & 				       & (Jy/beam$_{\rm 500}$)         &  (Jy)  		  &  (\arcsec)  	      &  (\arcsec)		  &  (\degr)	     \\       
  (15)           &      (16) ~ $\pm$ ~ (17)     &   (18)                                 &  (19)   			 &    (20) ~ $\pm$ ~ (21)   &  (22)                     & (23)                      & (24)             & (25)             &    (26) ~ $\pm$ ~ (27)    &  (28)                                  &  (29)                         &   (30) ~ $\pm$ ~ (31)    &  (32)                     &  (33)                     &  (34)            \\        
\hline
$\cdots$         &                              &                                        &                               &			    &				&			    &		       &		  &			      & 				       &                               &			  &			      & 			  &		     \\
7.977           & 2.36e-01 ~ 4.4e-02       &     0.32       &  8.10e-01                     &  1.39e+00 ~ 2.6e-01       &   37          & 24            & 20            &  14.2           &  6.41e-01 ~ 5.1e-02       &    0.45                  & 1.13e+00                     &  2.27e+00 ~ 1.8e-01    &  31               & 27             & 18 \\     
$\cdots$         &                              &                                        &                               &			    &				&			    &		       &		  &			      & 				       &                               &			  &			      & 			  &		     \\

24.410          & 9.10e-01 ~ 3.2e-02       &     1.23               &  9.91e-01             &  7.52e-01 ~ 2.7e-02       &  14           & 14            &  $-46$        &  15.3           &  4.12e-01 ~ 2.6e-02       &     0.33                 &  3.95e-01                    &  3.24e-01 ~ 2.1e-02     &  18              & 18             & $-31$ \\     
$\cdots$         &                              &                                        &                               &			    &				&			    &		       &		  &			      & 				       &                               &			  &			      & 			  &		     \\

15.510          & 4.83e-01 ~ 3.2e-02       & 0.63                   &  5.38e-01             &  3.64e-01 ~ 2.4e-02       & 14            & 14            & 47            &  15.2           &  4.19e-01 ~ 5.8e-02       &    0.31                  &  4.02e-01                    &  3.07e-01 ~ 4.3e-02     &  18              & 18             & $-87$ \\      
$\cdots$         &                              &                                        &                               &			    &				&			    &		       &		  &			      & 				       &                               &			  &			      & 			  &		     \\
\hline
\end{tabular}
}

\scalebox{1.2}{$\sim$}
{\renewcommand{\arraystretch}{0.5}
\begin{tabular}{cccccccccccccccc}  
\hline
\hline
Sig$_{\rm 350}$  &  $S^{\rm peak}_{\rm 350}$ &  $S^{\rm peak}_{\rm 350}$/$S_{\rm bg}$ &  $S^{\rm conv,500}_{\rm 350}$ &  $S^{\rm tot}_{\rm 350}$  &  FWHM$^{\rm a}_{\rm 350}$ &  FWHM$^{\rm b}_{\rm 350}$ &  PA$_{\rm 350}$   &  Sig$_{\rm 500}$ &  $S^{\rm peak}_{\rm 500}$ &  $S^{\rm peak}_{\rm 500}$/$S_{\rm bg}$ &  $S^{\rm tot}_{\rm 500}$ &  FWHM$^{\rm a}_{\rm 500}$ &  FWHM$^{\rm b}_{\rm 500}$  &  PA$_{\rm 500}$   \\
                 &  (Jy/beam)                &                                        & (Jy/beam$_{\rm 500}$)         &  (Jy)			  &  (\arcsec)  	      &  (\arcsec)		  &  (\degr)	      & 		 & (Jy/beam)		     &  				      &  (Jy)                    &  (\arcsec)                &  (\arcsec)                 &  (\degr)          \\	       
  (35)           &   (36) ~ $\pm$ ~ (37)     &  (38)                                  &  (39)                         &    (40) ~ $\pm$ ~ (41)    &  (42)                     &  (43)                     &  (44)             &  (45)            &    (46) ~ $\pm$ ~ (47)    &  (48)                                  &    (49) ~ $\pm$ ~ (50)   &  (51)                     &  (52)                      &  (53)             \\	
\hline
$\cdots$         &                           &                                        &                               & 			  &			      & 			  &		      & 		 &			     &  				      &                          &                           &                            &                   \\
15.490      &6.57e-01 ~ 6.9e-02       &    0.49            &  7.88e-01                      &  1.33e+00 ~ 1.4e-01   &  33                & 28         &  25          &     13.0         &   4.94e-01 ~ 8.4e-02      &    0.41                   & 6.37e-01 ~ 1.1e-01   &  38                & 36                & 64 \\              
$\cdots$         &                           &                                        &                               & 			  &			      & 			  &		      & 		 &			     &  				      &                          &                           &                            &                   \\

6.156       & 1.57e-01 ~ 3.3e-02      &     0.14                    &  1.55e-01             &  1.17e-01 ~ 2.5e-02   &   25               & 25          &  0           &  0.0             &   1.82e-02 ~ 2.0e-02      &    0.02                   &  2.58e-03 ~ 2.8e-03  &   36               & 36                &  1 \\      
$\cdots$         &                           &                                        &                               & 			  &			      & 			  &		      & 		 &			     &  				      &                          &                           &                            &                   \\

9.497       &  2.74e-01 ~ 5.1e-02    &   0.21                       & 2.72e-01              &  1.85e-01 ~ 3.5e-02   &   25               & 25          &      $-27$   &      6.2         &  1.64e-01 ~ 4.4e-02       &    0.14                   & 9.88e-02 ~ 2.6e-02   &  36                & 36                & $-75$ \\      
$\cdots$         &                           &                                        &                               & 			  &			      & 			  &		      & 		 &			     &  				      &                          &                           &                            &                   \\
\hline
\end{tabular}
}
\scalebox{1.2}{$\sim$}
\vspace{0.2cm}

\scalebox{1.2}{$\sim$}
{\renewcommand{\arraystretch}{0.5}
\begin{tabular}{cccccccccccccccc|}  
\hline
\hline
Sig$_{\rm N_{H_2}}$& $N^{\rm peak}_{\rm H_2}$ &  $N^{\rm peak}_{\rm H_2}$/$N_{\rm bg}$ &  $N^{\rm conv,500}_{\rm H_2}$ &  $N^{\rm bg}_{\rm H_2}$  &  FWHM$^{\rm a}_{\rm N_{H_2}}$  &  FWHM$^{\rm b}_{\rm N_{H_2}}$ & PA$_{\rm N_{H_2}}$  &  N$_{\rm SED}$  &   Core type      &  Comments	\\
                   & (10$^{21}$ cm$^{-2}$)    &                                        & (10$^{21}$ cm$^{-2}$)         &  (10$^{21}$ cm$^{-2}$)   &  (\arcsec)   &  (\arcsec)                    &  (\degr)            &                 &  		           &              \\ 
 (54)              & (55)                     &  (56)                                  & (57)                          &  (58)                    &  (59)                          &  (60)                         &  (61)               &    (62)         &   (63)	     & (64)		\\
\hline													        
$\cdots$           &      $\cdots$            &      $\cdots$                          &       $\cdots$                &    $\cdots$              &                  $\cdots$       &        $\cdots$               &         $\cdots$    &      $\cdots$   &  	$\cdots$  &  $\cdots$ \\
27.020             &    2.521                 & 0.79                                   &    1.07                       &  3.182                   &  30                             & 25                            & 11 &4               &    0            &  prestellar  &                      \\ 
$\cdots$           &                          &                                        &                               &                          &                                 &                               &                     &                 &  		      &  		      		\\
0.094              &   -0.003                 & -0.00                                  & -0.001                        &  2.167                   &  18                             & 18                            & 53                  & 4               &   starless   &                      \\                        $\cdots$           &                          &                                        &                               &                          &                               &                               &                     &                 &  		     &  		      		\\

13.110            &  0.963           &   0.33                      &   0.267               &  2.920           &  21              & 18                 &$-68$        &  5      &   protostellar   &   2MASS J18280848-0001065\\                 
$\cdots$           &       $\cdots$                   &        $\cdots$                                &     $\cdots$                          &       $\cdots$                   &         $\cdots$                      &          $\cdots$                     &      $\cdots$               &    $\cdots$             &  $\cdots$		     &  	$\cdots$	      &	$\cdots$	\\
\hline
\end{tabular}
}

 \caption{Catalog of dense cores identified in HGBS maps of Serpens (template, full catalog only provided online). \protect \footnote{Catalog entries are as follows: 
{(1)} Core running number;
{(2)} Core name $=$ HGBS\_J prefix directly followed by a tag created from the J2000 sexagesimal coordinates; 
{(3)} and {(4)}: Right ascension and declination of core center; 
{(5)}, {(15)}, {(25)}, {(35)}, and {(45)}: Detection significance from monochromatic single scales, in the 70~$\mu$m, 160~$\mu$m, 250~$\mu$m, 350~$\mu$m, and 500~$\mu$m maps, respectively. 
(NB: the detection significance has the special value of $0.0$ when the core is not visible in clean single scales); 
{(6)}$\pm${(7)}, {(16)}$\pm${(17)} {(26)}$\pm${(27)} {(36)}$\pm${(37)} {(46)}$\pm${(47)}: Peak flux density and its error in Jy/beam as estimated by \textsl{getsources};
{(8)}, {(18)}, {(28)}, {(38)}, {(48)}: Contrast over the local background, defined as the ratio of the background-subtracted peak intensity to the local background intensity ($S^{\rm peak}_{\rm \lambda}$/$S_{\rm bg}$); 
{(9)}, {(19)}, {(29)}, {(39)}: Peak flux density measured after smoothing to a 36.3$\arcsec$ beam; 
{(10)}$\pm${(11)}, {(20)}$\pm${(21)}, {(30)}$\pm${(31)}, {(40)}$\pm${(41)}, {(49)}$\pm${(50)}: Integrated flux density and its error in Jy as estimated by \textsl{getsources}; 
{(12)}--{(13)}, {(22)}--{(23)}, {(32)}--{(33)}, {(42)}--{(43)}, {(51)}--{(52)}: Major \& minor FWHM diameters of the core (in arcsec), respectively, 
as estimated by \textsl{getsources}. (NB: the special value of $-1$ means that no size measurement was possible); 
{(14)}, {(24)}, {(34)}, {(44)}, {(53)}: Position angle of the core major axis, measured east of north, in degrees; 
{(54)} 
Detection significance in the high-resolution column density image;  
{(55)} 
Peak H$_{2}$ column density in units of $10^{19}$ cm$^{-2}$ as estimated by \textsl{getsources} in the high-resolution column density image; 
{(56)} 
Column density contrast over the local background, as estimated by \textsl{getsources} in the high-resolution column density image;
{(57)} 
Peak column density measured in a 36.3$\arcsec$ beam; 
{(58)} Local background H$_{2}$ column density 
as estimated by \textsl{getsources} in the high-resolution column density image; 
{(59)}--{(60)}--{(61)}: Major \& minor FWHM diameters of the core, and position angle of the major axis, respectively,
as measured in the high-resolution column density image; 
{(62)} Number of {\it Herschel} bands in which the core is significant (Sig$_{\rm \lambda} >$ 5) and has a positive flux density, excluding the column density plane; 
{(63)} Core type: 1-starless, for gravitationally unbound cores; 2-prestellar, 3-protostellar, or 0-tentative core.; 
{(64)} Comments: ``N region''.}} 
\label{tab:es1}
\end{sidewaystable*}

\clearpage

\begin{sidewaystable*}[htb]\tiny\setlength{\tabcolsep}{6.8pt}
\label{tab2cat}
{\renewcommand{\arraystretch}{0.9}
\begin{tabular}{|r|c|c c c c c c c c c c c c|} 
\hline \hline
 rNO     & Core name         &  RA$_{\rm 2000}$ &  Dec$_{\rm 2000}$        & $R_{\rm core}$        &  $M_{\rm core}$   &  $T_{\rm dust}$   &  $N^{\rm peak}_{\rm H_2}$ &  $N^{\rm ave}_{\rm H_2}$   &   $n^{\rm peak}_{\rm H_2}$  &  $n^{\rm ave}_{\rm H_2}$   &  $\alpha_{\rm BE}$  & Core type  & Comments  \\
         & HGBS\_J*          &  (h m s)         &  (\degr~\arcmin~\arcsec) & (pc)                  &  ($M_\odot$)      &  (K)              &  (10$^{21}$ cm$^{-2}$)    &  (10$^{21}$ cm$^{-2}$)     &   (10$^{4}$ cm$^{-3}$)      &  (10$^{4}$ cm$^{-3}$)      &                     &		  &	       \\   
 (1)     & (2)               &  (3)             &   (4)                    & (5) ~~~~~~ (6)        &  (7) $\pm$ (8)    &  (9) $\pm$ (10)   &  (11)                     &  (12) ~~~~ (13)            &   (14)                      &  (15) ~~~ (16)             &         (17)        &    (18)    &    (19)    \\   
\hline
$\cdots$ &   $\cdots$                &  $\cdots$                &    $\cdots$                      &        $\cdots$               &       $\cdots$            &    $\cdots$               &    $\cdots$                       &    $\cdots$                        &	     $\cdots$                     &      $\cdots$                      &        $\cdots$             &   $\cdots$         &	 $\cdots$      \\ 
    2	 &  182713.5+003453  & 18:27:13.55      & +00:34:52.9             & 4.06e-02~5.50e-02      &  0.352 ~ 0.037    & 13.5 ~  0.3       &   2.166               & 3.03  ~  1.65              &  0.892                       &    1.81 ~   0.73           &  3.085              &   prestellar &      tentative \\
    $\cdots$ &   $\cdots$                &$\cdots$                  &   $\cdots$                       &    $\cdots$                   &    $\cdots$               &          $\cdots$         &  $\cdots$                         &   $\cdots$                         &	 $\cdots$                         &    $\cdots$                        &     $\cdots$                &     $\cdots$       &	  $\cdots$     \\ 

    6	 & 182731.6+010601   & 18:27:31.61      & +01:06:01.3             & 1.24e-02 ~ 3.71e-02    & 0.007 ~ 0.001     & 24.5 ~ 0.1        &   0.009               & 0.66 ~ 0.07                &  0.004                       &   1.29 ~ 0.05              &  83.763             & starless     &           -\\
    $\cdots$ &    $\cdots$               &$\cdots$                  &   $\cdots$                       &   $\cdots$                    &     $\cdots$              &        $\cdots$           &   $\cdots$                        &   $\cdots$                         &	 $\cdots$                         &              $\cdots$              &  $\cdots$                   &     $\cdots$       &	 $\cdots$      \\ 

    9	 & 182808.5-000107   & 18:28:08.50      & $-00$:01:07.0           & 1.43e-02 ~ 3.97e-02    &    0.002 ~ 0.001  & 29.4 ~ 0.7        &   0.551               & 0.13 ~ 0.02                &  0.227                       &   0.23 ~ 0.01              & 432.324             &  protostellar&           -\\
$\cdots$ & $\cdots$                  & $\cdots$                 & $\cdots$                         &    $\cdots$                   &    $\cdots$               &    $\cdots$               &    $\cdots$                       &    $\cdots$                        &	              $\cdots$            &                       $\cdots$     & $\cdots$                    &    $\cdots$        &	  $\cdots$     \\ 
\hline
\end{tabular}
}

\caption{Derived properties of dense cores identified in HGBS maps of Serpens (template, full table only provided online).
\protect \footnote{Table entries are as follows: {(1)} Core running number; {(2)} Core name $=$ HGBS\_J prefix directly followed by a tag created from the J2000 sexagesimal coordinates; 
{(3)} and {(4)}: Right ascension and declination of core center; 
{(5)} and {(6)}: Geometrical average between the major and minor FWHM sizes of the core (in pc), as measured in the high-resolution column density map after deconvolution from the 18.2$\arcsec$ HPBW resolution of the map and before deconvolution, respectively.
(NB: Both values provide estimates of the object's outer radius when the core can be approximately described by a Gaussian distribution, as is the case 
for a critical Bonnor-Ebert spheroid); 
{(7)} Estimated core mass ($M_\odot$) assuming the dust opacity law advocated by \citet{roy14}; 
{(9)} SED dust temperature (K); {(8)} \& {(10)} Statistical errors on the mass and temperature, respectively, including calibration uncertainties, but excluding dust opacity uncertainties; 
{(11)} Peak H$_2$ column density, at the resolution of the 500~$\mu$m data, derived from a 
modified black-body SED fit to the core peak flux densities measured in a common 36.3$\arcsec$ beam at all wavelengths; 
{(12)} Average column density, calculated as $N^{\rm ave}_{\rm H_2} = \frac{M_{\rm core}}{\pi R_{\rm core}^2} \frac{1}{\mu m_{\rm H}}$, where $M_{\rm core}$ is the estimated core mass (col. { 7}), $R_{\rm core}$ the estimated core radius prior to deconvolution (col. { 6}), and $\mu = 2.8$;
{(13)} Average column density calculated in the same way as for col. { 12} but using the deconvolved core radius (col. { 5});  
{(14)} Beam-averaged peak volume density at the resolution of the 500~$\mu$m data, derived from the peak column density (col. { 11}) assuming a Gaussian spherical distribution: $n^{\rm peak}_{\rm H_2} = \sqrt{\frac{4 \ln2}{\pi}} \frac{N^{\rm peak}_{\rm H_2}}{\overline{FWHM}_{\rm 500}}$; 
{(15)} Average volume density, calculated as $n^{\rm ave}_{\rm H_2} = \frac{M_{\rm core}}{4/3 \pi R_{\rm core}^3} \frac{1}{\mu m_{\rm H}}$, using the estimated core radius prior to deconvolution; 
{(16)} Average volume density, calculated in the same way as for col. { 15} but using the deconvolved core radius (col. { 5}); 
{(17)} Bonnor-Ebert mass ratio: $\alpha_{\rm BE} = M_{\rm BE,crit} / M_{\rm obs} $ (see text for details); 
{(18)} Core type: 1-starless, for gravitationally unbound cores, 2-prestellar, 3-protostellar, or 0-tentative core. The latter is a likely extragalactic source (see comments);
{(19)} Comments: ``no SED fit'', ``tentative bound'', ``N region''.
}}
\label{tab:es2}

\end{sidewaystable*}

\bsp	
\label{lastpage}
\end{document}